\documentclass[mathleft
]{an}
\usepackage{graphicx}
\usepackage{times}
\begin{document}

% The following seven commands are intended for editorial usage and should be ignored by
% the author(s).
\Pagespan{1}{}% Document's page range.
% If second parameter is left empty, the last page is computed automatically.
\Yearpublication{}%
\Yearsubmission{2011}%
\Month{}%
\Volume{}%
\Issue{}%
% \DOI{This.is/not.aDOI}%

\title{Studying the dwarf galaxies in nearby groups of galaxies: spectroscopic and photometric data}

\author{U. Hopp\inst{1,2}\fnmsep\thanks{\email{hopp@usm.lmu.de}\newline}
\and  J. Vennik\inst{3}
}
\titlerunning{Dwarf galaxies in nearby groups of galaxies}
\authorrunning{U. Hopp \& J. Vennik}
\institute{
Universit\"ats--Sternwarte M\"unchen, Scheiner Str. 1, D 81679 M\"unchen, Germany
\and
Max--Planck--Institut f\"ur Extraterrestrische Physik,D 85741 Garching, Germany
\and
Tartu Observatory, 61602 T\~oravere, Tartumaa, Estonia}
\received{}
\accepted{}
%\publonline{later}

%\keywords{galaxies: dwarf -- galaxies: distances and redshifts -- galaxies: clusters: individual
%(IC 65 group, NGC 6962 group, NGC 5005/5033 group)}

\abstract{
%{\it Context} 
Galaxy evolution by interaction driven transformation is probably 
highly efficient in groups of galaxies. Dwarf galaxies with their 
shallow potential are expected to reflect the interaction most 
prominently in their observable structure. 
%{\it Aims} 
The major aim of this series of papers is to establish a data base 
which allows to study the impact of group interaction onto the 
morphology and star-forming properties of dwarf galaxies.
%{\it Methods} 
Firstly, we present our selection rules for target groups and the 
morphological selection method of target dwarf member candidates. 
Secondly, the spectroscopic follow-up observations with the HET 
are present. Thirdly, we applied own reduction methods based on 
adaptive filtering to derive surface photometry of the candidates.
%{\it Results} 
The spectroscopic follow-up indicate a dwarf identification success 
rate of roughly 55\%, and a group member success rate of about 33\%. 
A total of 17 new low surface brightness members is presented. For 
all candidates, total magnitudes, colours, and light distribution
parameters are derived and discussed in the context of scaling 
relations. We point out short comings of the SDSS standard pipeline 
for surface photometry for these dim objects.
%{\it Conclusion}
We conclude that our selection strategy is rather efficient to 
obtain a sample of dim, low surface brightness members of groups of 
galaxies within the Virgo super-cluster. The photometric scaling 
relation in these X-ray dim, rather isolated groups does not 
significantly differ from those of the galaxies within the Local Volume.
}
\keywords{galaxies: dwarf -- galaxies: distances and redshifts -- 
galaxies: clusters: individual
(IC~65 group, NGC~697 group, NGC~6278 group, NGC~6962 group, 
NGC~5005/5033 group)}

\maketitle

\section{Introduction}

Groups of galaxies are excellent places to study the evolution of galaxies 
triggered by external processes. This includes all kind of interactions
with the surroundings, starting from wide to nearby fly-bys of (or: encounters with) other
group galaxies, the fly-bys of in-falling galaxies, the merging of equally
sized galaxies and the accretion of dwarf galaxies (hereafter DGs) by massive galaxies. This
also includes interaction of the individual galaxies with the overall group 
gravitational potential or with a homogeneously distributed gas\-eous component. 
Several of these processes like e.g. ram pressure stripping or harassment 
(e.g. Moore et al. \cite{moore98}) have been discussed in the theoretical 
literature, mostly ba\-sed on detailed simulations (see earlier reviews e.g. 
in Bar\-nes \& Hernquist \cite{bh92}; Mamon \cite{mamon93}, \cite{mamon07}; 
Lake \& Moore \cite{lake99}). \\
The individual galaxies can respond to these interactions in various ways,
among them bursts of star formation if the gaseous content gets condensed
and shocked, or quenching of star formation after a removal of that gas 
(Mayer et al. \cite{mayer00}). The shape of the galaxy can be temporarily 
or permanently be changed, yielding a more irregular appearance, tidal 
features like arms, higher order structural parameters up to the complete 
disruption of a galaxy and/or formation of tidal DGs 
and of the intergalactic matter (IGM), depending on the mass ratios of 
interacting galaxies (e.g. Toomre \& Toomre \cite{toomre1972},
Barnes \& Hernquist \cite{bh92}, Weinberg \cite{weinberg97}, 
Henriques et al. \cite{henriques08}, or for a more recent view the 
contributions in Smith et al. \cite{smith2010}).  

In general, the low-luminosity dwarf galaxy population is outnumbering their brighter counterparts 
by far (e.g. Ei\-gen\-thaler \& Zei\-lin\-ger~\cite{ez09}). The number of 
DGs in groups and clusters is puzzling, however. Lambda-Cold-Dark-Mat\-ter 
($\Lambda$CDM) cos\-mo\-lo\-gy suggests more DGs than observed, an issue of the 
so-called missing satellite problem (Klypin et al.~\cite{klypin99}). 
Due to the much smaller gravitational potential, DGs are 
normally expected to respond stronger on 
interactions than massive ga\-la\-xies. One of the most drastic Local Group 
examples is the Sgr dwarf (SagDEG) which is in a stage of disruption and 
accretion by the Milky Way galaxy (Johnston et al. \cite{johnston95}, 
Ibata et al. \cite{ibata01a}). Another illustrative example of a later stage 
of disruption is probably the great stream in the Andromeda galaxy (see e.g. 
Fardal et al. \cite{fardal08} for a recent discussion and references therein). 
The Magellanic stream between the LMC and SMC
and its long tail are normally understood as a response of these relatively
large dwarf galaxies to the interaction with the Milky Way. Despite these
 clear examples, observations of the - meanwhile roughly fifty 
(Whiting et al. \cite{whiting07}, Belokurov et al. \cite{belokurov07}) - 
dwarf galaxies of the Local Group yielded relatively little direct evidence
for interaction driven response of DGs. There exist also only
few indirect hints like the local mor\-pho\-logy-density relation (or 
morphological segregation) of the dwarf spheroidal and irregular galaxies 
found around the Milky Way and M31 (first discussed by 
Einasto et al. \cite{je1974}, \cite{je1975}, for a recent discussion 
see e.g. Grebel et al. \cite{grebel03} ).  
More distant groups can step in to provide good examples for the
interaction history of small galaxies, see e.g. the recent discussion by
Gallagher (\cite{gal2010}) who focuses on minor interactions (mass ratios
below 1 to 10). A special aspect of dwarf galaxy interactions within groups
is the relatively high probability of multiple interactions (D'Onghia \&
Lake \cite{dong2008}).

The current galaxy evolution paradigm of hierarchical clustering in a
$\Lambda$CDM universe rises the expectation of relatively frequent interactions
in groups of galaxies and a large number of accretions of small galaxies,
including the accretion of the early building blocks of the galaxy population. 
That so relatively few examples can be easily found inside our very local 
surrounding possibly points - beside the observational difficulties with some more
modest responses - to short time scales of many transformations triggered by
galaxy interactions. To enlarge the sample of responses which can be studied
in detail, one thus has to leave the Local Group despite the fact that this
shrinks the resolution and sets the flux limits to higher absolute values
(for given instrumentation). Recently, a number of detailed studies of more
distant individual groups have been presented (e.g. Chiboucas et al. \cite{chib09} 
on the M81 group, Cote et al. \cite{cote09}, and Crnojevic et al. \cite{crn09} on 
the Cen A group, Trentham \& Tully \cite{tt09a} on the NGC~1023 group, 
Karachentsev et al. \cite{kara13} on the groups in Ursa Majoris,
Tully \& Trentham \cite{tt08} on the NGC~5353/4 group, 
Mahdavi et al. \cite{mah05} on the NGC~5846 group, 
Gr\"utzbauch et al. \cite{gruetz05} on the NGC~4756 group).  

The detection of dwarf satellite population in groups and studying their 
photometric and structural properties, is an essential part of each of 
such group studies. However, detailed photometry and spectroscopy of 
low-luminosity ga\-la\-xies is challenging. Only the groups within the 
Local Volume ($\sim 10$ Mpc) and a few other nearby galaxy 
aggregates to some extent have been studied photometrically 
(e.g. Bremnes et al.~\cite{bremnes98}, \cite{bremnes99}, 
\cite{bremnes00} Parodi et al.~\cite{parodi02}, Jerjen et al.~\cite{jerjen00}) 
and kinematically (e.g. Karachentsev et al.~\cite{ikar99}, \cite{kara13}). 

Besides studying the galaxy properties in possibly part\-ly 
dynamically relaxed central regions of dense groups, extended infall 
regions around the groups should also be investigated for signs of 
the group-growth-history through infalling sub-units.
The mass of an infall region could be of order 20-120\% of the virial mass 
(Rines et al.~\cite{rines03}). Wetzel et al. (\cite{wetzel13}) 
argue that the environmental effects on the galaxy' properties can be traced 
on 'ejected' or 'back-splash' galaxies out to several virial radii around 
massive clusters as well as on the DGs around loose groups. 
Recently, several examples of dwarf galaxies in 
such in-falling sub\-groups have been detected and investigated 
(Tully et al \cite{tt08}, Cortese et al. \cite{A1367}).
In the extended regions around the groups significant perturbations 
to the global Hubble flow have been found, which could be interpreted 
either as an effect of deceleration due to the group mass, in 
the $\Omega_0$=0 ($\Lambda$=0) cosmology (Sandage~\cite{sandage86}), 
or as an effect of local dark energy, in the $\Lambda$CDM cosmology 
(Teerikorpi et al.~\cite{teeri08}, Hartwick~\cite{hart11}, 
Chernin et al.~\cite{chernin12}). 

This is the second paper in a small series, where we pre\-sent our study of 
a small sample of relatively nearby groups of galaxies. The first paper 
of this series (Vennik \& Hopp \cite{vh08}) presented a detailed surface photometry of
the known and probable members of the compact group of galaxies around IC~65. 
The paper included a kinematical study of that group. The goal of present paper 
is to detect new (main\-ly dwarf) group members, verify their true group 
membership by means of spectroscopic redshift measurements and constrain 
their photometric properties as a function of environment.
We present the selection of the groups in chapter 2 and the selection 
of additional dwarf galaxy candidates in chapter 3. In chapter 4, we 
discuss our attempts to confirm these candidates based on radial velocities 
with dedicated observations at the Hobby-Eberly telescope (hereafter HET).
The fifth chapter discusses and summarises the results of our spectroscopic 
and photometric (i.e. observational) studies, while we leave the detailed 
analysis of the studied (five) groups for the next paper.

Throughout this paper, we use the concordance cosmology ($\Omega_{m}$ = 0.3,
$\Omega_{\Lambda}$ = 0.7, and h = 0.7), and all magnitudes are given in the 
AB system unless explicitly noted. 

\section{Group selection}

\begin{table*}[t]
\caption{List of studied groups. Columns contain the following data: (1) group name, (2) name of the parent galaxy, (3) number of
group members from the original catalogue, (4, 5) R.A. and DEC of the group centre, (6) the mean velocity corrected for the LG motion,
 (7) dispersion of radial velocities within the virial radius, (8) the group virial radius,
 (9) number of bright galaxies with $M_{g'} < -17$ and total number of galaxies 
(in brackets) within the group virial radius, 
(10) number density of $n_{200}$ galaxies within the group volume, which was approximated by a triaxial spheroid, 
as defined in Tully (\cite{LScl}),    
(11) virial mass of the group, (12) zero-velocity radius of the group. }
\protect\label{tab:groups}
\begin{center}
\begin{small}
\begin{tabular}{lcccccccccccc}
\hline
Group & Parent galaxy & n$_{gal}$ & R.A. & DEC & $v_{LG}$ & $\sigma_{v,200}$& $R_{200}$ & $n_{200}$ & $\rho_{200}$ & $M_{200}$ & $R_0$\\
                 &&& [2000]  & [2000] &  \multicolumn{2}{c}{[km s$^{-1}$]} & [Mpc]&& [Mpc$^{-3}$] & [$10^{12} M_\odot$] & [Mpc]  \\
\hline
 (1) & (2) & (3) & (4) & (5) & (6) & (7) & (8) & (9) & (10) & (11) & (12) \\
\hline
\\
LGG 016 & IC 65   &  3 & 01:00:30 & 48:48.6 & 2965 & 75 & 0.20 & 3~(6) & 1.1~(2.2) & 0.8  & 0.8 \\
LGG 034 & NGC 697 & 10 & 01:50:37 & 21:54.2 & 3046 & 160 & 0.40 & 8~(10) & 0.2~(0.3) & 7.1 & 1.7 \\
LGG 334 & NGC 5005/33 & 16 & 13:11:42 & 36:42.5 & 975 & 95 & 0.26 & 5~(15) & 0.2~(0.5) & 1.6 & 1.0 \\
WBL 629 & NGC 6278 & 3 & 17:00:42 & 23:01.7 & 3036 & 133 & 0.33 & 4~(8) & 0.5~(0.9) & 4.1 & 1.4 \\
WBL 666 & NGC 6962 & 7 & 20:47:21 & 00:21.1 & 4115 & 224 & 0.55 & 14~(32) & 0.4~(0.8) & 19.2 & 2.3 \\
\hline
\end{tabular}
\end{small}
\end{center}
\end{table*}

\begin{figure*}[t]
\vspace{-30mm}
\hspace{-13mm}    
\resizebox{0.37\textwidth}{!}{\includegraphics{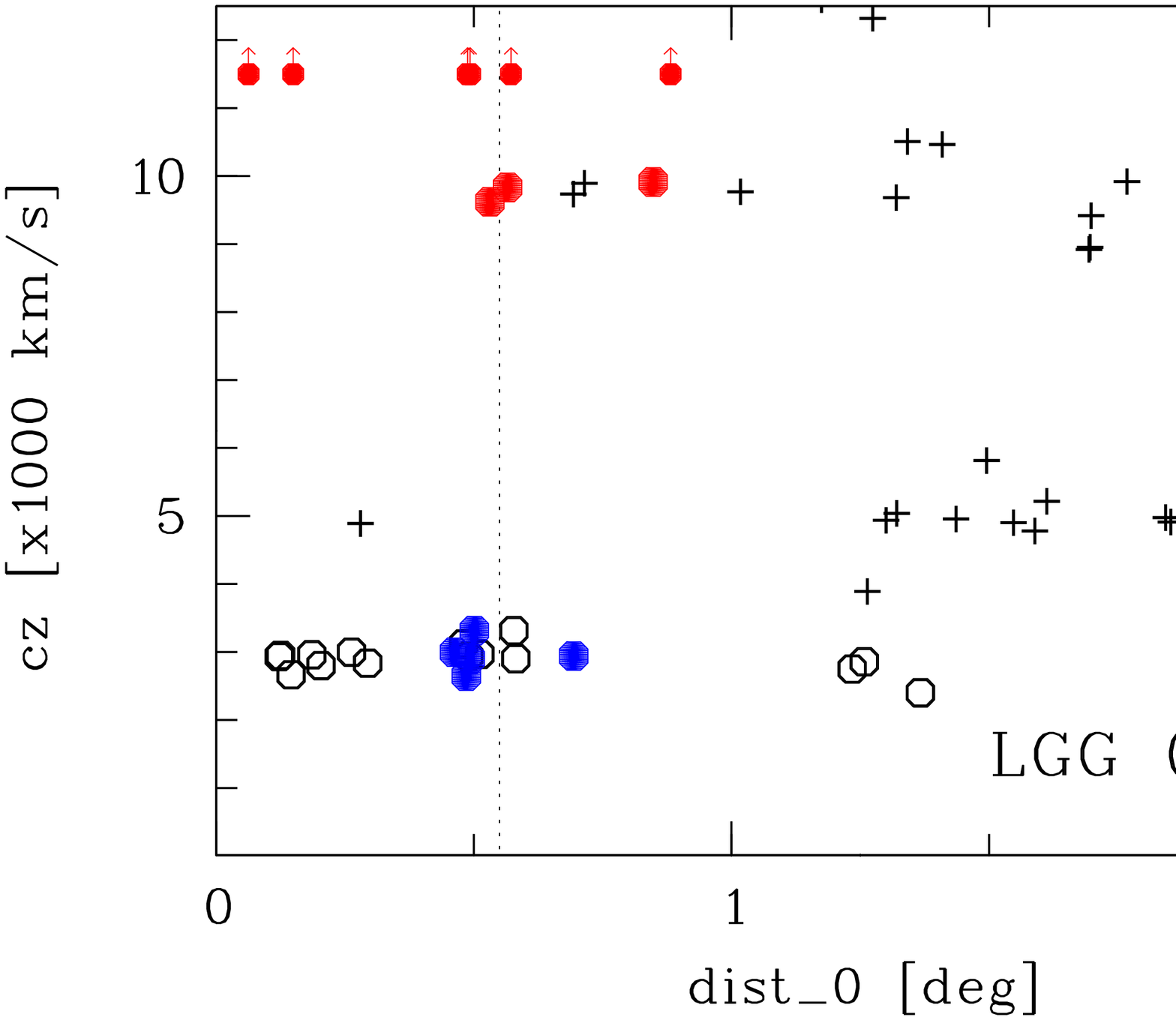}}
\hspace{1mm}  
\resizebox{0.37\textwidth}{!}{\includegraphics{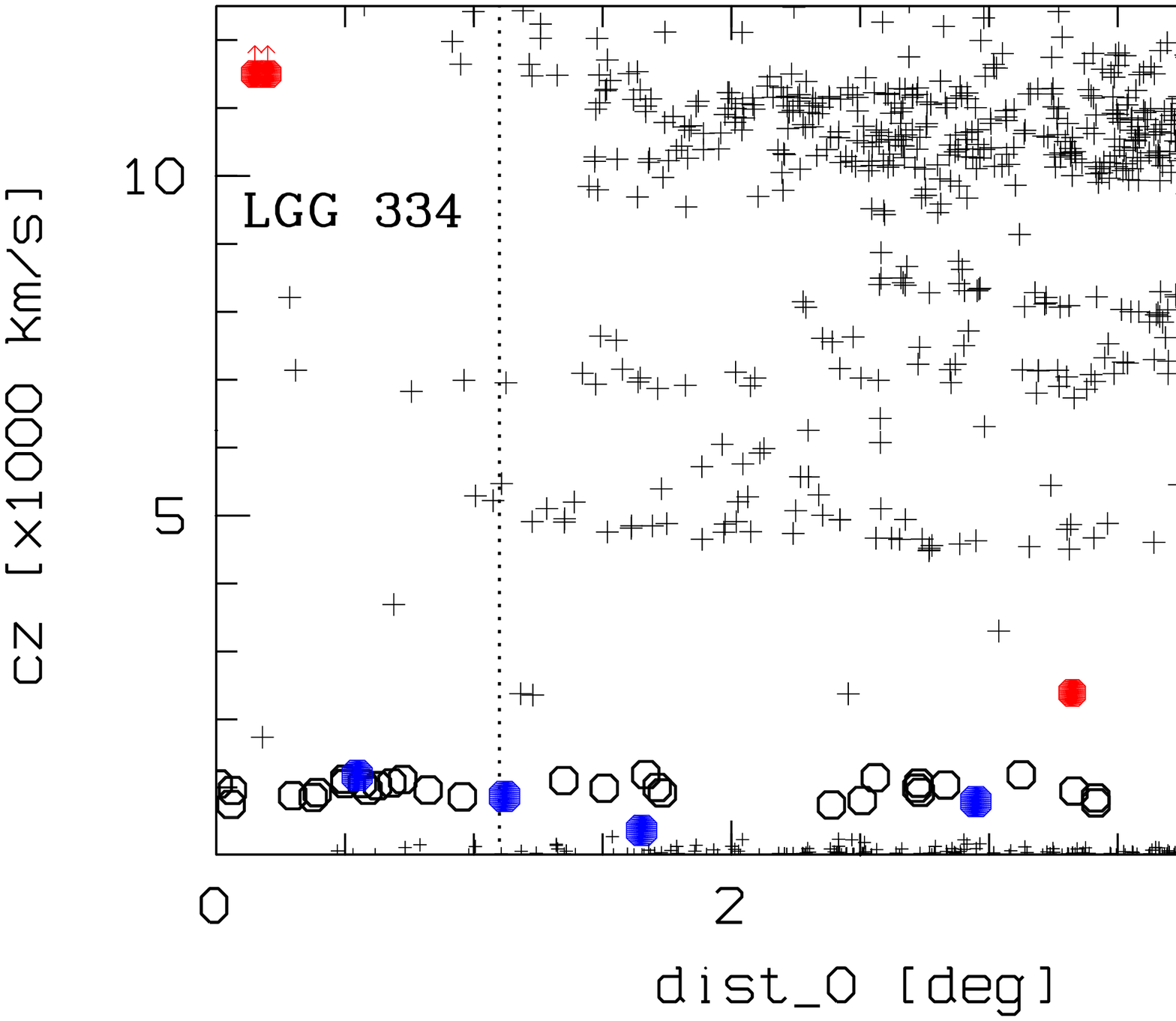}}
\vspace{-32mm}
\hspace{3mm}           
\resizebox{0.37\textwidth}{!}{\includegraphics{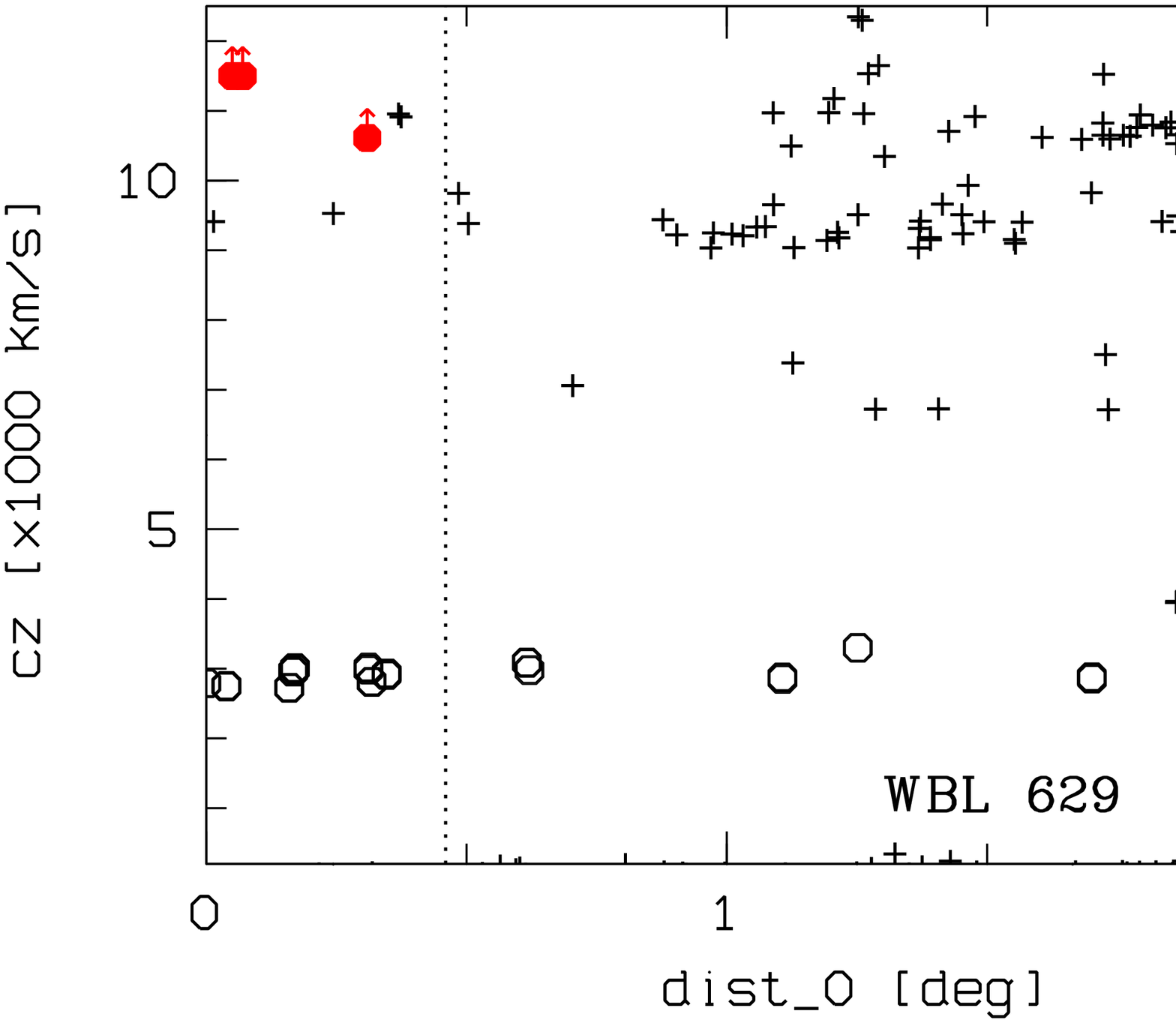}}
\hspace{22mm}
\resizebox{0.37\textwidth}{!}{\includegraphics{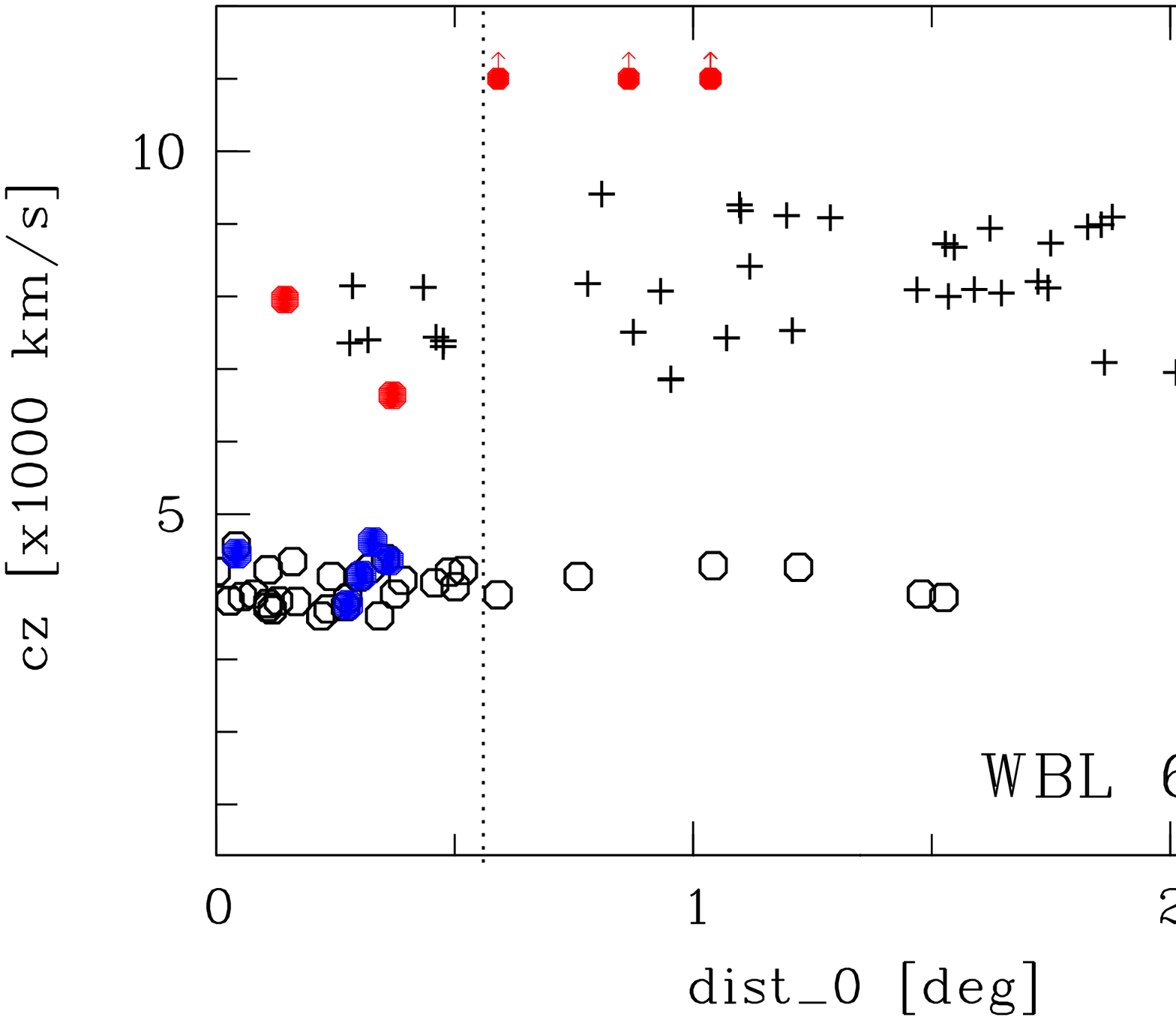}}
\caption{Distribution of galaxies with known redshifts (cz) in dependence 
of their group-centric angular distance (dist\_0) towards the NGC~697 
group (or LGG~034, top left), NGC~5005/5033 group (or LGG~334, top right), 
NGC~6278 group (or WBL~629, lower left), and NGC~6962 group (or WBL~666,
lower right). The group members 
within the virial radius (dotted line) and potential infalling galaxies 
within the zero-velocity radius (dashed line) are marked with hexagons. 
The galaxies seen in projection to the group are marked with crosses. 
New redshifts, measured in the present study, are marked with blue 
(new group members) and red (background galaxies) filled hexagons 
(in on-line version).
}
\label{fig:4groups}
\end{figure*}

Most of the galaxies and galaxy groups in and near
to the Local Super-cluster (hereafter Scl) are distributed in sheets and filaments. Thus, problems
arise from contamination between separate structures along a common line
of sight. Since this kind of background contamination effects the
iden\-ti\-fi\-ca\-tion of faint dwarf galaxies most prominently, we have preferentially selected
target groups where the group foreground is almost empty and the background galaxies 
are clearly separated in redshift by $\Delta cz \geq 2000~km~s^{-1}$ (see Fig.~1), 
i.e. we are seeing the large-scale structures (sheets) nearly face-on. 
The isolation of selected groups on the sky is less clearly defined. 
The number density of galaxies in the potentially virialized central regions of the groups 
$\rho_{200} \simeq 0.5 Mpc^{-3}$ (Table 1) exceeds the median galaxy density in the Nearby Galaxy Catalogue (Tully \cite{NBG}, NBG)
as well as the mean density of the clouds in the Local Scl (Tully~\cite{LScl}) by a factor of about two. 
However, there are galaxies with concordant redshifts, sparsely populating the group surroundings within the 
zero-velocity radius. These galaxies are potentially in-falling to the group along a sheet/filament. 
With the aim of studying environmentally conditioned galaxy pro\-perties in and around the groups, 
we avoid to set strict constraints on the group isolation on the sky, and will consider  
all galaxies within the group zero-velocity radius.

As further group selection criteria, the groups should be (1) located preferentially in the redshift range
between $2000 - 4000~km~s^{-1}$, in order to be able to cover the group
area by imaging studies with reasonable number of pointing's, but still
in the distance-range, where dwarf galaxies could have been distinguished
and classified using structural and photometric criteria, and (2) prefer groups less studied or un-studied earlier.

Here we describe a set of five optically selected groups, which all remain undetected as sources of 
diffuse (extended) X-ray emission, when cross-correlating their coordinates with the ROSAT pointed mode 
(PSPC) observations. That means, the groups have a diffuse X-ray flux less than $3 \sigma$ above the 
background level, which nearly corresponds to the upper flux limit of $\log L_X  \leq 40.7~ erg~s^{-1}$ 
(Osmond \& Ponman \cite{GEMS}, Mulchaey et al. \cite{MDMB}).
The lack of group-scale X-ray emission is common for a class of spiral-rich groups and could probably mean that 
those groups are not yet fully collapsed, but are sites of recent dynamical evolution (Osmond \& Ponman \cite{GEMS}).

We note that a few galaxies in selected groups are detected in the ROSAT all-sky survey as X-ray point sources, 
i.e. they probably host AGNs\footnote{E.g. the luminous galaxies with an AGN in the LGG~334 group - NGC~5005 (LINER) and NGC~5033 (Sy 1.9), and the 
compact pair of galaxies NGC~6962/6964 in the WBL~666, which has a bolometric X-ray luminosity of 
$L_X = 1.03~10^{41}~erg~s^{-1}$ (Henriksen et al. \cite{henriksen99}.)}. 

Initial group membership has been obtained 
from published catalogues by Vennik (\cite{jv84}), Garcia (\cite{LGG}, LGG), 
White et al. (\cite{WBL}, WBL). The lists have been supplemented with new 
galaxies with concordant redshifts from the SDSS and/or from the NED, located 
within the zero-velocity radius, which separates the group from the 
Hubble flow. 

The zero-velocity (or infall) radius of each group was estimated, according 
to Sandage (\cite{sandage86}) as:
%\noindent
$$R_0[Mpc] = (\frac{8 G T_0^2}{\pi^2}M_{\rm vir})^{1/3} \simeq 0.88 (M_{200} [10^{12}M_{\odot}])^{1/3},$$
where $T_0 = 13.7$ Gyr is the age of the Universe, and 
$$M_{200} = 3~G^{-1}~R_{200}~\sigma_{v,200}^2$$ 
%\noindent
 is the mass within the standard (virial) radius 
%\noindent
$$R_{200} = \sqrt{3} \sigma_{v,200}/(10 H_0),$$ 
within which the (galaxy) density exceeds the mean density by a factor of 200 
(Carlberg et al. \cite{carl97}). 
The standard radius $R_{200}$ and the velocity dispersion within that 
radius $\sigma_{v,200}$ were estimated iteratively. The characteristics 
discussed above for selected groups are summarised in Table \ref{tab:groups}.          
The distribution of galaxies with known redshifts around the four groups 
is shown in Fig.~\ref{fig:4groups} (the LGG 016/IC 65 group has been 
studied earlier in Vennik \& Hopp~\cite{vh08}). Isolation properties of 
each group are shortly described below.   

{\bf The NGC~697 group (LGG~034)} (Garcia~\cite{LGG}) 
is a local density enhancement in a filamentary structure, which appears to
connect the (canonical) Pisces-Perseus Super-cluster 
(PPS, Haynes \& Giovanelli \cite{PPS}) ridge with the Local Scl, and the 
group is located to the South of the PPS main ridge.
The group is behind of a local mini-void with the depth of $\sim 5$ Mpc 
(Karachentsev et al. \cite{kara02}). The major portion of the foreground 
between the Local Scl and PPS is found to be empty and the NGC~697 group 
area is essentially devoid of galaxies with measured redshifts
up to $cz \sim$ 10000 km/s (see top left of Fig.~\ref{fig:4groups}).

{\bf The NGC~5005/5033 group (LGG 334)}. 
This group is a locally isolated condensation of late-type galaxies in a 
sheet-like structure in the central region of the Local Scl. Despite of a 
crowded background (with $cz \ge$ 4500 km/s), the group itself is well-defined 
along the line-of-sight (see Fig.~\ref{fig:4groups}, top right). There are 
several sub-condensations of galaxies with concordant redshifts within the 
group infall region.

{\bf The NGC~6278 group (WBL~629)} (White et al.~\cite{WBL})  
is located behind the Local Void (Tully \& Fisher \cite{LVoid}), known also 
as the Hercules-Aquila Void in the direction of RA = $18^h 38^m$, DEC = +18$^o$ 
(Karachentseva et al. \cite{kara99}). This void has been found to be 
completely empty of known galaxies, including dwarf galaxies, 
with $cz < 1500$ km/s, and containing only few known galaxies in between 
1500 $< cz <$ 2500 km/s. Unfortunately, this group is partly projected to 
the outskirts of the Hercules Scl ($cz \ge$ 9000 km/s), which 
potentially rises the contamination problem from\\ crowded background 
(but see Sec. 3.2.2).

{\bf The NGC~6962 group (WBL~666)} 
is a conspicuous condensation in a ridge or filament of galaxies, 
surrounded by several local ($z \simeq$ 0.015) mini-voids, e.g. Aquila, 
Pegasus and Microscopum voids (Fairall et al. \cite{fair94}). The group 
foreground is almost empty as the group is projected onto the outskirts 
of the Local Void. Another sheet/filament is crossing the  group background 
at $cz \simeq$ 8000 km/s. This group is favourably located within 
the $\sim 3^o$ wide Stripe 82 of the SDSS, which provides co-added images 
of $\sim$ 2 mag deeper than SDSS in general. 

{\bf The IC~65 group (LGG~16)} 
is a poor compact assembly of four late-type galaxies, located $\sim 15^o$ 
to the North of the PPS main ridge. There are several filaments extending 
to the North of the PPS main ridge, but the group background is nearly 
empty of galaxies with measured redshift. However, our observations have 
shown that there are a number of nearby DGs, 
loosely distributed in front of this group, and probably associated with 
the luminous spiral galaxy NGC~278.

\section{Dwarf galaxy candidate selection}
\label{sec:sec_sele}

The first step to supplement the membership of the original groups 
was to query the SDSS\footnote{http://www.sdss.org/} and the 
NED\footnote{http://ned.ipac.caltech.edu/} for all galaxies within 
the group infall radius $R_0$ and with redshifts $\pm$~500 km/s around the 
mean velocity of the corresponding group. As a result the membership 
of the NGC~6278 and the NGC~5005/5033 groups, as well that of 
the central area of the NGC~6962 group within $\sim R_{200}$ should 
be almost complete within the completeness limit of the SDSS 
spectroscopic survey of $r' \le 17.77$ mag. Thus, the completeness 
of the group galaxy membership lists extend into the domain of dwarf 
galaxies ($M_{r'} \ge -17$) by $\sim 1$ mag (NGC~6962 group) 
to $\sim 4$ mag (NGC~5005/5033 group). However, the NGC~697 group 
is only partly covered by the SDSS DR8 imaging survey, and its 
redshift data are not yet available for the full area of the group. 
Finally, the IC~65 group area is outside the SDSS survey footprint 
and its members could be only selected from the inhomogeneous 
NED database. \\
Next, we have searched for new  dwarf member candidates using the 
DPOSS\footnote{http://www.astro.caltech.edu/~george/dposs/} blue and 
red frames, SDSS multicolour imaging data (and frames), as well as 
our own imaging studies of the IC~65 group. The various imaging data 
bases for\-ced slightly different selection techniques as described in 
the following.

\subsection{Candidate selection on the DPOSS (IC~65 group)}

We carried out a parametric search for new dwarf galaxies within 
an area of 1$^o \times 1^o$ around the centre of the IC 65 group 
applying the SExtractor software (Bertin \& Ar\-nouts~\cite{Sxtr}) 
to the DPOSS blue and red frames, which were previously linearised 
and photometrically calibrated. To di\-sentangle the possible dwarf 
members of the group from distant galaxies in SExtracted catalogue
we used two criteria, firstly the Binggeli's (\cite{bing94}) 
empirical relation between the central surface brightness (SB) 
and absolute magnitude, common both for the dE and dIrr galaxies, 
and secondly, the empirical light concentration parameter 
(Trentham et al.~\cite{trent01}), versus colour index. More details 
about the dwarf selection and classification procedure in the IC~65 
group is given in Vennik \& Hopp (\cite{vh08}). As result, we have 
selected four LSB irregular galaxies with lowest light concentration 
and the bluest colour (0.45 $< B-R<$ 1.05) as the highest rated 
dwarf member candidates for the follow-up spectroscopy with the HET. 

\subsection{Candidate selection with the SDSS data} 

The other four groups of our small sample are either completely, 
i.e. within the infall radius $R_0$ (NGC~6278 and \\ NGC~5005/5033), 
or partly, i.e. within the virial radius $R_{200}$ (NGC~697 and NGC~6962), 
visible in the SDSS DR7. We have searched for new dwarf member 
candidates of these groups using the homogeneous multi-colour imaging 
data of the SDSS.

\subsubsection{Automatic searches towards NGC~6962}
 
First we have queried the $PhotoObjAll$ catalogue within a radius 
up to $R_0 \simeq 2.4^o$ around the centre of the NGC~6962 group, 
for objects, which have: (1) low surface brightness with 
$\mu_{\mathrm {ef}, g'} > 22.0$ g~arcsec$^{-2}$; (2) low light concentration 
with $petroR50/petroR90 > 0.4$; (3) isophotal diameter of $isoA > 15''$, 
which corresponds to the linear diameter $\sim$ 4 kpc at the distance 
of the NGC~6962 group. All pre-selected dwarf candidates have been 
visually inspected on the SDSS and DPOSS frames and final rating 
of the group membership probabilities (rated \lq 1 \rq - probable group member, 
rated \lq 2 \rq - possible member, and rated \lq 3 \rq - probable 
background galaxy) has been made on their morphological and colour grounds. 

We have retrieved 29 probable and 44 possible members of the NGC~6962 
group within $R_{200} \simeq 0.55$ Mpc around the parent galaxy NGC~6962.
We have studied the clustering properties among various sub-samples  
comparing the angular 2-point correlation properties of true (confirmed 
by redshift) and candidate members of the NGC~6962 group. We found 
that the "reference correlation" for 22 true members is significantly 
more peaked than the functions describing correlation properties for 
different sub-samples of the candidates, even that for the ensemble 
of rated \lq 1 \rq candidates. A simple model sample consisting of a mix of 
55\% of galaxies in the "reference sample" and 45\% of randomly 
distributed objects was a good description of the correlation properties 
of rated \lq 1 \rq galaxies (Vennik \& Tago~\cite{vt07}). Therefore, we admit 
that even among the highest rated candidates there may be nearly 50\% 
of projecting background galaxies. However, the results obtained by 
the HET spectroscopy (Section 4), confirm our ability to distinguish 
between the dwarf group members and distant LSB galaxies even in 
outskirts the Local Scl.

\subsubsection{Visual candidate selection for the remaining groups}

Our exercises with the parametric search for the LSB dwarf galaxies 
in the IC~65 and NGC~6962 groups have shown that, despite of the 
advantage of having the well-defined selection criteria, they will 
inevitably result in an automatic selection of a large number of 
different kinds of spurious objects and defects, which could have 
been rejected only by a careful visual follow-up inspection of blue 
and red DPOSS or SDSS frames. Furthermore, the parametric (automatic) 
search in the other three groups would have been more complicated 
because: \\
(1) the NGC~697 group is only partly covered by SDSS DR7 survey, 
and combination of parametric searches on the SDSS and DPOSS would 
introduce additional uncertainties;\\ 
(2) the NGC~6278 group has a crowded background due to projection 
to the outskirts of the Hercules Scl, which again complicates the 
automatic selection procedure;\\
(3) the nearest group around the NGC~5005/5033 is rather extended 
in the sky and, consequently, a much too large number of candidates 
would have been selected by automatic procedure.

Therefore, we decided for the visual search for LSBGs, as a pilot 
project, preceding the more sophisticated parametric selection 
(with well defined completeness limits), which would result in 
selection of a reasonable number of most probable dwarf galaxy 
candidates for the follow-up long-slit spectroscopy.

We carried out a visual search for the LSB dwarf galaxy candidates of these three groups in two iterations.
Firstly, we have selected the most probable (highest rated) dwarf candidates on the SDSS frames (supplemented by DPOSS frames 
for the NGC~697 group) and compared their photometric characteristics (colour, surface brightness, light concentration), 
obtained from the SDSS photometric catalogue, with those of the full sample with $cz \le 15000~km~s^{-1}$ within this area.
We found that the dwarf galaxy candidates, selected in the first round, are best separated from the background (field) 
objects in the colour index - surface brightness parameter space, where the dwarf candidates are distributed in a sector of blue 
colours ($g'-r' < 0.5$) and of low surface brightness ($<\mu_{ef}>~ \ge 22.5~g'~arcsec^{-2}$). 
Specifically, for the NGC~6278 group surroundings, only $\sim 7\%$ of the Sloan galaxies within the survey area and with $cz < 15000~km~s^{-1}$, 
are distributed in this parametric range. This indicates that the contamination from the Hercules Scl is not particularly 
serious. 

In the second round of selecting dwarf galaxy candidates, we scrutinised other galaxies within this parametric range, which were not selected 
in the first iteration. The final decision of their membership probabilities was based on their morphological appearance.

As a result, we have selected 6 targets in the central 
$\sim 0.5^o$ (0.4 Mpc) area of the NGC~6278 group, which  
include only one probable dwarf galaxy with several luminous clumps, 
identified earlier as kkr~31 by Karachentseva et al.~(\cite{kara99})), 
and other five selected objects being classified as possible 
(small and diffuse but nucleated) dwarf galaxies.\\
For the NGC~5005/5033 group of galaxies, our survey resulted in 
selection of five probable and eight possible dwarf galaxy 
candidates within a radius of $\sim 6^o$ (1.4 Mpc) around the group centre.\\
In the NGC~697 group we have classified 13 targets within the 
search radius of $R_{200} \simeq 0.5^o$ (0.4 Mpc) as follows: 
three LSB irregular galaxies were rated \lq 1\rq, six compact, 
bluish targets - rated \lq 2\rq, and four small, diffuse but 
nucleated targets - rated \lq3\rq. 

\section{Spectroscopic confirmation}

\subsection{SDSS and NED}

As already discussed, the original group member lists were 
supplemented by querying the SDSS and NED data base for additional 
further members of the five groups. This process was repeated 
after we finished the selection of member candidates from
imaging data as described above (chap. \ref{sec:sec_sele}). 
Objects with literature redshifts available form the two archives 
were not included into the follow-up observing program.

\subsection{Hobby-Eberly telescope observations}

\begin{table}[t]
\caption{Used setup of the Marcario Low-Resolution Spectrograph 
(Hill et al.~\cite{LRS}). Column (3) lists the resolution of the
setup and column (4) the rms scatter of the linear dispersion coefficient
of the wavelength fit.}
\protect\label{tab:LRS}
\begin{center}
\begin{small}
\begin{tabular}{lccc}
\hline
Setup    & Property & Resolution & rms scatter\\
         &          &  [nm]      &    [nm]     \\
\hline
 (1) & (2) & (3) & (4)\\
\hline
\\
Long slit       & 4 arc min   &  & \\
Filter          & GG385       &  & \\
\hline
Grism g1$^a$    &  410 - 800 nm&  & \\
slit width      &  2.0 arcsec & 1.67 & 0.002 \\
\hline
Grism g2$^a$    &  430 - 730 nm&  & \\
slit width      &  1.0 arcsec & 0.41 & 0.0001 \\
                &  2.0 arcsec & 0.82 & \\     
\hline
\end{tabular}
\end{small}
\end{center}
$a$ used wavelength range given in column (2)\\
\end{table}

The prime focus station Low-Resolution Spectrograph (Hill et al.~\cite{LRS}, LRS) of the 
Hobby-Eberly Telescope (HET) (Ramsey et al.~\cite{HET}) 
was used through several observing seasons to
verify the group membership of dwarf galaxy candidates. A few objects
which we had classified as background objects were also included.
The LRS is a focal reducer system which enables pre-imaging for the 
acquisition of the objects and the slit centring and offers various
spectral resolution (see Tab.~\ref{tab:LRS}). 
We started the program with the higher resolution grism 
to obtain also kinematical parameters of 
the galaxies. This turned out to be too ambitious for many of our 
low surface brightness targets and we changed to lower resolution
(see Tab.~\ref{tab:LRS} and Tab.~\ref{tab:speclog}).

Finally, the most efficient set-up was the one with grism g1 which
take the full advantage of the HET queue scheduling 
(Shetrone et al.~\cite{shetrone07}). Its relatively low resolution
and thus less efficient OH-sky line subtraction in the CCD near infrared
was not balanced by the higher resolution of the other grisms.
In most cases, the long-slit was oriented along the
apparent major axis of the candidate to collect as many photons as possible. 
In some case, this strategy was somehow compromised to either include 
bright knots seen in the relevant candidate, or to set a nearby foreground
star onto the slit which ease the observational procedure and thus helps
to keep the overhead time low. 
Here, the correct positioning of the
slit onto the brightest parts of the candidate by fixing the slit position
on stars easily visible in the short acquisition images of the LRS helps
a lot to improve the signal-to-noise ratio.
With the typical apparent extension of the 
objects and their relatively smooth light distributions, seeing (or 
image quality due to the segmented mirror stack) influences the 
observations only marginally, thus, an upper limit of 3.0$''$ was set. 

\begin{figure*}
%\resizebox{0.48\textwidth}{!}{\includegraphics{Hopp_Vennik_fig2a.ps}}
\resizebox{0.48\textwidth}{!}{\includegraphics{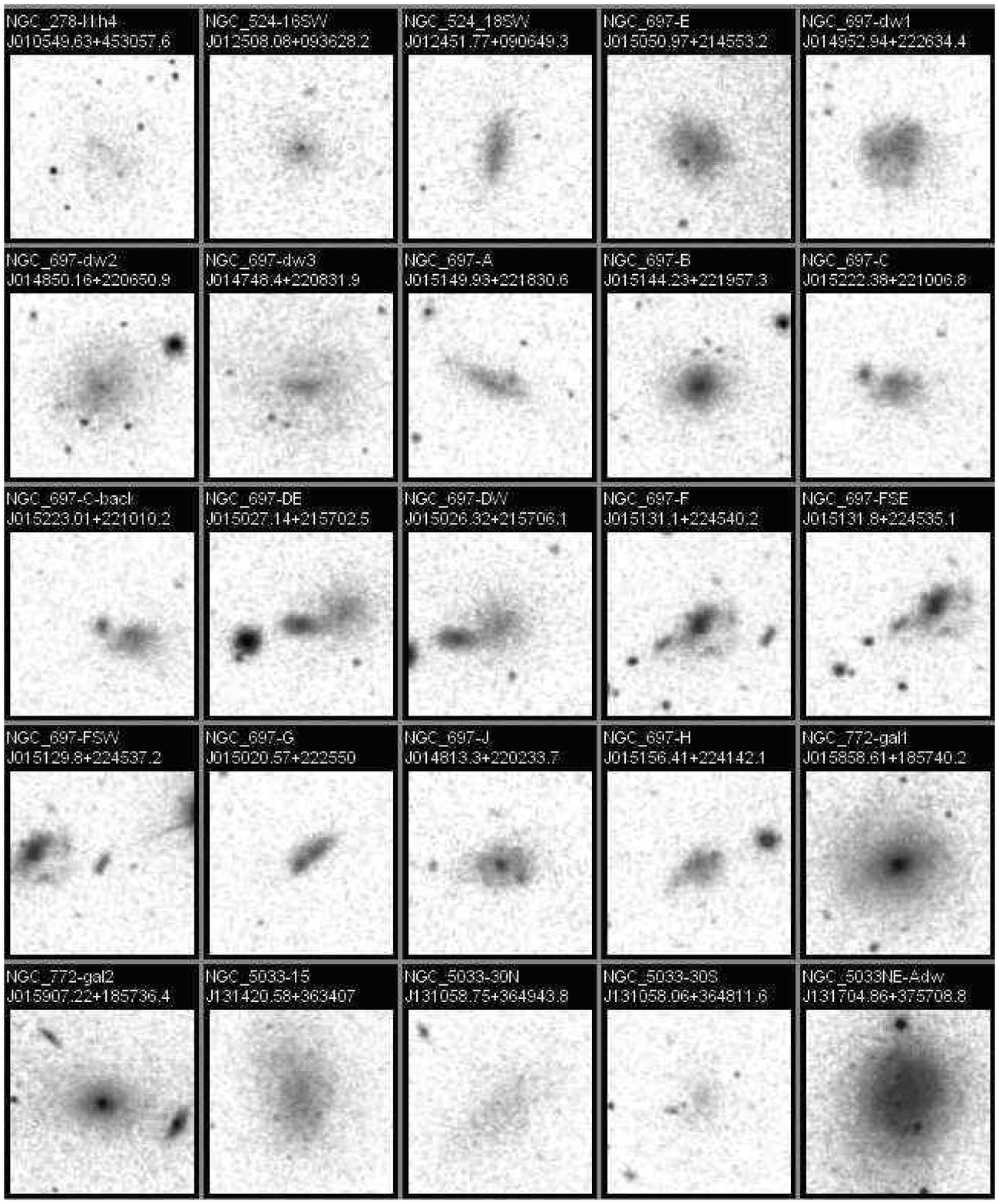}}
\hspace{3mm}
%\resizebox{0.48\textwidth}{!}{\includegraphics{Hopp_Vennik_fig2b.ps}}
\resizebox{0.48\textwidth}{!}{\includegraphics{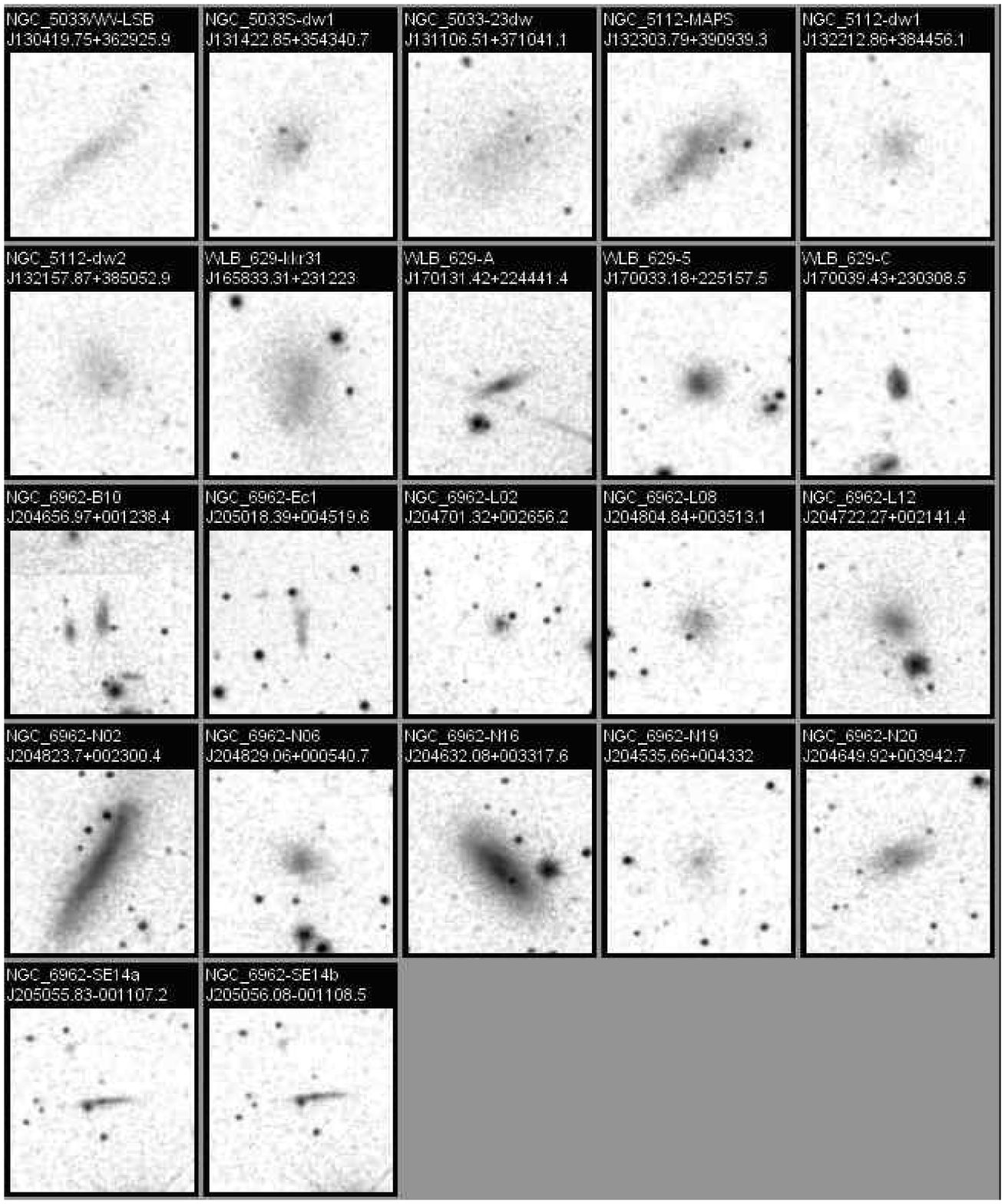}}
\caption{A mosaic of the (47) studied galaxies, which are visible in 
the SDSS. NGC~697-D is represented by two components - 'DE' and 'DW', 
and for the NGC~6962-SE14 we disentangle the 'a' and 'b' components. 
The NGC~697-F has two faint companions, identified as 'FSE' and 
'FSW'. Each sub-frame is centred on the corresponding galaxy and has 
a size of $\approx$ 1 $\times$ 1 arc-minutes; North is top
and East is to the left. For the better visibility, the contrast of the 
images has been enhanced.
}
\label{fig:galfigs}
\end{figure*}

\subsection{Data reduction technique}

The LRS data were reduced with standard reduction tools inside the ESO
MIDAS package LONG (see e.g.  Grosbol et al. \cite{midas1}, Warmels \cite{midas2}). Bias pattern and offset as well as pixel-to-pixel
variation (flat-fielding) were corrected and all frames were wavelength
calibrated by Cd-Ne-lamp exposures. The wavelength calibration was
double-checked with the night-sky lines and offset corrected where necessary
(see also Tab.~\ref{tab:LRS}).
The night-sky air glow was interpolated along the co\-lumns of the frames 
and removed. Finally, the objects were extracted applying the Horn 
algorithm for optimal extraction (Horne~\cite{horne1986}). 
In a few cases, other galaxies appeared 
projected on the long-slit (by chance or on purpose), and those were 
extracted in the same manner and are also listed in Table~\ref{tab:speresults}. 

In half of the nights, flux standard stars from the HET calibration 
list were observed (stars from the lists 
of Oke~\cite{oke1990}, Massey \& Gronwall~\cite{mass1990}, 
and Fukugita et al.~\cite{fukugita96}). We used these observations 
to establish a common relative flux calibration for all nights 
to correct the instrumental profile before measuring
line positions, line ratios, and comparison to template spectra.

\begin{figure}
\resizebox{0.48\textwidth}{!}{\includegraphics{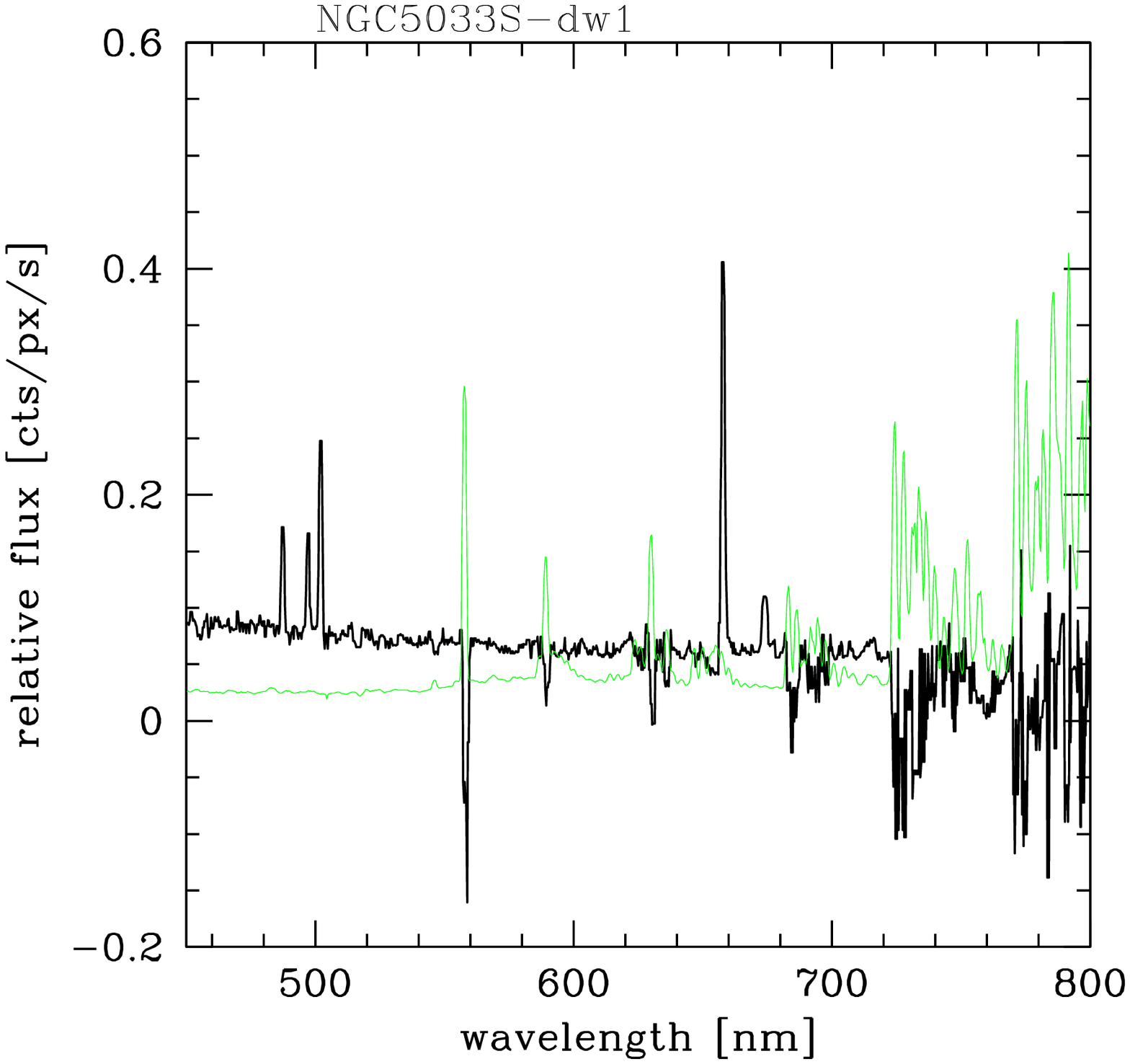}}\\
\resizebox{0.48\textwidth}{!}{\includegraphics{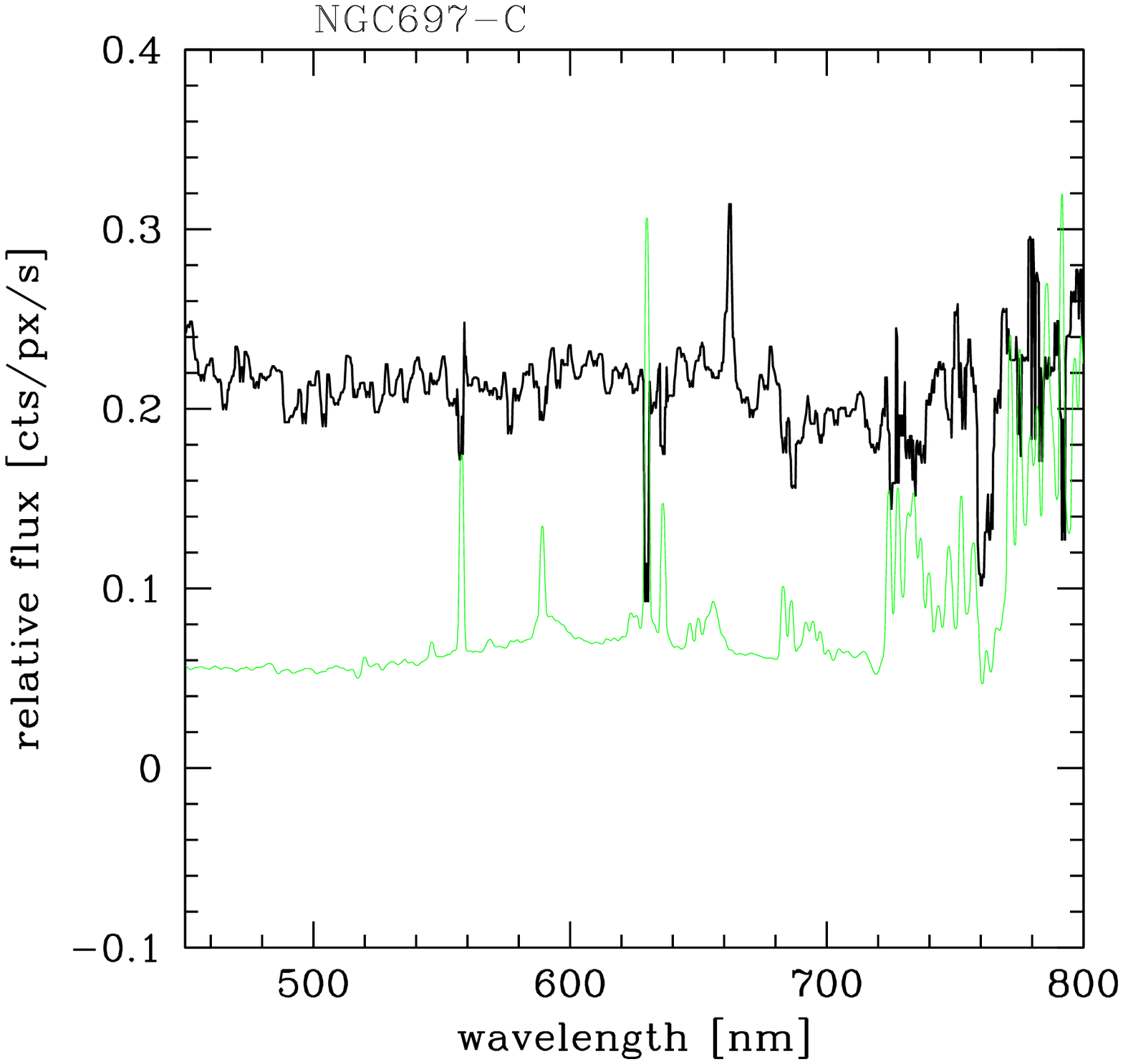}}
\caption{Two examples of spectra obtained with the HET LRS. The upper part
shows a typical HII galaxy, the type which by far dominates the newly 
confirmed group members, showing the classical bright emission lines of
Hydrogen, Oxygen, and Sulfur and a just tracable faint continuum (black line).
To easily indicate the night sky residulas in the object spectrum, 
the green line shows the night sky spectrum traced in that
observation, but scaled down by a factor 1000 for plotting purposes. The
lower part shows with NGC~697-C one of the brighter new drawf members, 
but a less typical case as only H$\alpha$+[NII]6584 are easily seen 
in emission. The [SII]6712+6731 doublet is just detectable as are the 
higher Balmer lines which appear in absorption. Again, the sky from the 
same observation is shown, here in the same scaling as for the object.
}
\label{fig:spectra}
\end{figure}

In most cases (see Fig.~\ref{fig:spectra} and Tab.~\ref{tab:speresults}), prominent HII emission lines
were visible and gaussian fits to them delivered the redshift of the galaxy.
As we used several observational setups, the derived data quality is
not homogeneous. The signal-to-noise ratios of the individual emission lines used to
measure the radial velocity vary between $\approx$~10 and $\approx$~150.
In few cases (one probable dwarf elliptical group member, several background
galaxies), no emission lines were visible, especially not at the location 
expected for a group member. The spectra of these low surface brightness
objects hardly show clear indications of absorption lines which would enable
a direct measurement of the redshift. We correlated the spectra in relative
flux units of these objects with the template spectra of 
Kinney \& Calzetti (\cite{kc96}) to estimate 
a redshift. While the redshifts obtained by emission lines normally
are accurate to $\pm~30$ km/s, the results of the
simple correlation technique applied to these low S/N spectra is in the
order of $\pm~300$ km/s, only. 

\subsection {Spectroscopic results}

\begin{figure}
\resizebox{0.45\textwidth}{!}{\includegraphics{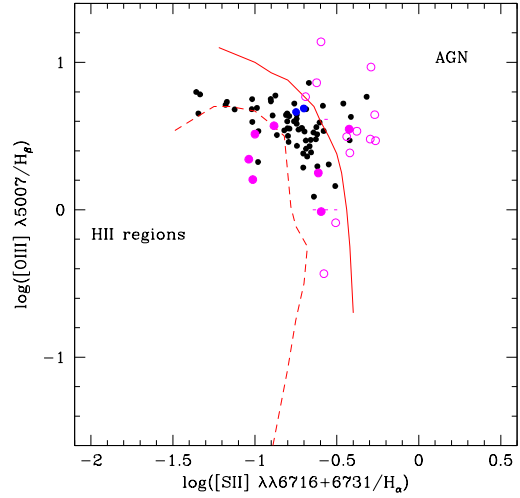}}
\caption{
The diagnostic line ratios of [OIII]~5007 to H$\beta$ and of
[SII]~6716+6731 to H$\alpha$ following 
Veilleux \& Osterbrock (\cite{veilost87}). Note that 
the H$\alpha$ and the [NII] lines are not resolved, therefore the flux is the
sum of the three lines. Open (red) dots are 
background galaxies of our observational sample, filled (red) dots
are new members of our target groups. The dashed line from Veilleux \&
Osterbrock indicate the distribution of normal HII regions driven by
photo-ionisation. The solid line (from Kewley \& Ellison \cite{kew08}) 
indicates the dividing line between dominant photo-ionisation 
and AGN-driven ionisation. For comparison, we show also the ratio 
for a pure HII sample (Popescu et al. (\cite{pope99})) where the 
ratios have been derived in a similar manner (small black dots).
}
\label{fig:diag}
\end{figure}

Table~\ref{tab:speresults} lists the individual redshifts together 
with the statistical errors, coordinates, type and membership 
classification. A total of 51 target galaxies were observed, another 
4 galaxies appeared projected on the long-slit. For 7 (14\%) of target 
galaxies, a redshift could not be extracted either due to low S/N or 
because of missing features or for both reasons. A mosaic of those 
47 observed galaxies, which are visible in the Sloan Survey (SDSS), 
has been composed with the help of the SDSS Visual Tools, and is 
presented in the Fig.~\ref{fig:galfigs}.  

The above described target selection aimed for DGs. Based 
on the redshifts and photometry (see below), 26 of the target 
galaxies are dwarfs in the sense that $M_{g'} \ge -17.3$. % (M$_B \ge -16.9$). 
This corresponds to a success rate of 51\%. We intentionally added 
two galaxies classified as back\-ground giants to the HET observations 
which were indeed verified as such. Including those brings the success 
rate of the morphological selection to 55\%.

The more ambitious aim of the selection is to identify new members of 
our target groups. Using a simple cut-off in the radial velocity of 
$\pm$~500~km/s around the group velocity (Tab.~\ref{tab:groups}), we 
classify 17 galaxies as new group members in the target groups.
This corresponds to a success rate of 33\%. It is worth mentioning 
that only 15 of these new members are dwarf galaxies while two have 
luminosities above the dwarf threshold and are in the range of 
LMC-M~33-like objects. Furthermore, the majority of the new dwarf
members are belonging to the brighter irregular dwarf galaxies, only 
three of them have absolute magnitudes fainter than -15 (in B-band). 
Among those faint new members is our only dE candidate galaxy 
(NGC~5033-30N). Two (brighter) dwarf members show Balmer absorption lines
(indicating recent star formation, so called star bursts, NGC~697-B,
and NGC~6962-L12) and all other member dwarfs show classical HII spectra 
pointing to ongoing star formation activity.

We found two DGs in the foreground of our target group IC~65. 
Using additional data from NED, we suggest that there is a small 
foreground group around NGC~278, which would fulfil our group 
selection criteria\footnote{The members of this foreground group
are NGC~278, UGC~672, and UGC~731 according to NED data, and 
IC~65-dw1 and IC~65-dw4 based on our own observations. We emphasise 
that this proposal needs further verification
as redshifts are only a poor distance measure below 1000 km/s.}. 
Adding these two galaxies brings the success rate to 37\%. NGC~278-kkh4 
would qualify as a member of the IC~65 group according to its redshift, 
but its projected distance of about 2.5~Mpc is too large to accept
this dwarf galaxy as a genuine IC~65 group member. It may belong to a 
neighbouring group or the host large scale structure of the group.
The remaining galaxies, including all chance projections onto the 
spectrograph slit, are background galaxies.

We notice that our success rate in identifying new dwarf members 
of our target groups neither depends on the Milky Way foreground 
extinction, nor on the velocity of the group, or on the richness 
of the group. Even a relatively nearby background structure might
or might not confuse the selection. We did not identify a single 
member in the group WLB~629, which has a background group of similar 
richness at about 9200 km/s. On the other hand, we found 6 new members 
in LGG~034, a group with almost identical foreground reddening, 
distance, and richness, which also appears in front of a small group
of galaxies at about 9800 km/s. Here our observations succeeded not only 
in the identification of six group members, but we were able to 
identify also three new members of the background group. This 
background group has a helio-centric radial velocity of 9794 $\pm$ 48 km/s,
the total of now six members allows a first estimate of the group 
velocity dispersion to 109 km/s
\footnote{The members of this background group are NGC~0695, LCSB~S0298P, 
and IC~1742 (according to data from NED), and NGC~697-dw1, 
NGC~697-G, and NGC~697-H (from our HET spectra).}.

In Table~\ref{tab:speresults}, we provide our final classification 
if the ga\-la\-xy belongs to a group, the fore- or background. % ({\bf bkg}). 
Galaxies are accepted as group members if their velocity does not 
deviate by more than 500 km/s from the mean of the group as given 
in Table~\ref{tab:groups}. Table~\ref{tab:groupsnew} presents the 
numbers of members identified from the NED data base together with 
the new members and the total number of objects probably belonging 
to the groups as well as the revised virial masses of target groups, 
which have changed by factors between 0.6 -- 2.0.

\begin{table}[h]
\caption{Revised richness  and virial masses of the studied groups}
\protect\label{tab:groupsnew}
\begin{center}
\begin{small}
\begin{tabular}{lcccc}
\hline
Group & n$_{NED}$ & n$_{new}$ & n$_{total}$ & $M_{200}^{new}$\\
 &&&& [$10^{12} M_\odot$] \\
\hline
LGG 016 &  4 &  2 &  6 & 0.7 \\
LGG 016$^a$ &  3 &  2 &  5 &  \\
LGG 034  &  9 &  5 & 16 & 9.1 \\
LGG 334 & 12 &  5 & 17 & 2.8 \\
WBL 629 & 10 &  0 & 10 & 2.3 \\
WBL 666 & 27 &  5 & 32 & 40.0 \\
\hline
\end{tabular}
\\
$^a$ Foreground group around NGC~278, see text for details.
\end{small}
\end{center}
\end{table}

For those galaxies where emission lines are present in the observed 
spectrum, the relative line flux in the emission lines was derived 
by fitting a gaussian to the line (MIDAS procedure {\it integrate/line}). 
The leading error of this measurement is the determination of the 
continuum level which in most of our spectra has a low S/N (the spectra 
were originally not taken for this purpose). If the relevant lines
are observed, one can calculate line ratios which allow to classify 
the dominant emission process in the galaxy following Veilleux \& 
Osterbrock (\cite{veilost87}). From multiple, independent measurements, 
we estimate a typical error of these ratios to be of order 0.1 dex.
We do not present line fluxes (and properties which might be derived 
from them like e.g. star formation rates) as many of our observations 
were taken under non-photometric conditions. Further, slit losses are 
not really under control for our sample galaxies. One has to take in 
mind for the following discussion that the [NII] lines are not
resolved from H$\alpha$, and the [SII] doublet is not resolved. The 
LRS at HET does not allow to observe low-redshift [OII]~3727, thus, 
an abundances estimate based on the R$_{23}$ method is not possible 
for the group members. Our spectra do not allow to independently
estimate the impact of the underlying stellar absorption features to
the H$\beta$ emission lines strength. Thus, we do not correct for
the absorption. Typically, the absorption lines of the early type
stars of the stellar population produce absorption lines of an equivalent 
width of about 0.2 nm
(Skillman \& Kennicutt \cite{Skillman1993}), correcting for an effect
of this magnitude would increase the H$\beta$ fluxes by about 13\% (or 0.05
in a log ratio as used later on, which does not impact the conclusions
we will draw from the data.
Finally, we calculated the (log of the) ratio of [OIII]~5007 to H$\beta$ 
and of [SII]~6716+6731 to H$\alpha$+[NII]6584. 

Figure~\ref{fig:diag} shows the [OIII]/H$\beta$ and of [SII]/H$\alpha$ ratios
derived from our spectra. As the lines used to derive the ratios are nearby 
in wavelength space, their differential extinction dimming is negligible 
within the accuracy of our data and we did not correct for that effect 
despite the available H$\beta$ to H$\alpha$ ratio as the correction adds 
further noise. The observations are compared to the model sequence of 
photo-ionisation following Veilleux \& Osterbrock (dashed line)
as well as to the observed quantities of an HII-galaxy sample 
(Popescu et al. \cite{pope99}). We further add the dividing line between 
the HII and AGN regime from Kewley \& Ellison (\cite{kew08}, solid line). 
For 19 of our targets, all 4 required fluxes could be measured (for the
remaining, at least one of those lines was not measurable). Only seven 
of those 19 galaxies are members of our target groups (shown as filled dots). 
All of them occupy the parameter space of normal HII (star forming) 
dwarf galaxies. We see no hint for unusual low or high metallicity. 
Comparing the spectra of all our group members does not indicate that 
those where all four lines could be measured differ in any sense from 
the others where one or more of those lines could not be measured due 
to bad S/N, night sky line residuals, and further similar technical 
reasons. Therefore, we draw the conclusion that all new group members 
with the spectral classification 'HII' are star forming dwarf galaxies,
in good agreement with the photometric results (see next chapter). 
We mention in passing that many of the twelve background galaxies 
with measured ratios spread into the AGN regime.

\section{Surface photometry of observed galaxies}
\subsection{Surface photometry method}

We have performed standard surface photometry of all observed galaxies 
with new redshifts on frames downloaded from the SDSS imaging database. 
For that exercise, we used the so-called corrected frames in the 
$g'$, $r'$, and $i'$ band of different data releases (DR7, DR8 and DR9). 
Most of our targets are LSB galaxies of irregular shape (Fig.~\ref{galfigs}). Therefore, 
to increase the S/N ratio and to extend the surface photometry on 
those relatively shallow SDSS frames as far into the galaxy's periphery 
as possible, we applied a space-variable (i.e. adaptive) filter 
(Lorenz et al.~\cite{afi}), as implemented in the ESO MIDAS environment 
({\it filter/adaptive} task). The adaptive filtering procedure is described 
in more detail in Vennik et al. (\cite{v96}, \cite{vh08}). 
After filtering, the local sky level around the target galaxy 
was fitted, using the least-square method of the MIDAS 
task {\it fit/flat-sky}. Then, the fo\-re\-ground stars and other 
disturbing objects on the galaxy image were masked and replaced 
using an interactive interpolation routine.

The main result of the surface photometry are the differential and 
cumulative SB profiles, as well as a set of isophotal, effective, 
and integral parameters, derived from these profiles. The SB 
distribution of galaxies was analysed in a traditional way, by 
means of fitting the galaxy isophotes with a set of model ellipses, 
using the {\it fit/ell3} task in MIDAS. First, we fitted ellipses 
to the adaptive-filtered $g'$, $r'$, and $i'$ band composite image. 
Higher S/N ratio of the composite image permits to trace the light 
distribution at larger radii. Next, the cleaned and background 
subtracted images in each filter were fitted with the same set 
of ellipses. Finally, the azimuthally averaged SB and colour index 
profiles were constructed as a function of the equivalent radius 
$R_\mathrm{eq} = \sqrt{ab}$, where $a$, $b$ are major and minor 
semi-axes of the fitted ellipses. 

The total magnitudes were estimated by asymptotic extrapolation 
of the radial growth curves. The effective (half-light) radius 
($R_\mathrm{ef}$) was measured on the growth curve and the effective 
SB at $R_\mathrm{ef}$ ($\mu_\mathrm{ef} = \mu(R_\mathrm{ef}$) was 
determined on the SB profile. One should keep in mind that 
the NGC~6962 group is located within the SDSS Stripe 82. Thus, 
we used the co-added $g'$, $r'$ and $i'$ band frames, which 
are about 2 magnitude deeper than normal SDSS frames. 

\subsection{The photometry errors}

\begin{figure*}[t]
\vspace{-27mm}
\hspace{-5mm}
\resizebox{0.37\textwidth}{!}{\includegraphics{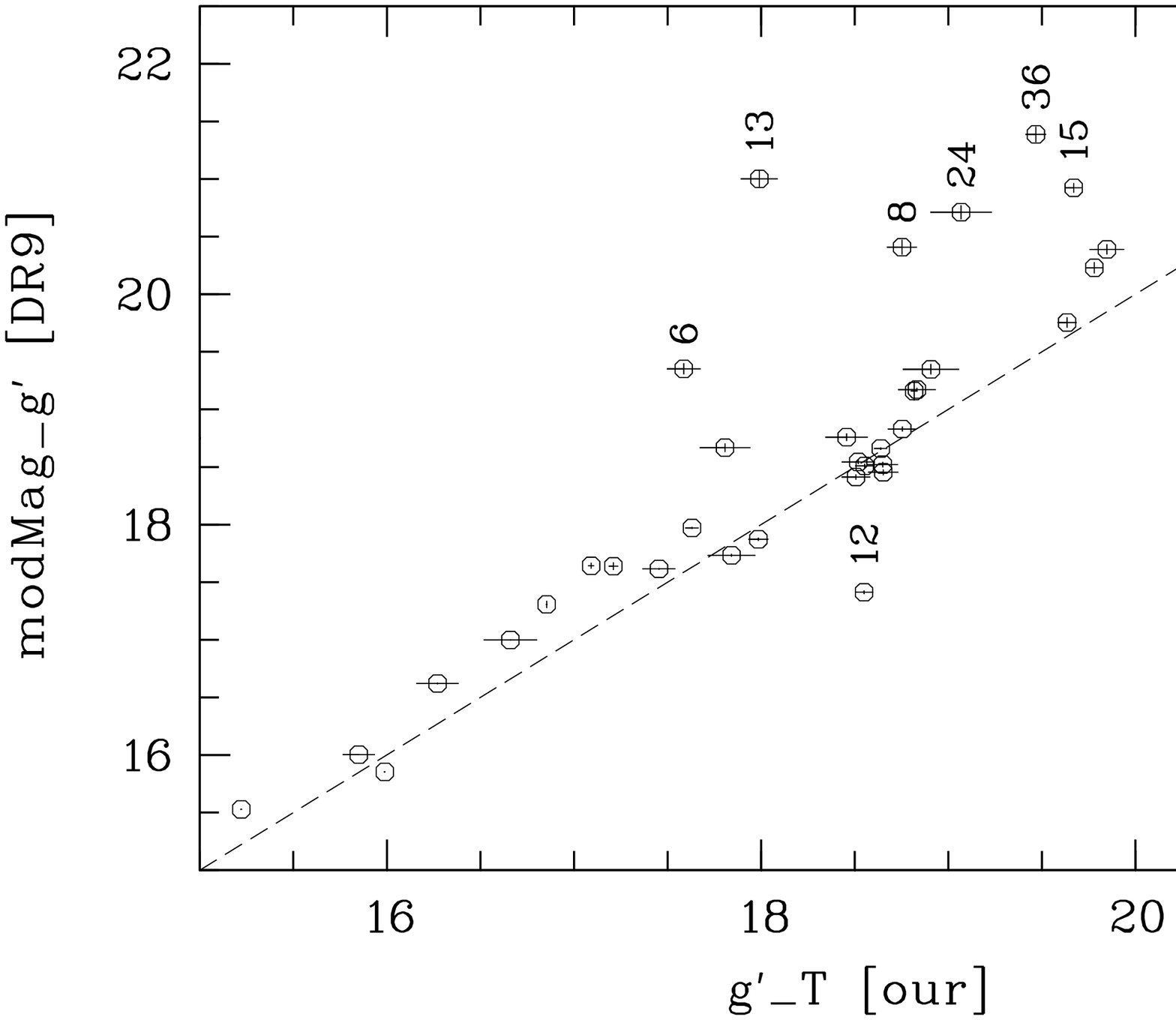}}
\hspace{22mm}   
\resizebox{0.37\textwidth}{!}{\includegraphics{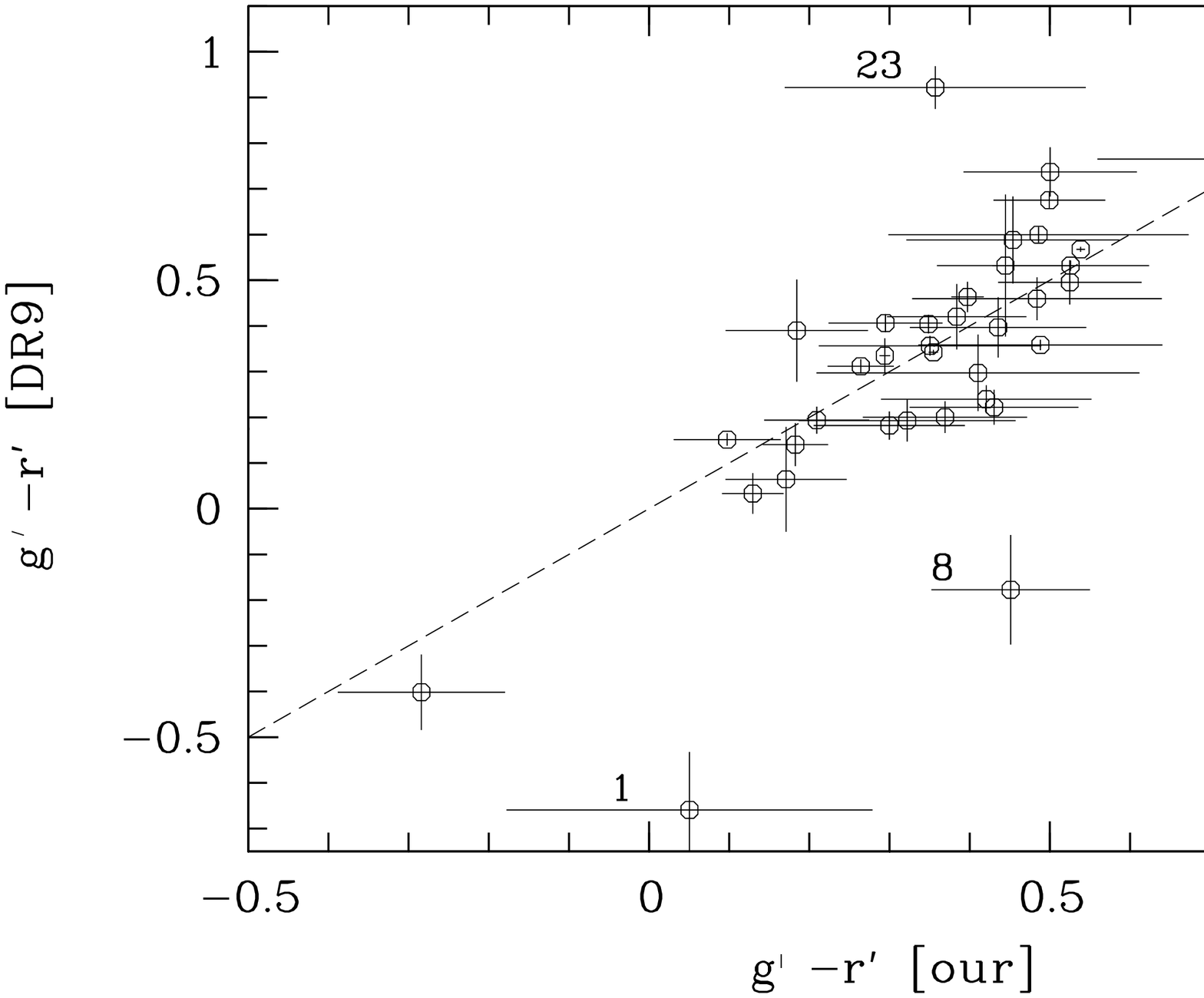}}
\caption{Our total (asymptotic) $g'$-magnitudes ($g'\_T$) {\it (left)} 
and $g'-r'$ colours {\it (right)} are compared to the model-magnitudes 
($modMag\_{g'}$) and colours of SDSS DR9. Big differences ($\Delta m > 1$ 
and $\Delta (g'-r') > 0.5$) are identified with galaxy numbers
in the Table~\ref{tab:phot-gri-err}. The dashed lines indicate 
the 1:1 relation, individual error bars are given.
}
\label{fig:grT-modMag}
\end{figure*}

The magnitude errors consist of internal and external components. The 
internal random component is dominated by the accuracy of the background 
estimates and, in addition, includes, the random count error determined 
from the photon statistics in the aperture measurements. Following 
Vader \& Chaboyer (\cite{vader94}), we calculated the internal errors 
in intensity as
%
%\begin{equation}
$$\Delta I = \sqrt{N_{\rm tot}+(\delta~n_{\rm sky}~A)^2},$$  
%\end{equation}
where $N_{\rm tot}$ is the total number of counts within aperture $A$, 
which is defined as the area between successive isophotes (in pixels), 
and $n_{\rm sky}$ is the mean sky counts per pixel, and $\delta$ the 
fractional error in the mean sky value.

The variations in the sky background, as obtained from the parameters 
(SkyIvar, Sky) in the SDSS DR9 PhotoObjAll catalogue, are typically 
0.17\% (in $g'$), 0.16\% ($r'$) and 0.25\% ($i'$). This yields the 
typical sky uncertainty in the range of 0.002 - 0.003 mag~arcsec$^{-2}$. 
We note, however, that the parameter SkyIvar in the SDSS PhotoObjAll 
catalogue is strongly varying for individual objects, indicating possibly 
problems with proper sky-level determination by automatic data processing. 
Since most of our targets are LSB galaxies, their photometry is 
particularly sensitive to the sky-level variations. Therefore, we have 
measured interactively the local sky-level in empty areas around each 
of the studied galaxies (on the frames of DR7 and earlier),
as well as re-measured the local sky level on the sky-subtracted and 
calibrated DR8 and DR9 frames. We applied the same approach to the 
co-added Stripe 82 frames. 

We found significant differences in sky-level values \\ 
($\Delta Sky = Sky_{\mathrm {our}}-Sky_{\mathrm {DR7}}$), 
when using the frames from the DR7 and earlier. With mean 
relative ($\Delta Sky/Sky_{\mathrm {DR7}}$) values of 
$+1.1 \pm 7 \%$ (in $g'$), $-0.6 \pm 9 \%$ (in $r'$), and $8 \pm 14 \%$ (in $i'$). 
On the opposite, we found only marginal sky-corrections, when using 
the re-processed sky-subtracted fra\-mes from the DR8 and DR9. 
Those differences are typically 0.13 \% (in $g'$), 0.07 \% (in $r'$) 
and 0.1 \% (in $i'$), with $1\sigma$ deviations of the same order. 
Therefore we conclude, based on our rather limited sample of $\sim$~50 
galaxies, that sky-subtraction on the re-processed frames since DR8 
is generally reliable.  

Using our interactively determined sky values and their 1$\sigma$ 
uncertainties, we have calculated the errors of total magnitudes, 
as those, corresponding to the 1$\sigma$ sky-level deviations. 
The 1$\sigma$ sky-error component in total magnitudes in the different 
filters are typically 0.071 mag (in $g'$), 0.069 mag (in $r'$), 
and 0.104 mag (in $i'$). The individual errors are listed in the 
Table~\ref{tab:phot-gri-err}. We note, that the model-magnitude 
errors of the studied galaxies, as listed in the DR9 catalogues, are 
about a factor of two smaller.

Besides of sky-subtraction issues there are various potential caveats 
related to the automatic pipeline data reduction in the SDSS 
which were discussed and flagged, e.g. in the NYU-VAGC 
(Blanton et al.~\cite{VAGC}). Most of these issues are related to poor 
de-blending of large and/or LSB galaxies with complicated morphology 
(e.g. star-forming regions, dust features etc.). At low redshifts a 
number of extended, knotty LSB galaxies have been found shredded, 
i.e. a large galaxy image is split by the target selection algorithm into
several sub-images (e.g. Panter et al.~\cite{panter07}, 
Tago et al.~\cite{DR5groups}). Therefore, the treatment of LSB galaxies 
requires special care.

In order to estimate the true (external) magnitude errors we need 
high-quality magnitudes for comparison. For the dwarf galaxies in the 
IC 65 group region we carried out error estimates in 
Vennik \& Hopp (\cite{vh08}). For other galaxies, which were studied 
on the SDSS frames, no independent comparison data are available.
Therefore, we apply for comparison the model magnitudes of the SDSS, 
which were derived during automatic pipeline data reduction.

In the Fig.~\ref{fig:grT-modMag} we compare our total magnitudes 
and $(g'-r')$ colours to the model magnitudes and colours as given in 
the SDSS DR9. Eight galaxies with large magnitude differences 
of $|\Delta m| = |g'_{\mathrm {T,our}}-modMag\_g'_{\mathrm {DR9}}| > 1.0$ 
are identified with the galaxy sequence number in the 
Table~\ref{tab:phot-gri-err}. To understand the reason of big 
magnitude differences we have checked all galaxy images using the 
SDSS Visual Tools option and found several cases of shredding - 
in galaxies 6, 8, 24, 29, 36, and poor de-blending of galaxies seen 
in projection to each-other (galaxies 12, 13, 15), and poor 
identification of very LSB irregular galaxies (e.g. the galaxy 1).
When neglecting those cases of poor galaxy identification in the DR9, 
we obtain small systematic total magnitude differences 
of $\Delta m = -0.20 \pm 0.25 (g'), -0.17\pm 0.21 (r'$), 
and $-0.19 \pm 0.25 (i'$). respectively. 

\subsection{The SB profile fitting}

The inspection of the derived  SB profiles (see Fig.~\ref{fig:profiles}) 
generally reveal a close agreement with pure exponential disk profile. 
Only minor deviations are observed in some cases, both, central flattening % or depressions
as well as central brigh\-tenings as can be caused by 
small bulges or nuclei. Depending on the particular profile shape, 
we fitted the SB profiles either by one single or two Sersic power laws
$$ \mu (r) = \mu_0 + 1.086~(r/h)^{1/n}. $$
First, we estimated the extent of outer linear part of the SB profile 
(comparing the profiles in different bands), fitted it with a pure 
exponential disk model ($n$ = 1) and determined the best-fitting model 
parameters ($\mu_0, h$). If the fit residuals show a reliable light excess 
over the exponential disk model, we fitted it with a second power law. 
In case of central light depression, as it is common for many LSB dwarf 
galaxies, we fitted the SB-profile with a single Sersic function 
with $n < 1$. The model-free effective (half-light) photometric 
parameters and exponential disk model parameters (if applicable) of 
galaxies, with new redshift measurements, are given in 
Table~\ref{tab:phot-gri-err}. 

\begin{figure*}[h]
\vspace{-27mm}
\hspace{-5mm}
\resizebox{0.37\textwidth}{!}{\includegraphics{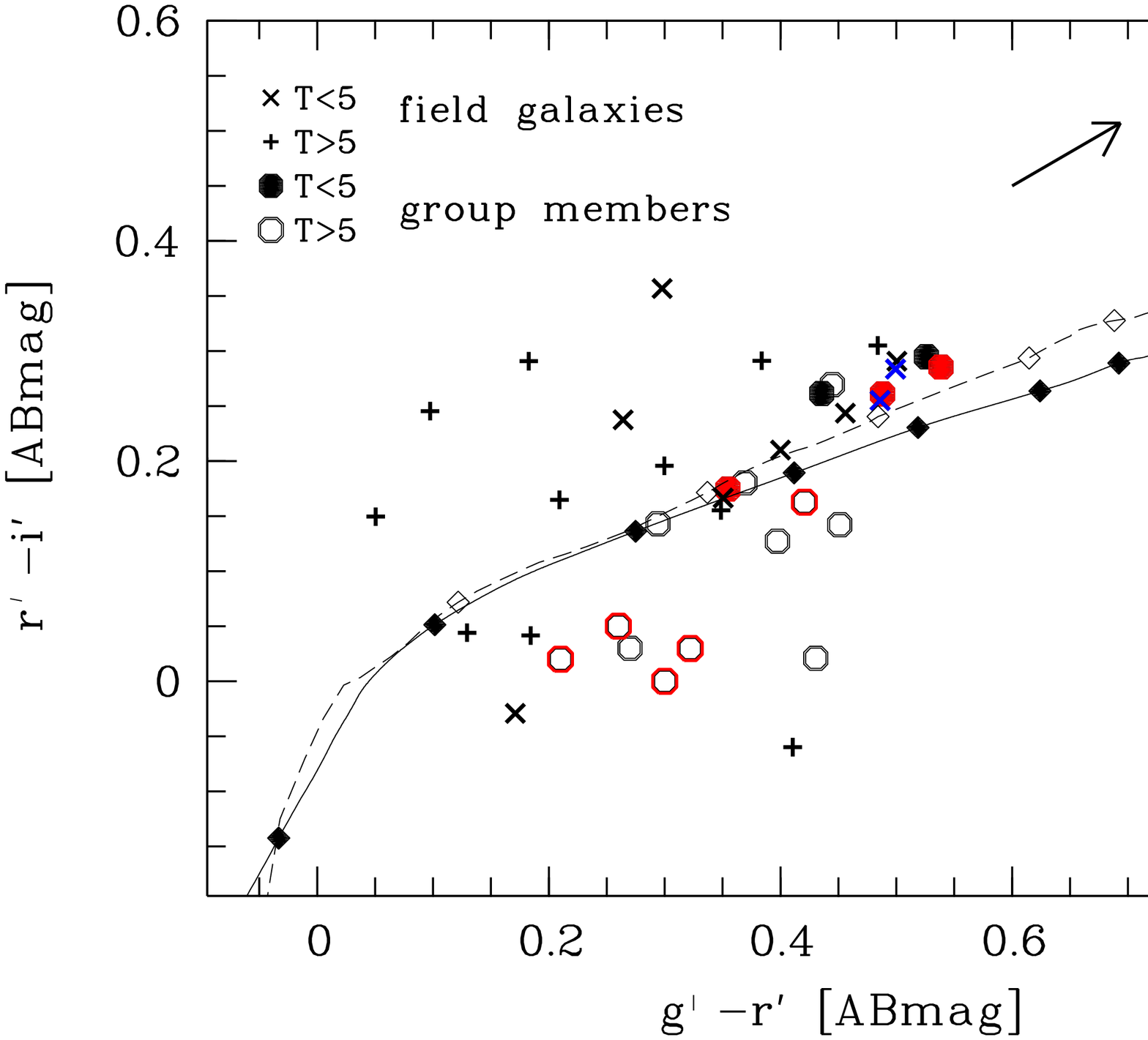}}  
\hspace{22mm}
\resizebox{0.37\textwidth}{!}{\includegraphics{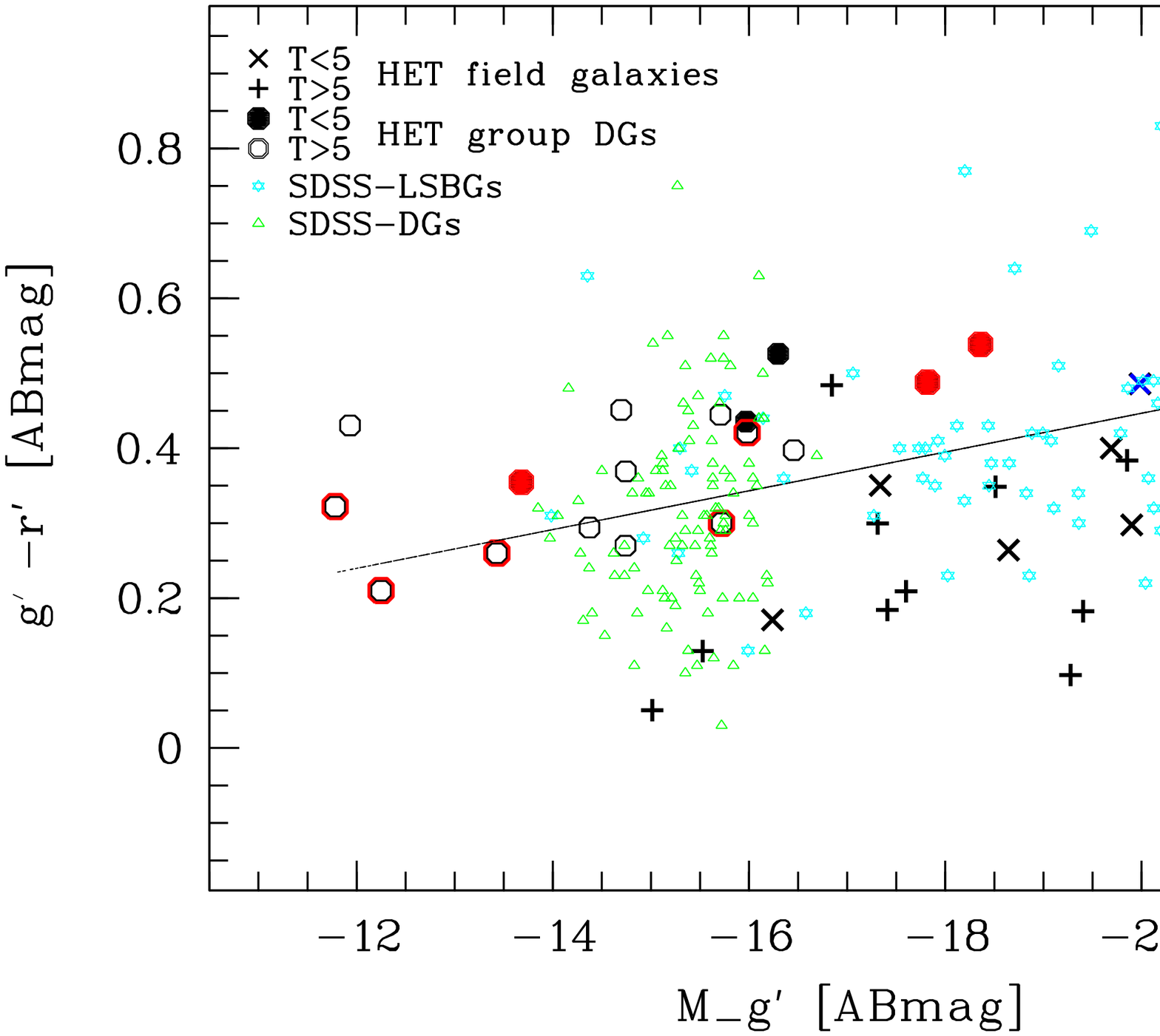}}  
\caption{Colour-magnitude (right) and colour-colour (left) diagrams of 
studied early (T$<$5) and late type (T$\ge$5) galaxies. Group members
 are shown with hexagons, field galaxies are designated with crosses. 
Blue symbols (in the on line article) are for galaxies with significant 
($3\sigma$) negative $(g'-r')$ radial gradients and red symbols denote 
galaxies with significant positive colour gradients.
The error bar (lower right) represents random errors of the magnitudes 
and colours.
{\it Left:}  
The observed colours are compared to the GALEV model evolutionary tracks 
with a standard Salpeter IMF with two different SF laws:
solid line traces the colour evolution of a Sd galaxy and combines a 
constant SFR with a single STB (burst onset at 1 Gyr, burst strength 
of 0.7 and $\tau_{burst}=10^9$ yrs);
dashed line shows a model with exponentially decreasing SFR 
($\tau_{decl}=10^9$ yrs, $Z = 0.5Z_\odot$), and reddened by $E(B-V)=0.25$).
Losanges indicate following time steps (from left to right): 
1, 2, 3, 4, 5, 7, 10 Gyr since the onset of SF.
The arrow indicates the effect of dust on the colours. 
{\it Right:}
The group members and the field galaxies appear to be (marginally) 
separated on the CMD. 
Linear regression (for HET galaxies) is shown with a continuous line. 
The magnitudes and colours of the studied galaxies (HET sample) 
are compared to those of the extremely low mass DGs 
(Geha et al.~\cite{geha06}, green triangles) and to those of the 
LSB galaxies (Kniazev et al.~\cite{kniazev04}, cyan stars), 
selected from the SDSS. 
}
\label{fig:CMD-grri-GALEV}
\end{figure*}

\subsection{Photometric properties of the studied galaxies}

According to our morphological classification the studied sample is 
largely dominated by late-type ($T\ge 5$) spiral and 
irregular galaxies, with a few examples of early-type dwarf galaxies. 
The studied galaxies can be naturally divided into two sub-samples. 
The first sub-sample consists of 17 new confirmed dwarf(ish) 
members of six (5 targeted + 1 foreground) groups - hereafter 'group DGs'. 
One very nearby dSph galaxy NGC~5033-Adw, located within the Local Volume, 
could be added to this sub-sample. The second sub-sample consists 
of 26 (mostly late-type) galaxies located in the background of the 
target groups. Those galaxies are main\-ly distributed in the lower 
density 'field' (sheets/fila\-ments, but also in, at least, one 
possible group) environment. The main characteristics of these two 
sub-samples are compared in the Table~\ref{tab:HET-gri}. 
The new group members are predominantly dwarf (with median luminosity of
$<M_g'>_\mathrm{med} = -15.3$) and LSB ($\mu_{\mathrm 0,g'}$ = 22.55 mag~arcsec$^{-2}$) 
galaxies. The background ('field') galaxies are $\sim$~4 mag more 
luminous than the group dwarfs, show typical Freeman's disk central SBs 
and median disk scale length of $\sim$~2 kpc. Interestingly, the group 
DGs and the field LTGs have similar median ($g'-r'$) and ($r'-i'$)
colours (which is close to the mean colour of the SDSS DGs of 
$<g'-r'>$ = 0.45 (ABmag), as given in Barazza et al.~\cite{barazza06}), 
however they are marginally separated in the colour-magnitude 
diagram (CMD) in the Fig.~\ref{fig:CMD-grri-GALEV}, 
and appear to trace different colour-magnitude correlations. 
The group members trace a narrow sequence on the CMD, possibly having 
(more) homogeneous stellar populations. The field galaxies of the same 
luminosity tend to have bluer $g'-r'$ colours by $\sim$ 0.2 mag, which could be 
an age and/or metallicity effect. The data of our two samples overlap
with the distribution presented by (Geha et al.~\cite{geha06}) for low
mass DGs and the one given by (Kniazev et al.~\cite{kniazev04} for
LSB galaxies (both derived from SDSS data). The literature data
show a wider spread in colour for given magnitude than our group sample
again supporting the view that those group members might have a more
homogeneous stellar population.

In order to study the colours and star-formation histories (i.e. to 
obtain a crude estimate of the stellar population ages) of the observed 
galaxies we compare their colours with the GALEV evolutionary synthesis 
models (Kotulla et al.~\cite{galev}), which were computed using the webform \\
{\it http://model.galev.org}.
Since our galaxy sample being dominated by late morphological types 
includes a few early ty\-pes too, we selected for comparison two (extreme) 
models - one with the exponentially declining SFR, typical for ETGs 
(ellipticals), and second one with constant SFR, proper for actively 
star-forming Sd galaxies. The models with constant SFR have been found 
to be typical for the field late-type dwarf galaxies (LTDGs) (e.g. 
van Zee 2001) and they can reproduce the colours of the bluest objects. 
The single burst followed by exponentially declining SFR  models are 
able to cover the colours of redder objects with no present time SF.
However, the SFHs of DGs in the Local Group, both the late and early 
types, are known to be more complex and constitute a mixture of 
multiple bursts and periods of constant SF 
(see e.g. Tolstoy et al \cite{tolstoy} for a recent review). In the (observed) 
colour-colour diagram (Fig.~\ref{fig:CMD-grri-GALEV}, left) we show 
for comparison two GALEV models. Solid line traces the chemically 
consistent colour evolution of a late-type (Sd) galaxy and combines 
a constant star-formation rate (SFR) with a single star-burst (STB)
at $T_{burst}=10^9$ yr ($\tau_{burst}=10^9$, burst strength 0.7).
Dotted line shows a model of a single (early) STB followed by an 
exponentially decreasing SF. (More details of both models are given in 
the figure caption). In order to reproduce the reddest observed colours 
we have to expect a modest reddening by dust in an amount of $\sim$ 0.25 mag. 
The mean stellar ages of the dwarf group members appear to be confined 
in a narrow range of 2 - 5 Gyrs. Colours of the field galaxies show a 
large dispersion and their age estimates are much more uncertain. 
The sequence of galaxies in the colour-colour plot is probably determined 
both by the most recent episode of star-formation and also by the 
luminosity, as indicated by the rather steep slope of the CMD in 
the Fig.~\ref{fig:CMD-grri-GALEV} (and its large scatter).
We can conclude that the range of observed colours is consistent with 
a SFH combining single (or multiple) star-burst(s) with intermediate 
periods of constant SFRs. A model with exponentially decreasing SFR 
can also reproduce the global colours (Barazza et al.~\cite{barazza06}). \\
We can (qualitatively) say that the stellar disks of late-type 
($T\ge 5$) field galaxies are, as a mean, marginally bluer (and 
probably younger), when compared to the colours of the early-type 
($T < 5$) field galaxies. The two reddest galaxies (with oldest 
stellar populations) are NGC~772-gal1 and -gal2. These qualitative 
age estimates are certainly hampered by the well-known age-metallicity 
degeneracy of the optical colours.

In addition to the common SB profiles we also constructed the colour 
index profiles and determined the radial colour gradients of 
34 galaxies with our new detailed photometry, which may give us 
information on the stellar content and other physical parameters 
as a function of radius. In about half (55\%) of the studied galaxies 
we found statistically significant (at $1\sigma$ level) 
($g'-r'$) colour gradients. Positive gradients (i.e. the galaxy is 
getting redder towards its periphery) are slightly dominating in our sample, 
in proportion 11 positive to 8 negative gradients. The dwarf group 
members (14) tend to show positive $(g'-r')$ gradients,
with $<\frac{d(g'-r')}{dR}>_{14} = +0.017 \pm 0.018$. The five steepest 
gradients, significant at $3\sigma$ level, are all positive.
The (more luminous) 'field' galaxies (20 of them with new photometry) 
tend to show negative gradients,
$<\frac{d(g'-r')}{dR}>_{20} = -0.023 \pm 0.050$, among them 6 (30\%) 
at $2\sigma$, and  4 (20\%) at $3\sigma$ are all with negative gradients.

There are observational indications that massive galaxies are growing 
inside-out fashion but the DGs show no gradients in age,  i.e. no gas 
accretion in dwarf satellites residing in the DM halo of the parent 
galaxy (Fossati et al.~\cite{fossati13}). Environmentally conditioned 
differences in SF properties have been noted between the cluster and 
field early-type disk galaxies, particularly, enhanced SF in central 
and truncated SF in outer disk of cluster galaxies 
(Bretherton et al.~\cite{bretherton13}).
Generally, most of LSBDGs have been found revealing flat colour profiles, 
however, both negative and positive colour gradients were noted for 
some dIrr's earlier (e.g. Patterson \& Thuan~\cite{PatThuan96}), and 
for LSBGs in nearby groups (Bremnes et al.~\cite{bremnes98}, \cite{bremnes99}).
A variety of scenarios exist to explain the transformation of 
spiral-rich field galaxy population into early-type rich (S0/dE/dSph) 
population in clusters and groups of galaxies. Removal of gaseous halos, 
outer and/or inner HI disks leads to SF strangulation in long time-scales 
or SF truncation on shorter ones (Kotulla et al.~\cite{galev}). 

Thus, in line with the above considerations, the marginal colour 
differences between the group and field/bkg galaxies in our small sample, 
could be interpreted as an effect of gas depletion predominantly in 
outer fragile disks of dwarf satellite galaxies in groups, which results 
in positive colour gradients. The field/background disk galaxies are 
evolving in common inside-out fashion.\\
However, the colour profiles and their gradients are highly sensitive 
to proper sky subtraction issues. Despite of special effort made 
previously by sky level determination, the reliability of detected 
(generally shallow) gradients should be confirmed by new  deep 
imaging studies.

\subsection{The photometric scaling relations}

\begin{figure*}
\vspace{-30mm}
\hspace{-5mm}
\resizebox{0.35\textwidth}{!}{\includegraphics{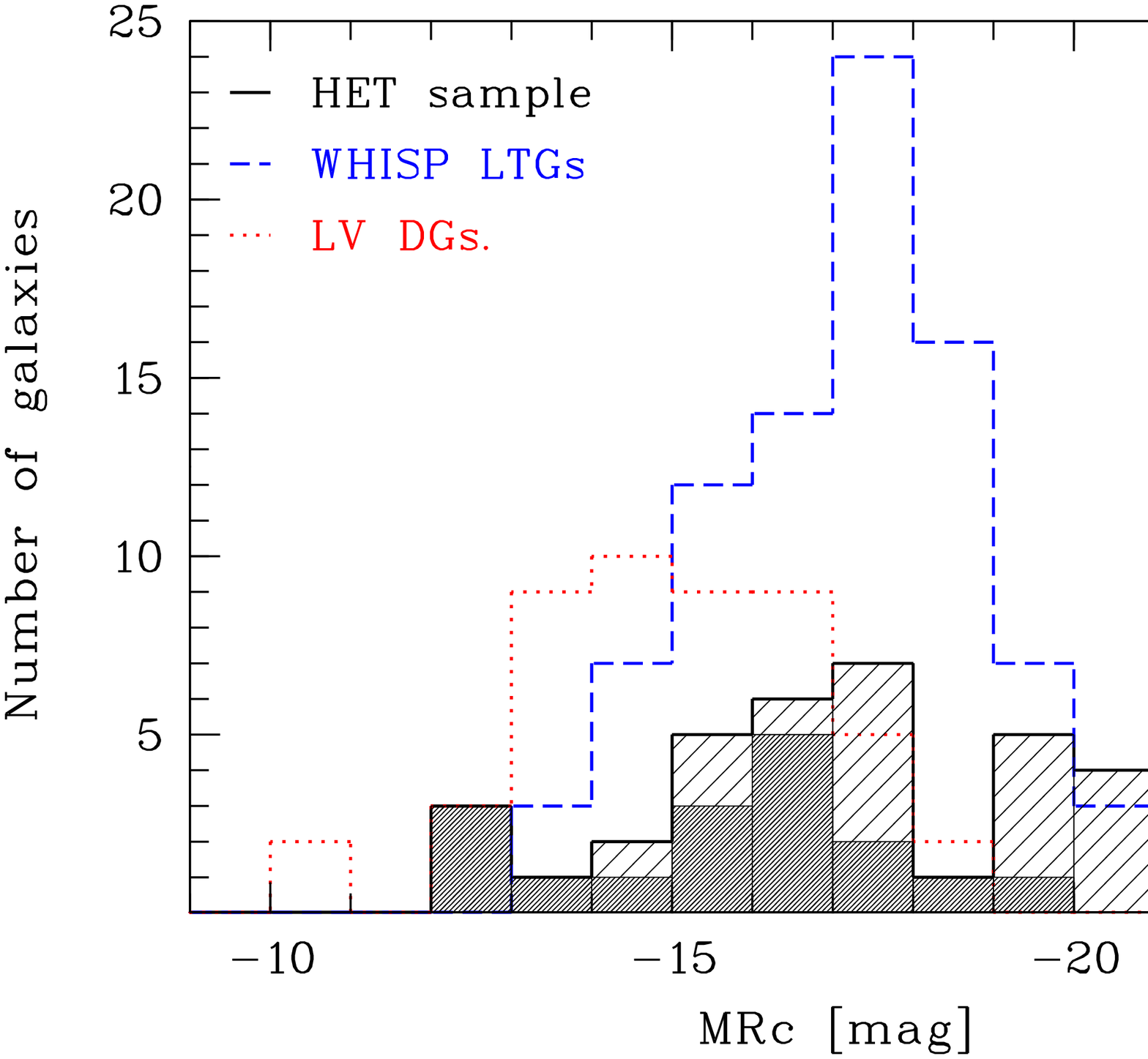}}
\hspace{22mm}
\resizebox{0.35\textwidth}{!}{\includegraphics{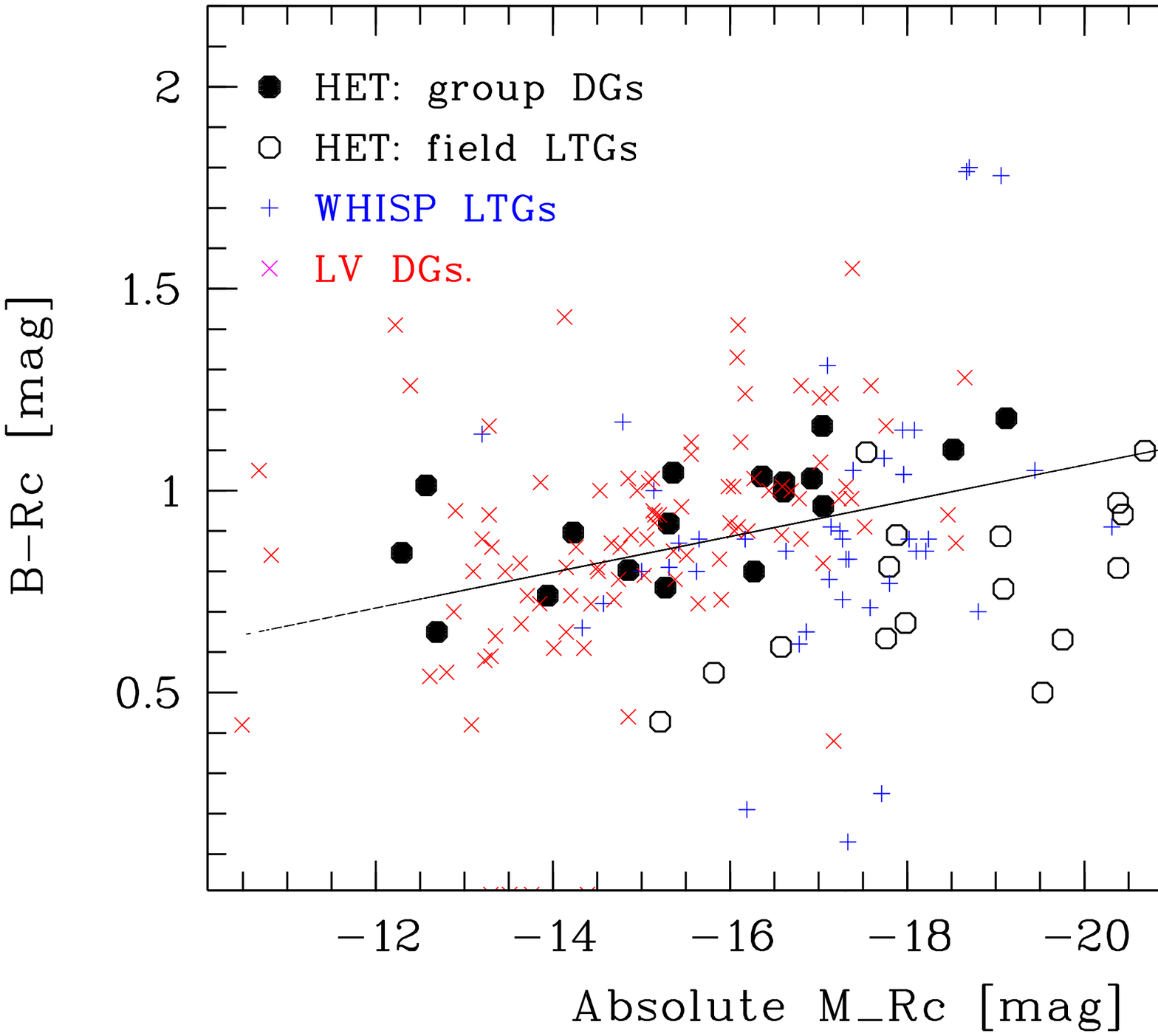}}
\caption{The $R_c$-absolute magnitudes and $B-R_c$ colours of studied 
galaxies (HET sample), compared to those of the DGs of the 
Local Volume (LV DGs) and to those of the LTGs of the WHISP 
sample (WHISP LTGs). {\it Left:} filled black histogram is for the 
HET group DGs, hashed black histogram is for the HET 'field' galaxies; 
dashed blue histogram is for the WHISP sample and dotted red
histogram is for the LV DGs. (Note: the LV and WHISP samples are scaled 
by a factor 0.5). {\it Right:} relation between the $(B-R_c)$ colours 
and absolute $R_c$-magnitudes is approximated by a linear regression 
with a slope -0.044$\pm$0.011 for the full combined sample (solid line). 
}
\label{fig:CI-MR}
\end{figure*}

\begin{figure*}
\hspace{-5mm}
\resizebox{0.37\textwidth}{!}{\includegraphics{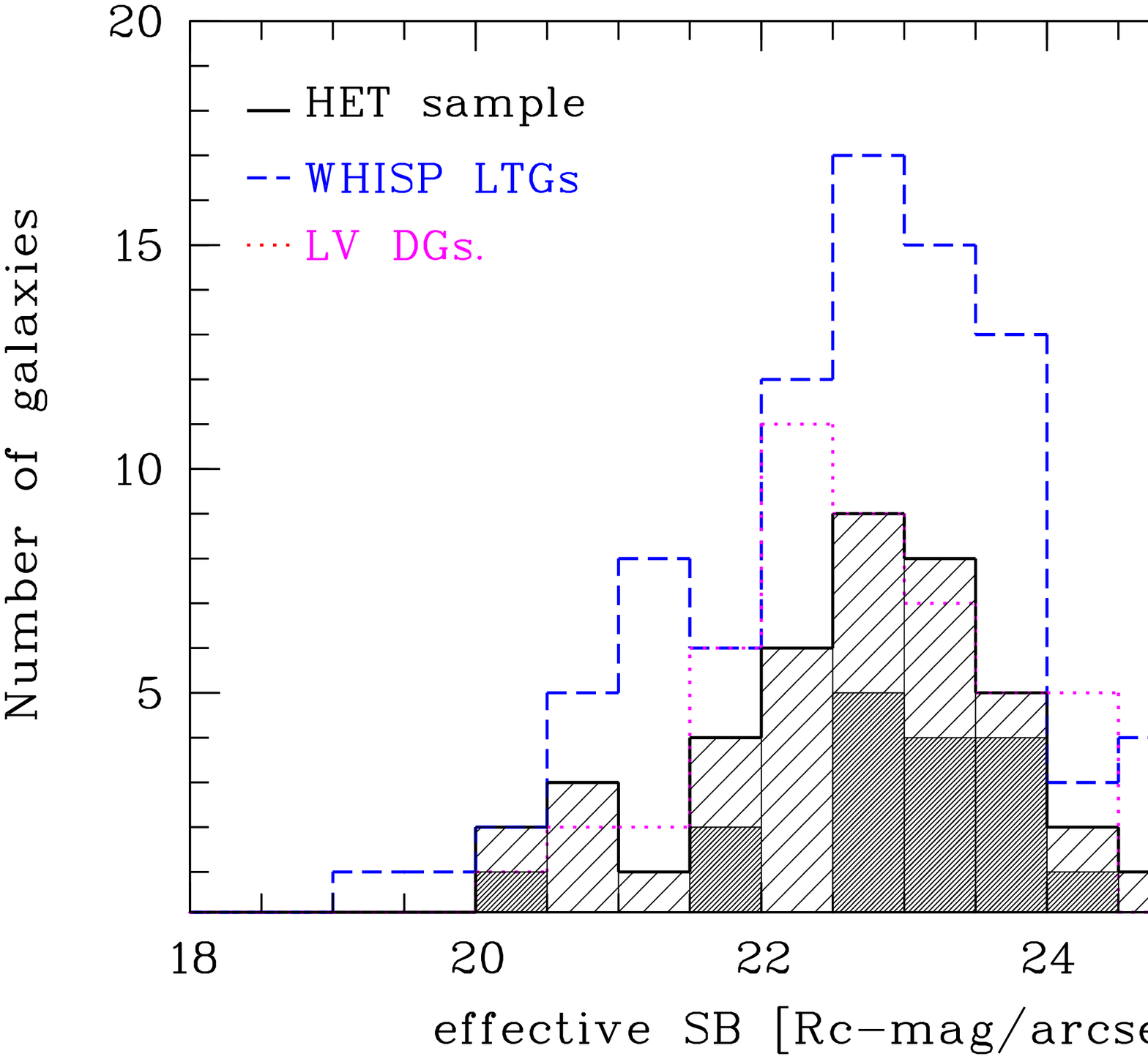}}
\hspace{22mm}
\resizebox{0.37\textwidth}{!}{\includegraphics{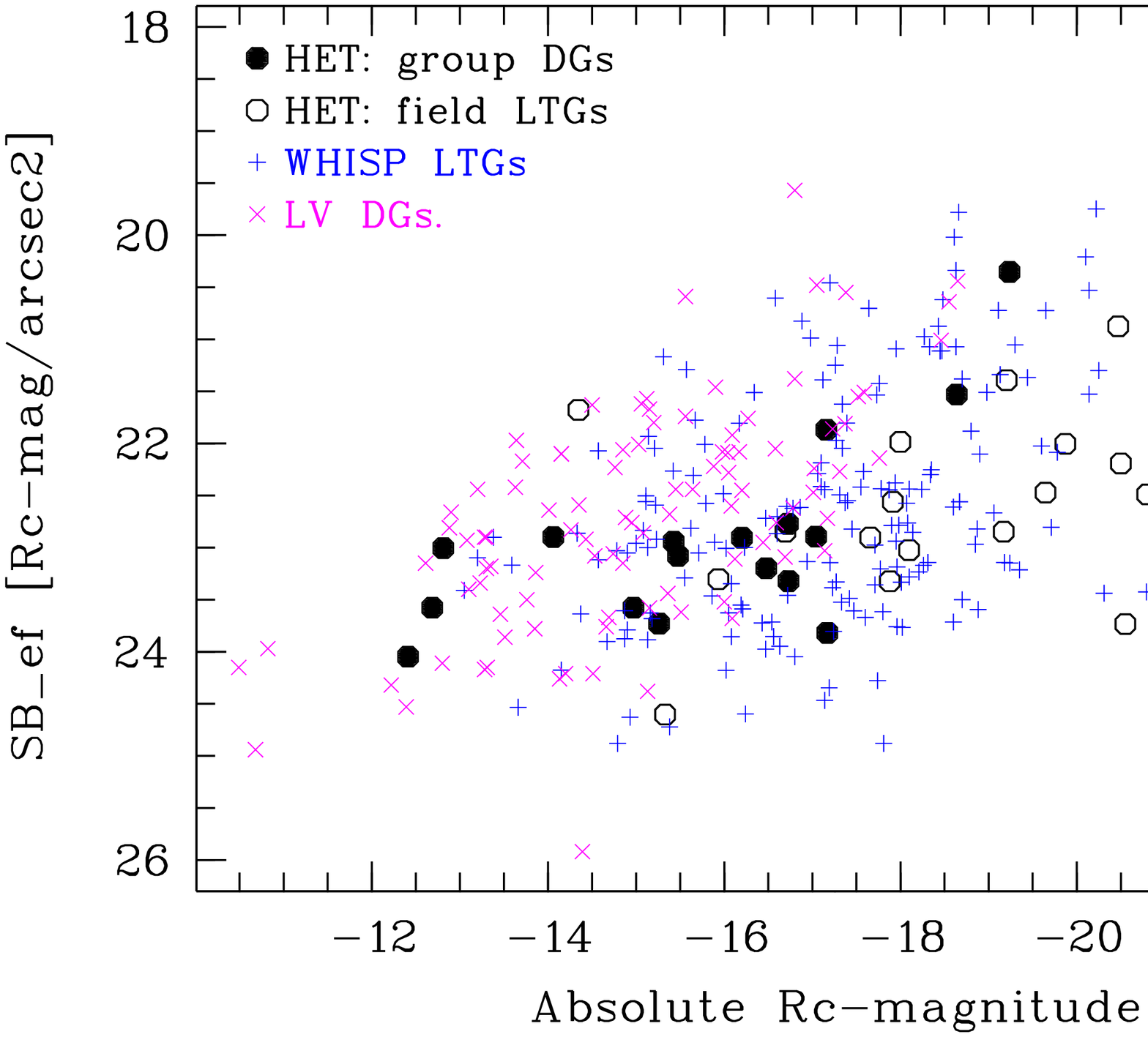}}
\caption{The mean effective SBs of studied galaxies (HET sample), 
compared to those of the DGs of the Local Volume (LV DGs) and to those 
of the LTGs of the WHISP sample (WHISP LTGs). {\it Left:} histograms 
for the HET sample(s), LV DGs, and WHISP LTGs. Coded as in 
Fig.~\ref{fig:CI-MR}. {\it Right:} relation between the mean effective 
SBs and absolute $R_c$-magnitudes.
}
\label{fig:mSBef-MR}
\end{figure*}

\begin{figure*}[h]
\hspace{-2mm}
\resizebox{0.35\textwidth}{!}{\includegraphics{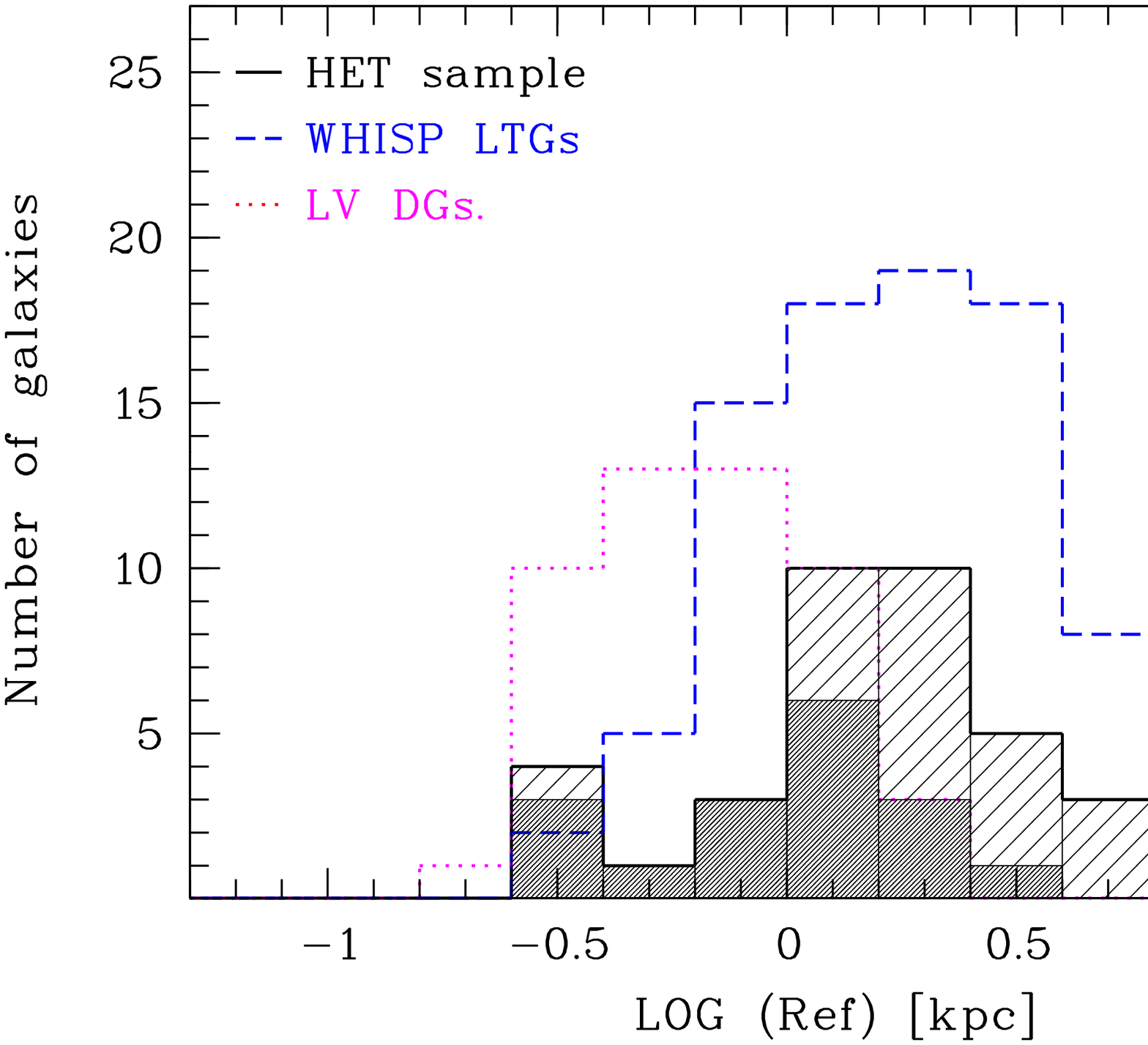}}
\hspace{25mm}
\resizebox{0.35\textwidth}{!}{\includegraphics{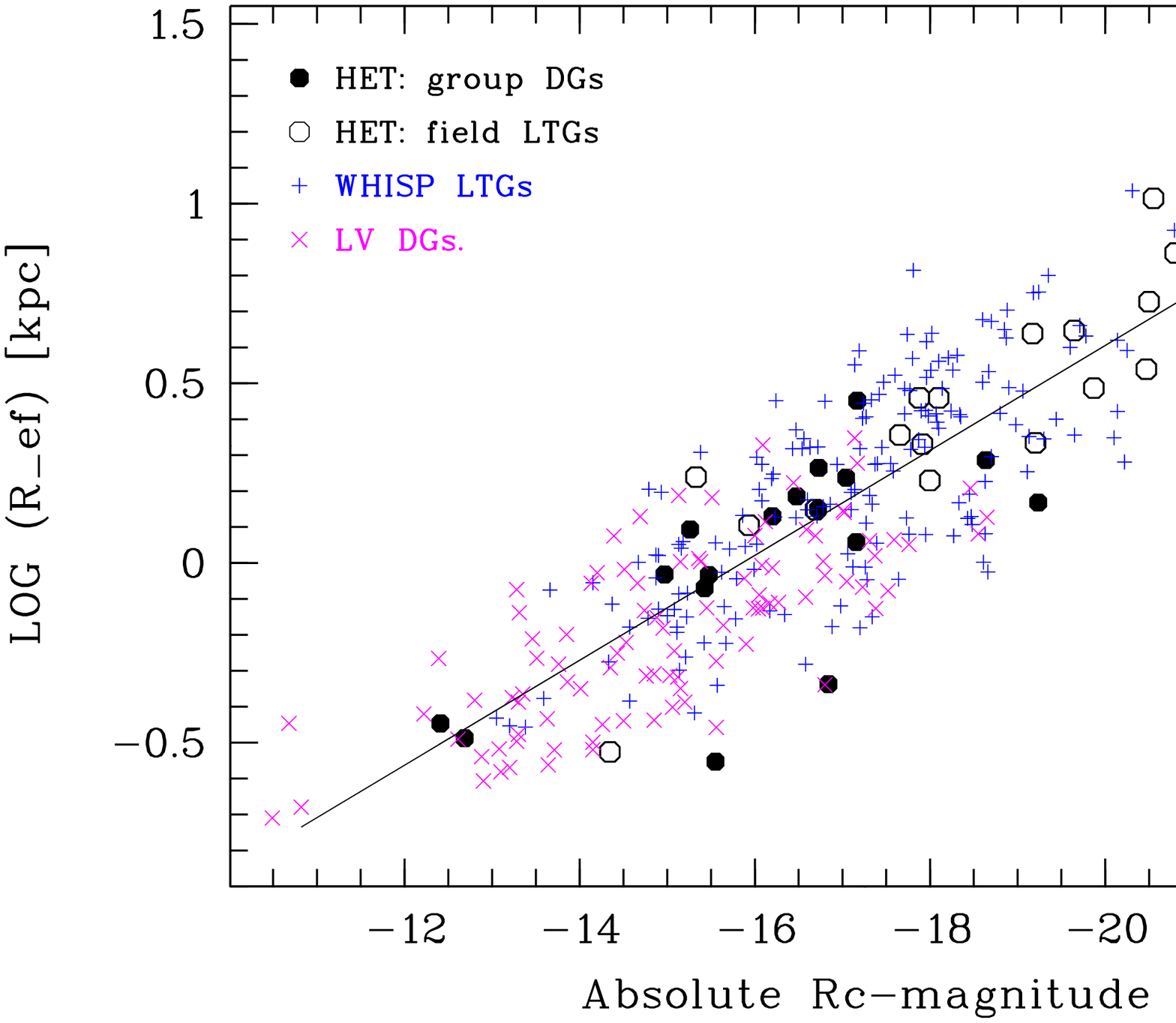}}
\caption{The effective radii ($R_\mathrm{ef}$ in kpc)  of studied 
galaxies (HET sample), compared to those of the DGs of the 
Local Volume (LV DGs) and to those of the LTGs of the WHISP sample 
(WHISP LTGs). 
{\it Left:} histograms, coded as in Fig.~\ref{fig:CI-MR}. 
{\it Right:} relation between the $\log R_\mathrm{ef}$ and $M_{Rc}$ 
is approximated by a linear regression with a slope -0.145$\pm$0.020 
(solid line).
}
\label{fig:logRef-MR}
\end{figure*}

Next, we analyse the relations of effective and exponential disk model 
and colour characteristics with the total luminosity, i.e. the 
photometric scaling relations. In order to put the photometric properties 
of the studied galaxies into a wider context, we compare our data with two 
representative samples of nearby (mostly late-type and/or dwarf) galaxies: 
(1) with the Westerbork HI survey of 171 spiral and irregular galaxies 
(WHISP project of Swaters \& Balcells \cite{WHISP}), and (2) with a 
combined sample of the 98 Local Volume (LV) late-type dwarf galaxies 
(LV-DGs) from Parodi et al.~(\cite{parodi02}), and Sharina 
et al.~(\cite{sharina08}). Since the data in the comparison samples 
are given in the Johnson-Cousins $B, R_c$, we have converted the 
SDSS $g', r'$, and $i'$ (AB-magnitudes) into the Johnson-Cousins 
$B, R_c, I_c$ (Vega-magnitudes) using the transformation equations 
given in the Cross et al. (\cite{cross04}).
 
\begin{figure*}
\hspace{-5mm}
\resizebox{0.37\textwidth}{!}{\includegraphics{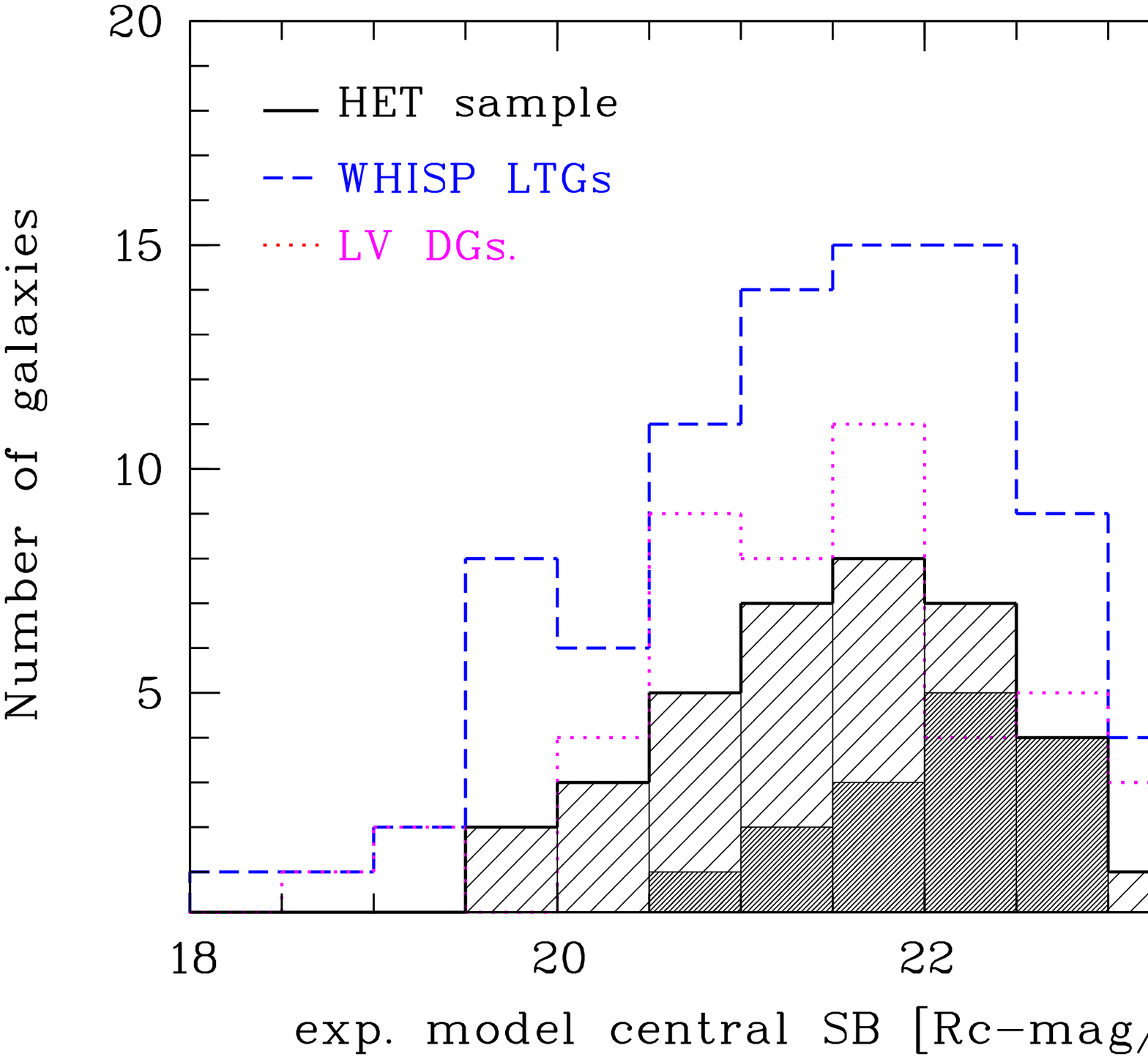}} 
\hspace{22mm}
\resizebox{0.37\textwidth}{!}{\includegraphics{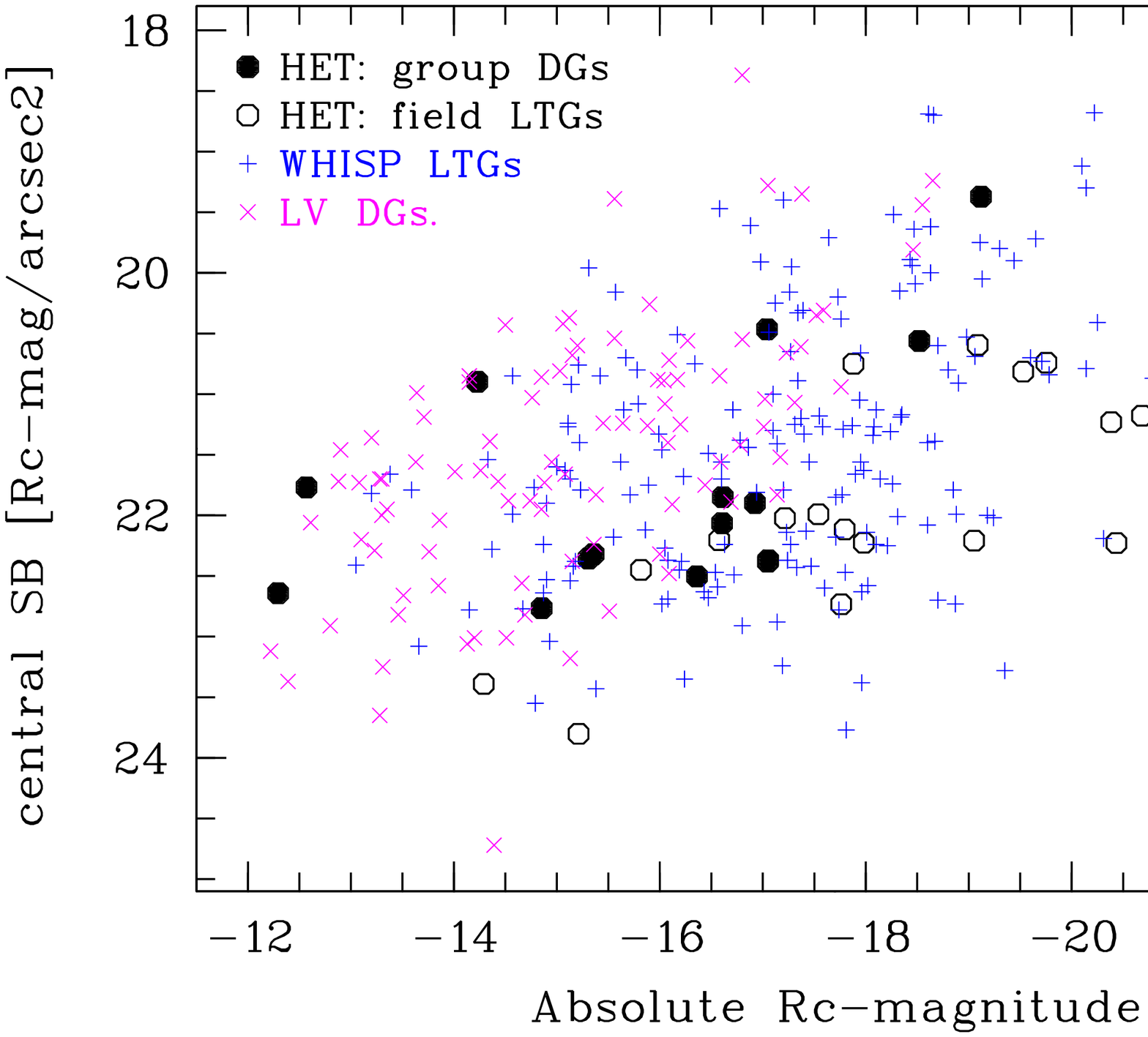}}
\caption{The disk central SB  of studied galaxies (HET sample), compared 
to those of the DGs of the Local Volume (LV DGs) and to those of the 
LTGs of the WHISP sample (WHISP LTGs).
{\it Left:} histograms, coded as in Fig.~\ref{fig:CI-MR}. 
{\it Right:} relation between the disk central SB and $M_{Rc}$.
}
\label{fig:exp0-MR}
\end{figure*}

\begin{figure*}[h]
\hspace{-5mm}
\resizebox{0.35\textwidth}{!}{\includegraphics{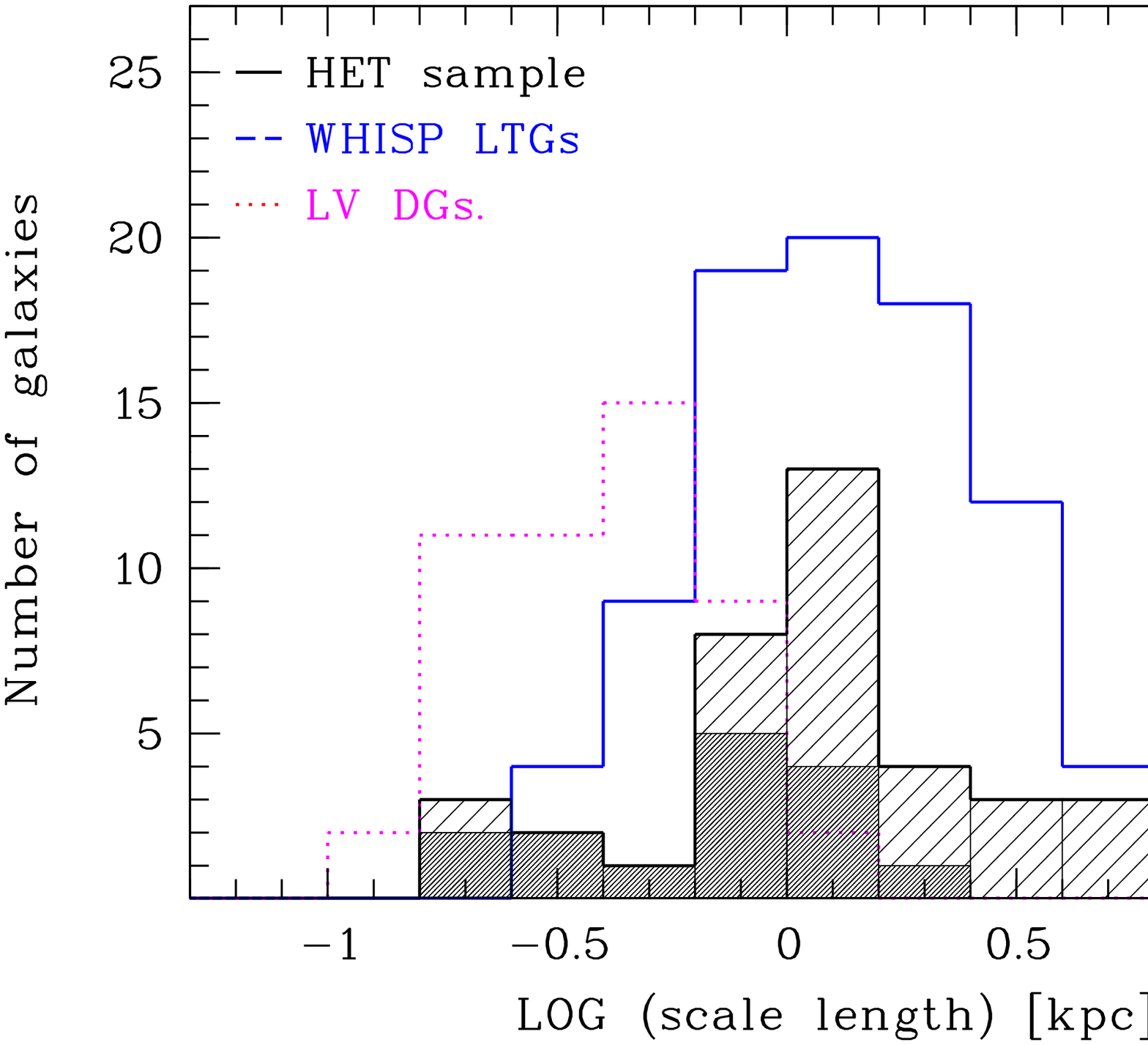}}   
\hspace{22mm}
\resizebox{0.38\textwidth}{!}{\includegraphics{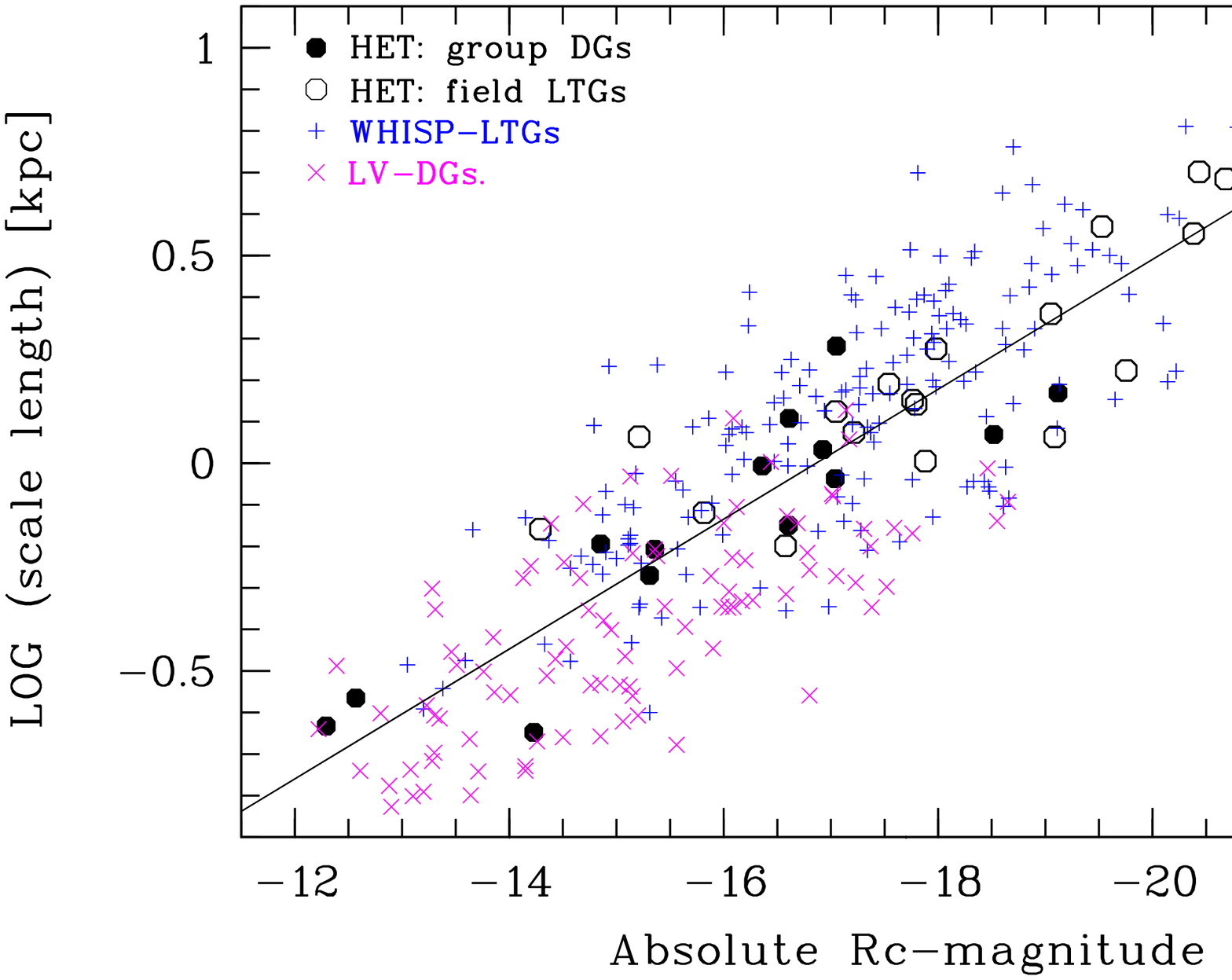}}       
\caption{The ($\log$) scale length ($h$ in kpc)  of studied galaxies 
(HET sample), compared to those of the DGs of the Local Volume (LV DGs) 
and to those of the LTGs of the WHISP sample (WHISP LTGs).
{\it Left:} histograms, coded as in Fig.~\ref{fig:CI-MR}. 
{\it Right:} relation between the $\log h$ and $M_{Rc}$ is approximated 
by a linear regression with a slope -0.156$\pm$0.006 (solid line).
}
\label{fig:scl-MR}
\end{figure*}

\begin{figure*}[h]
\vspace{-20mm}
\hspace{-10mm}
\resizebox{0.27\textwidth}{!}{\includegraphics{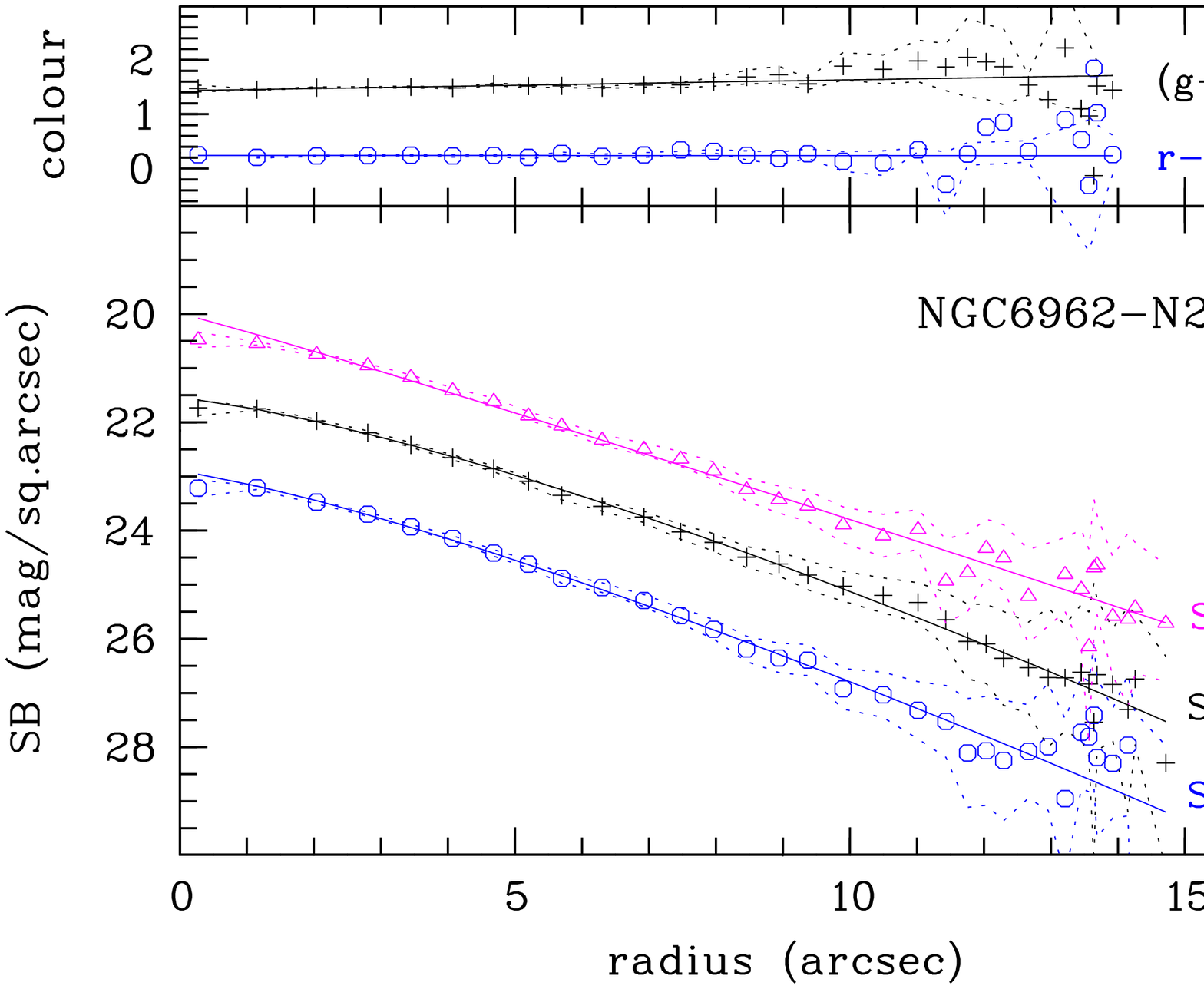}}
\resizebox{0.27\textwidth}{!}{\includegraphics{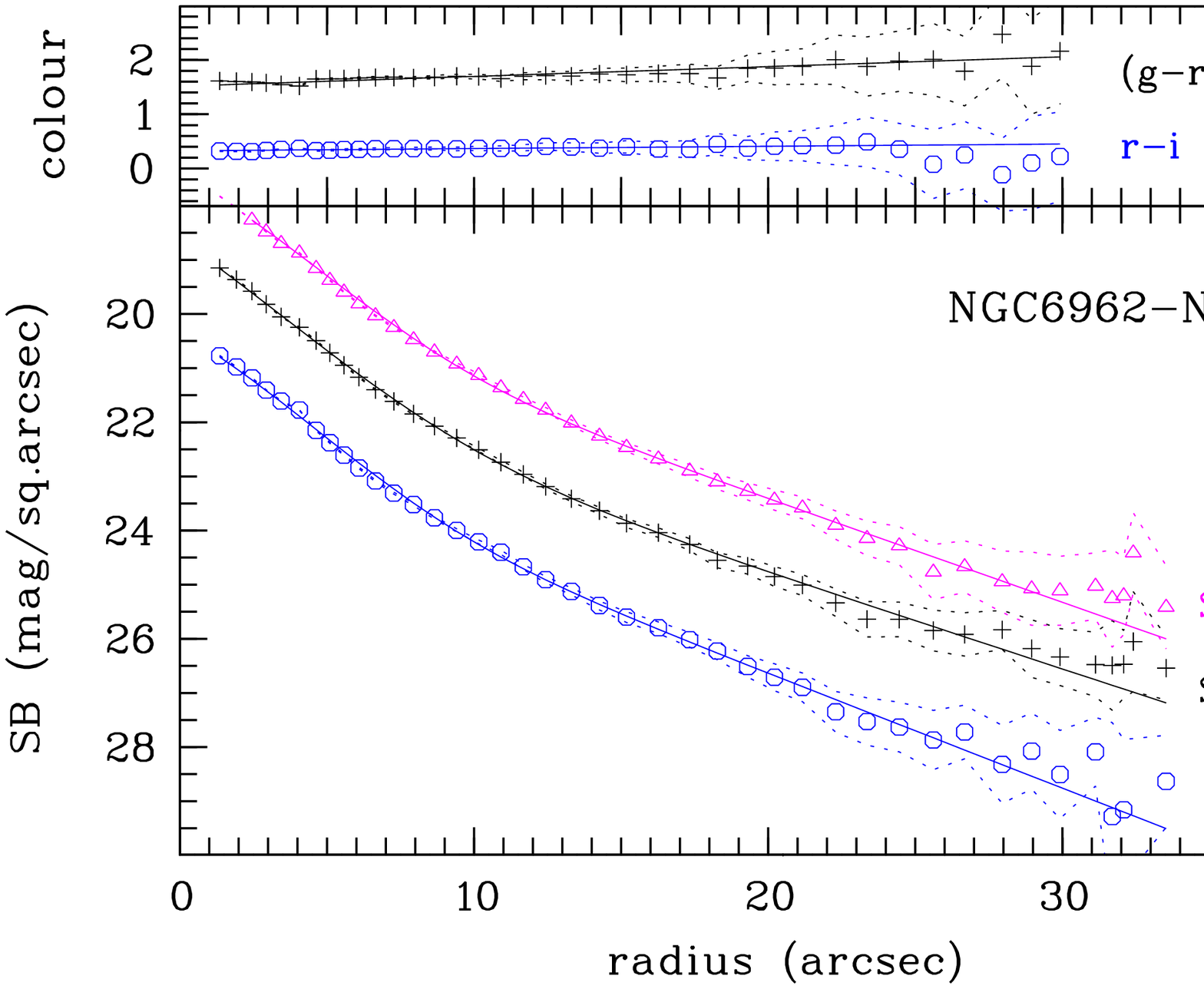}}
\resizebox{0.27\textwidth}{!}{\includegraphics{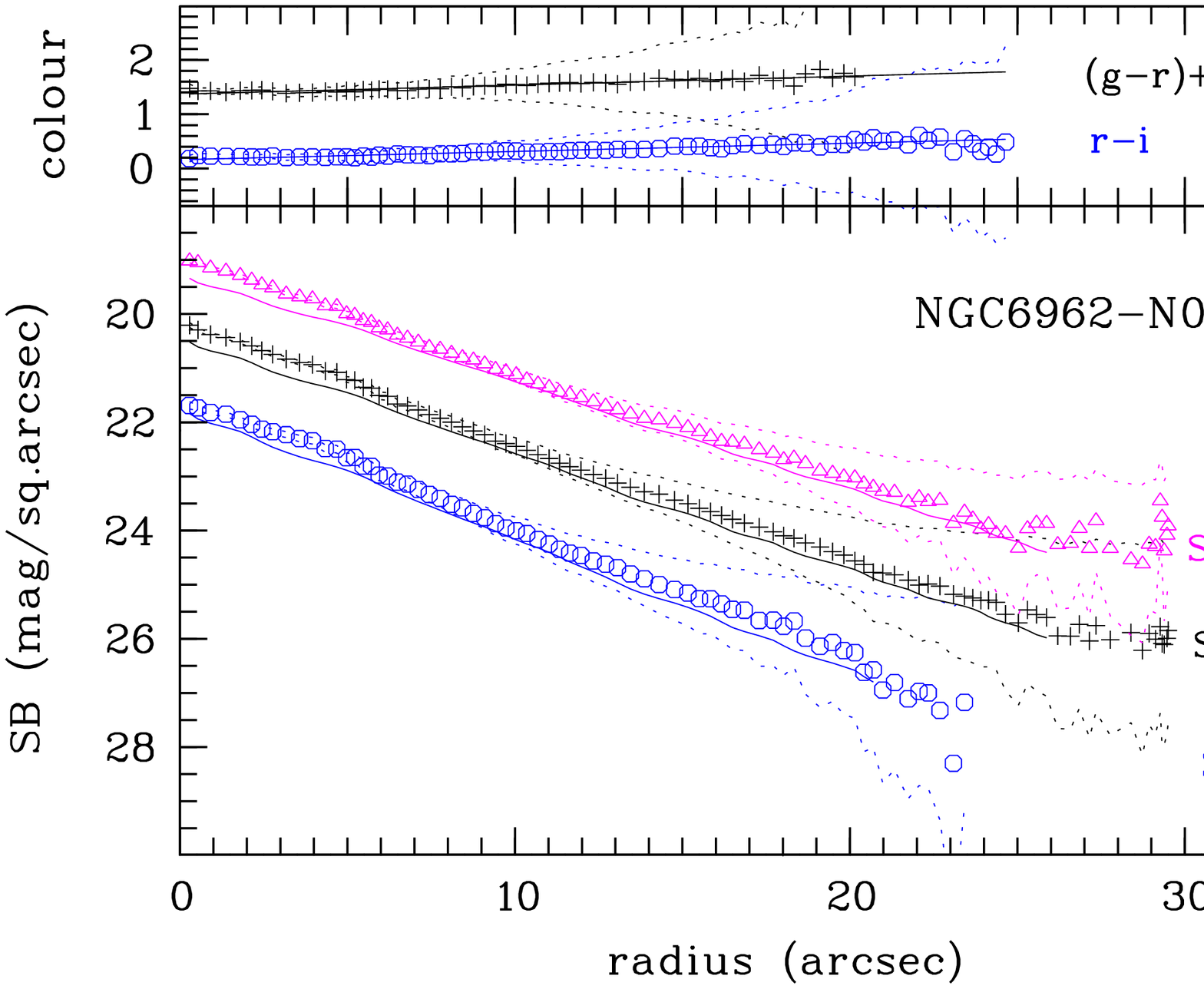}}
%\begin{figure*}[h]
\vspace{-25mm}
\hspace{-5mm}
\resizebox{0.27\textwidth}{!}{\includegraphics{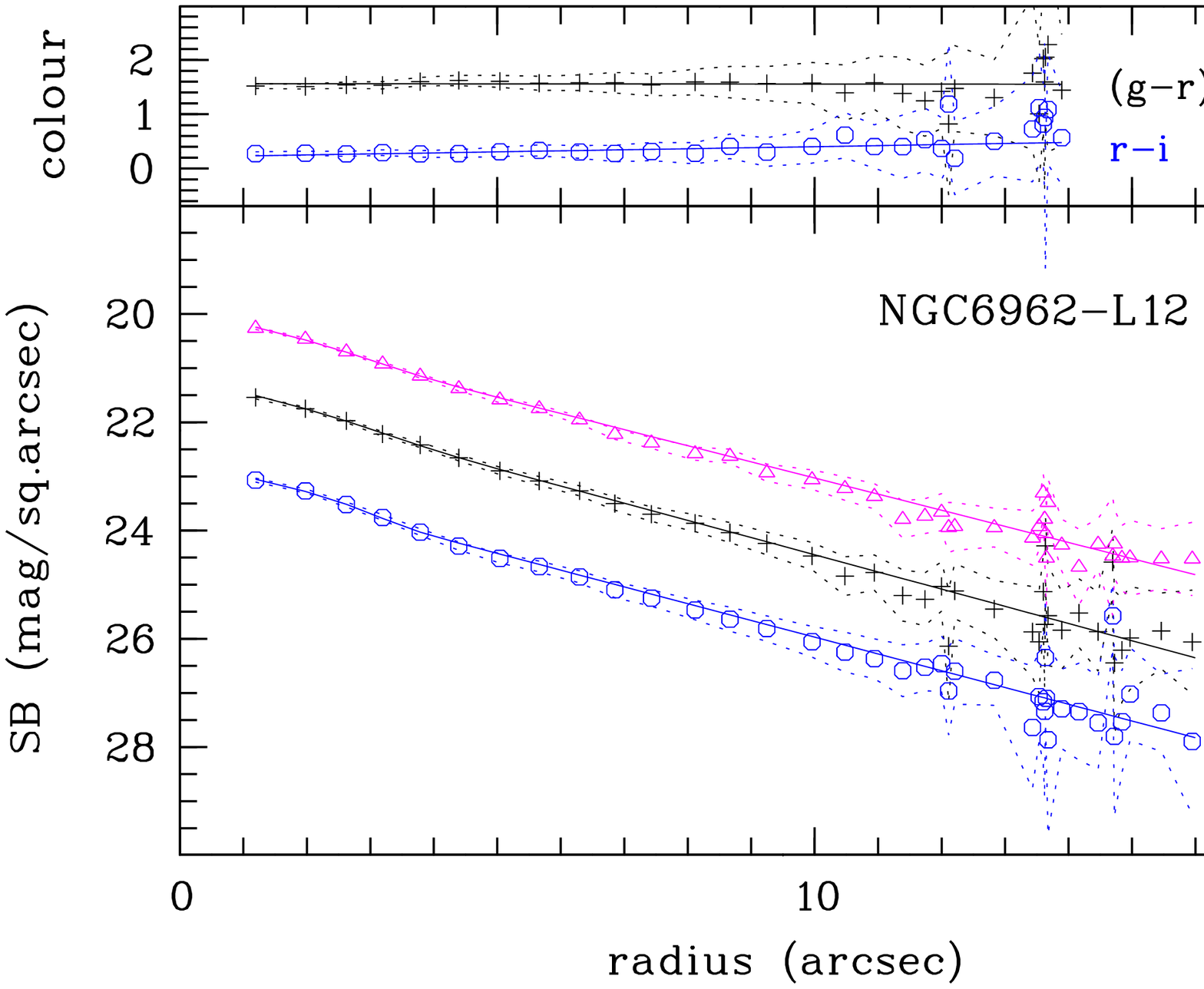}}
\resizebox{0.27\textwidth}{!}{\includegraphics{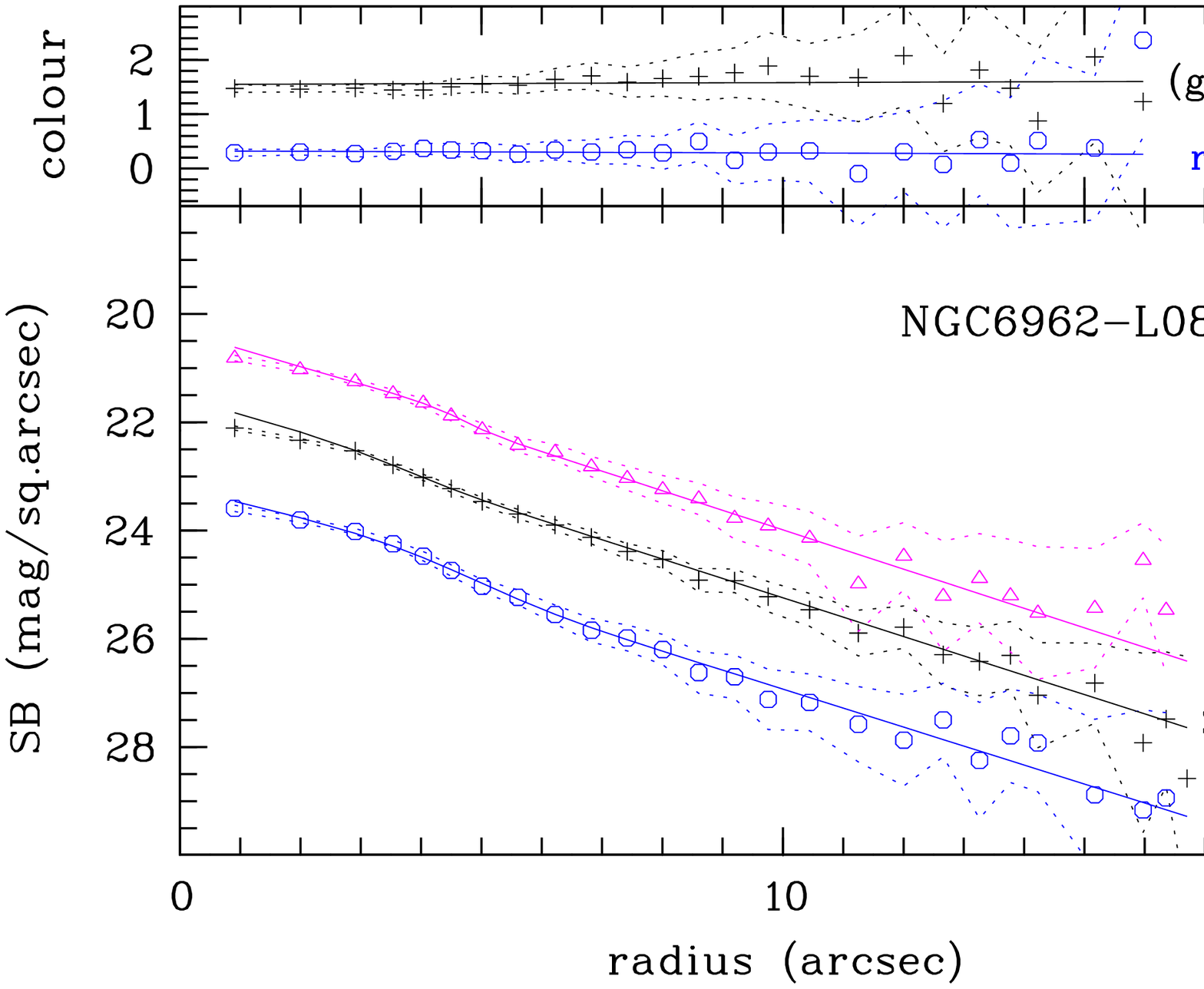}}
\resizebox{0.27\textwidth}{!}{\includegraphics{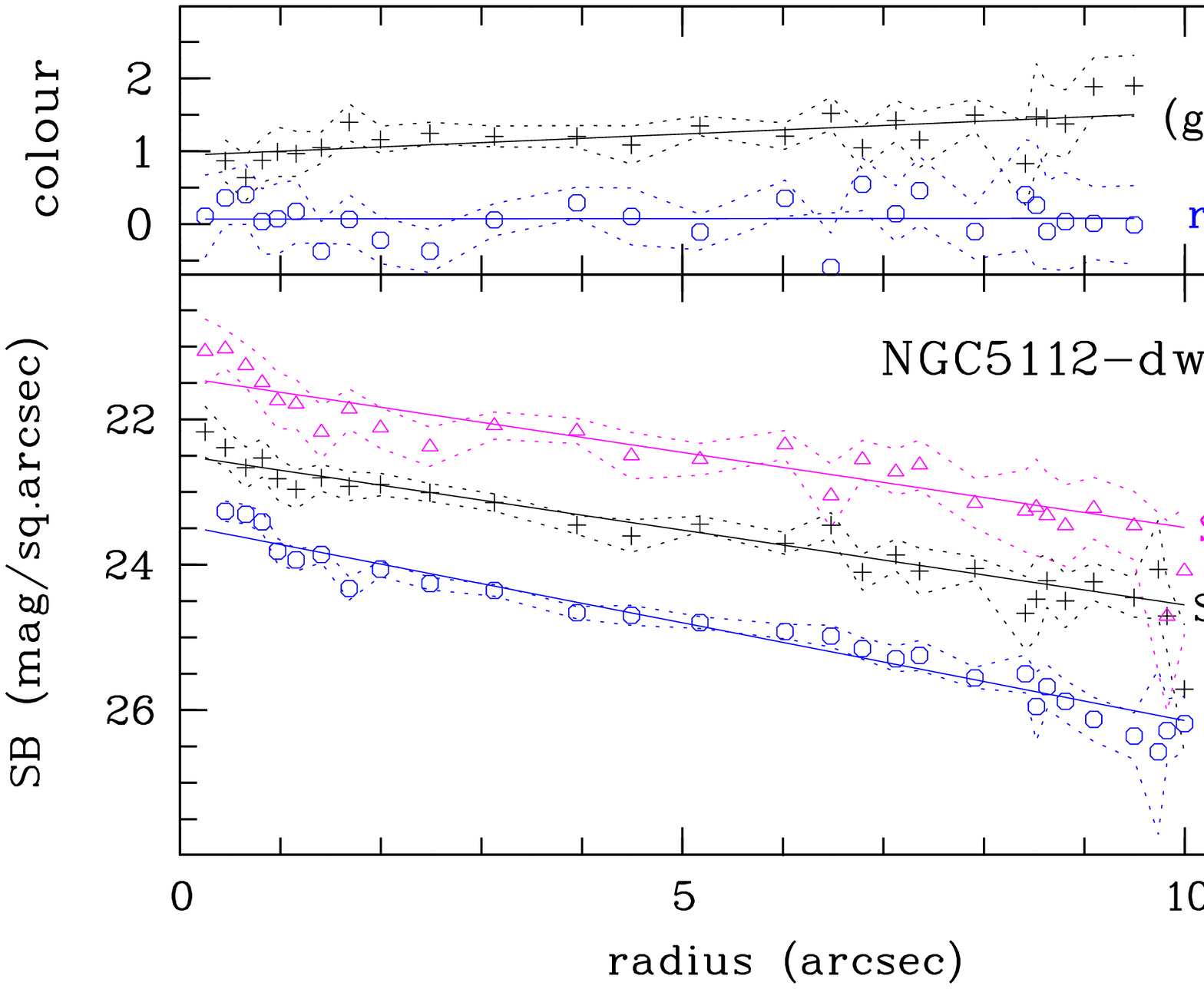}}
%\begin{figure*}[h]
\vspace{-25mm}
\hspace{-5mm}
\resizebox{0.27\textwidth}{!}{\includegraphics{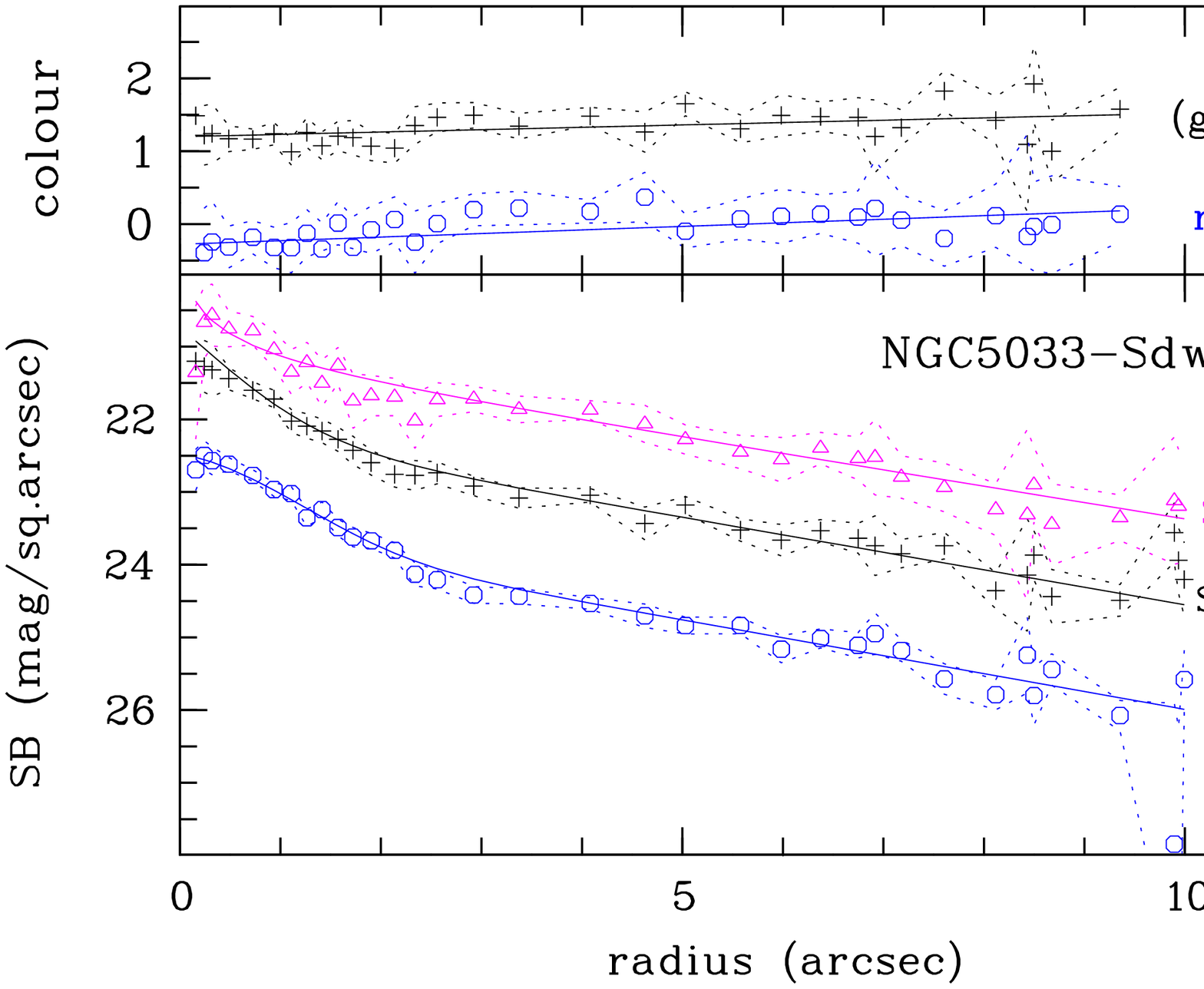}}
\resizebox{0.27\textwidth}{!}{\includegraphics{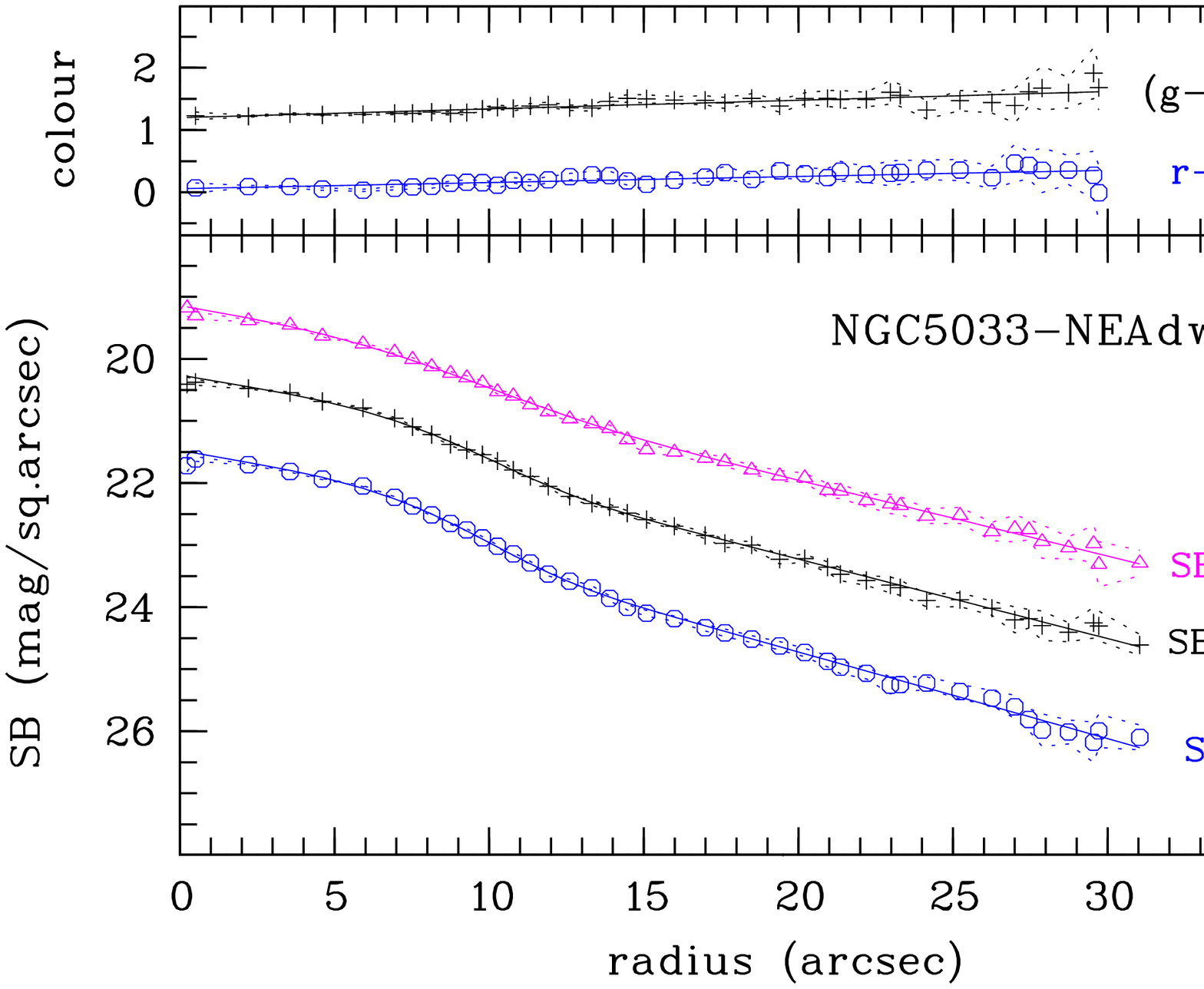}}
\resizebox{0.27\textwidth}{!}{\includegraphics{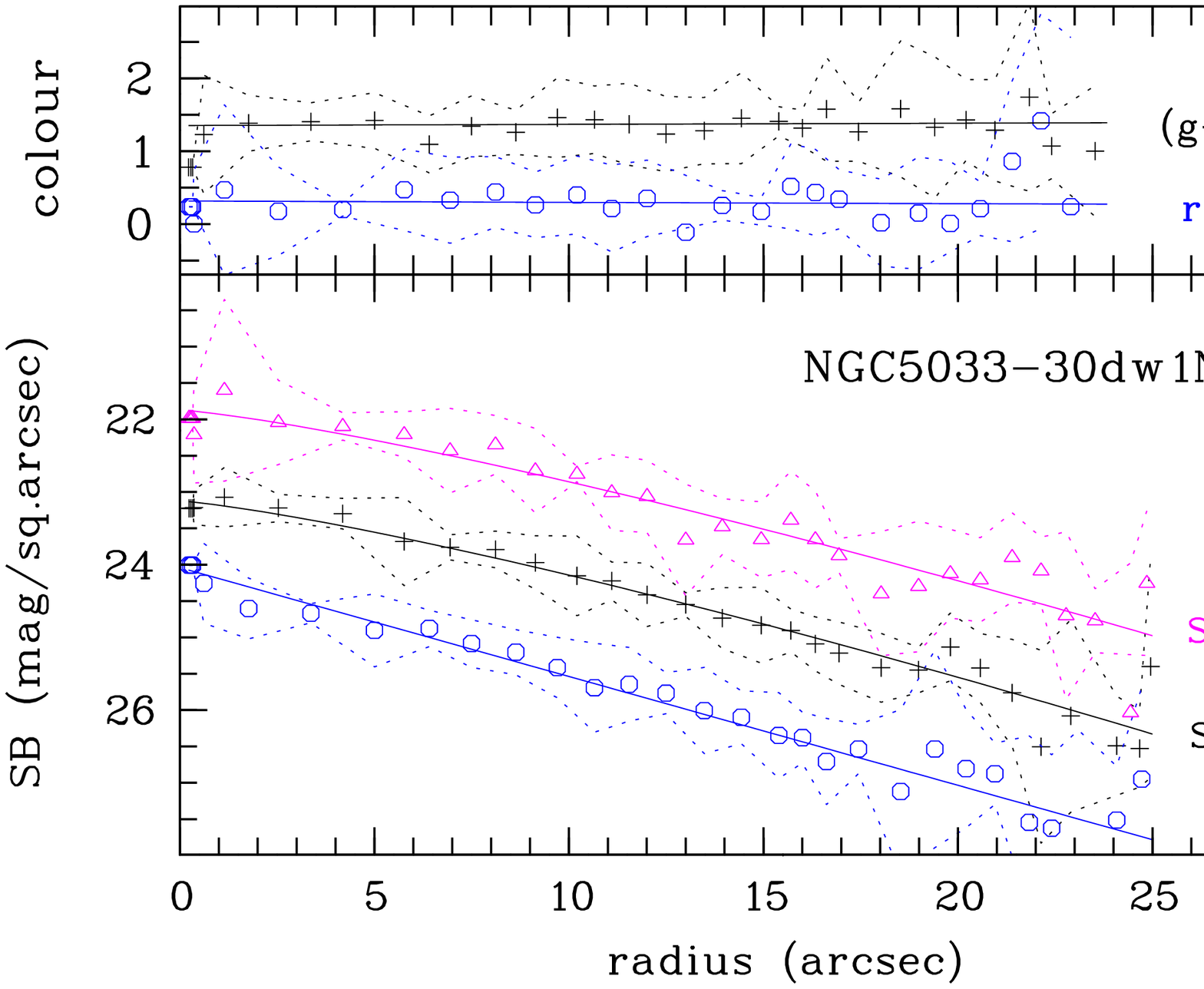}}
%\begin{figure*}[h]
\vspace{-25mm}
\hspace{-5mm}
\resizebox{0.27\textwidth}{!}{\includegraphics{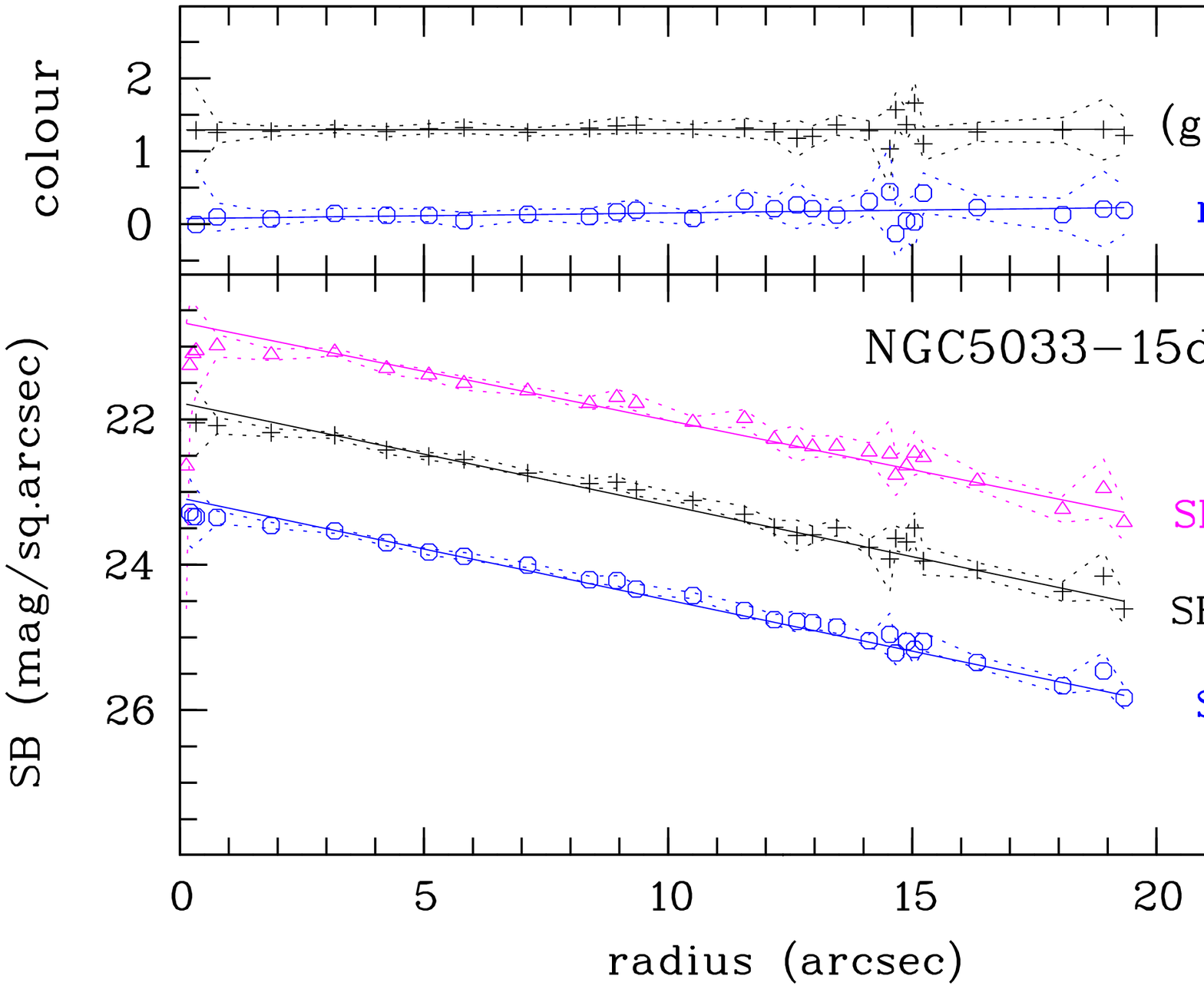}}
\resizebox{0.27\textwidth}{!}{\includegraphics{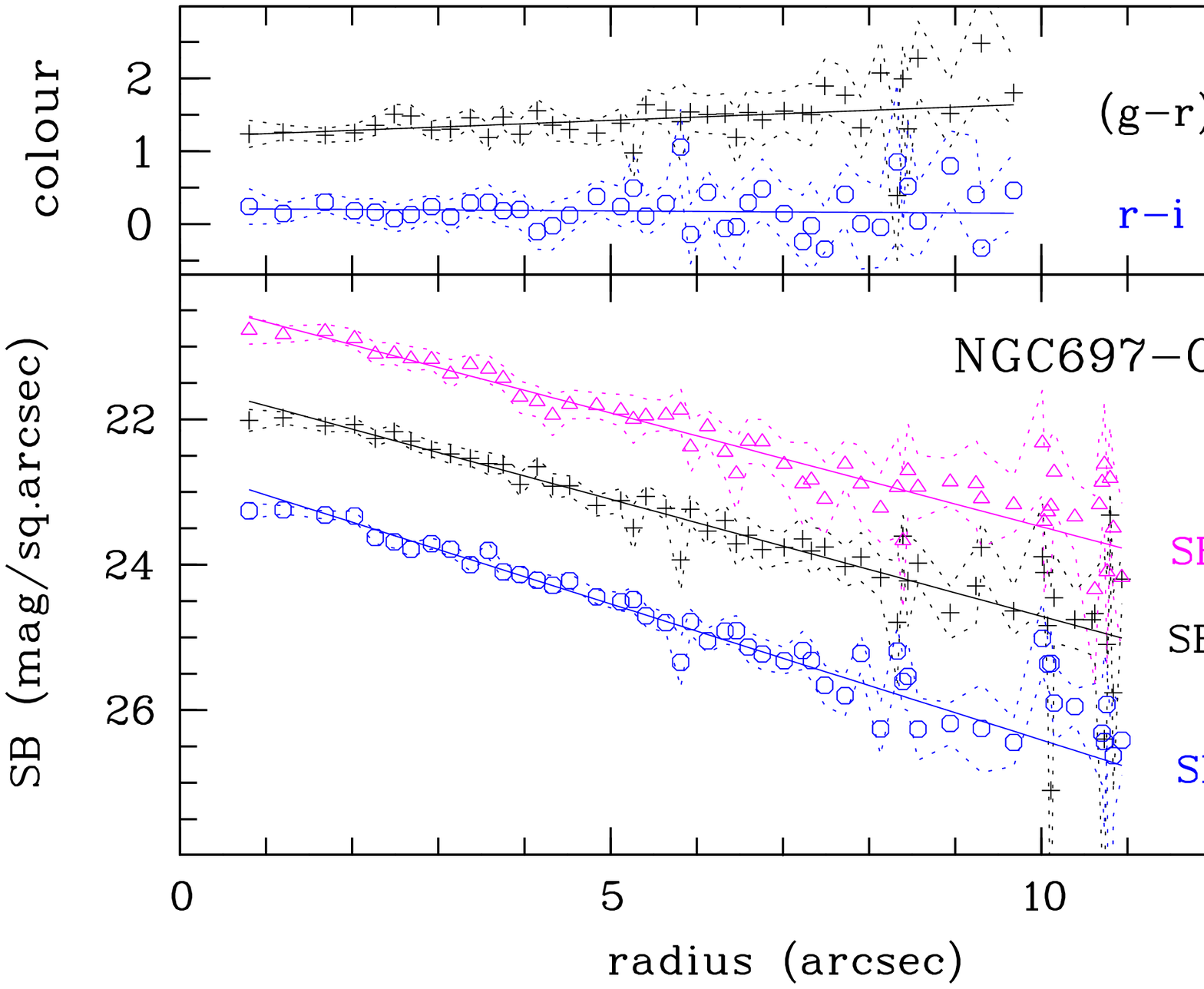}}
\resizebox{0.27\textwidth}{!}{\includegraphics{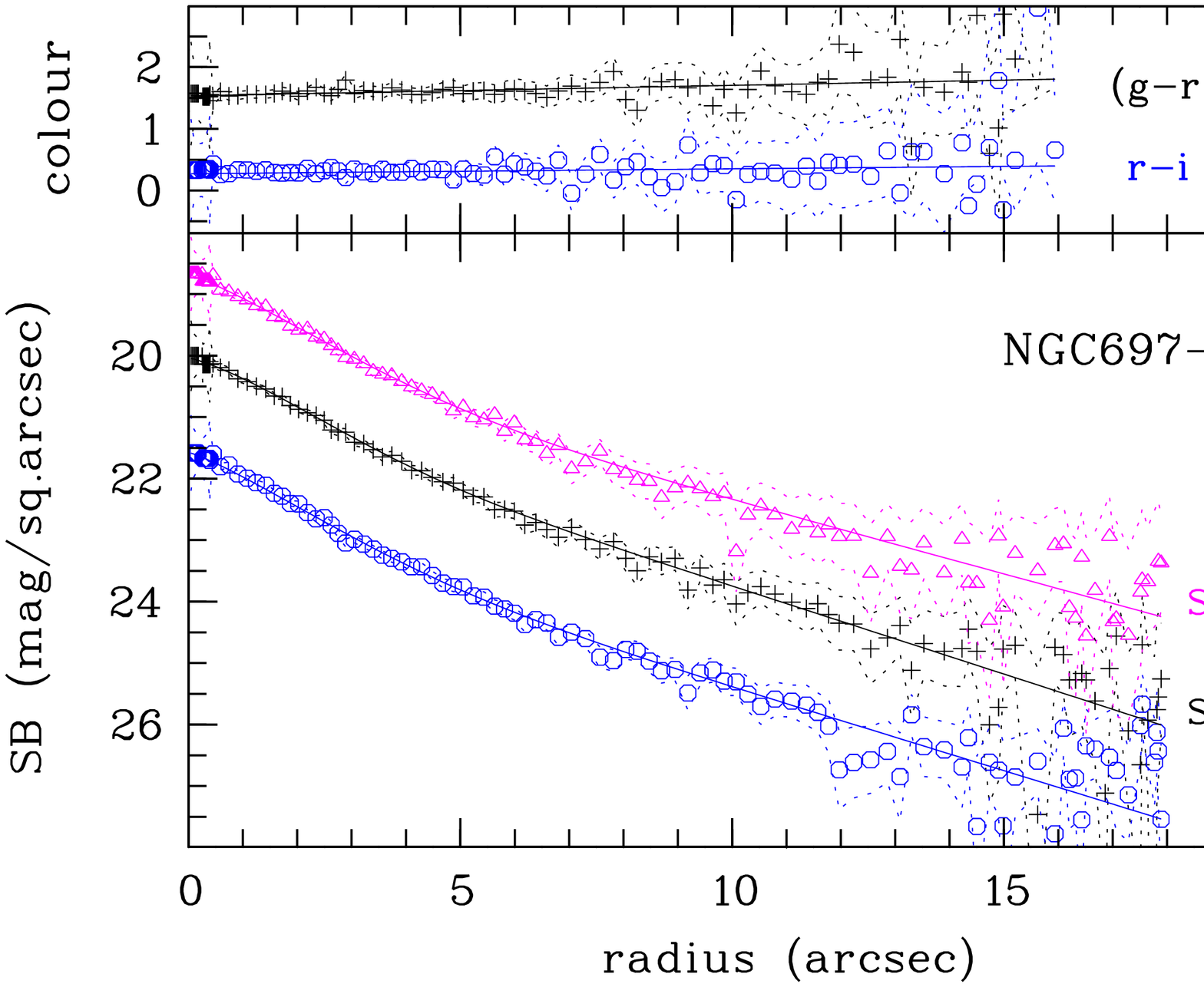}}
%\begin{figure*}[h]
\vspace{-25mm}
\hspace{-5mm}
\resizebox{0.27\textwidth}{!}{\includegraphics{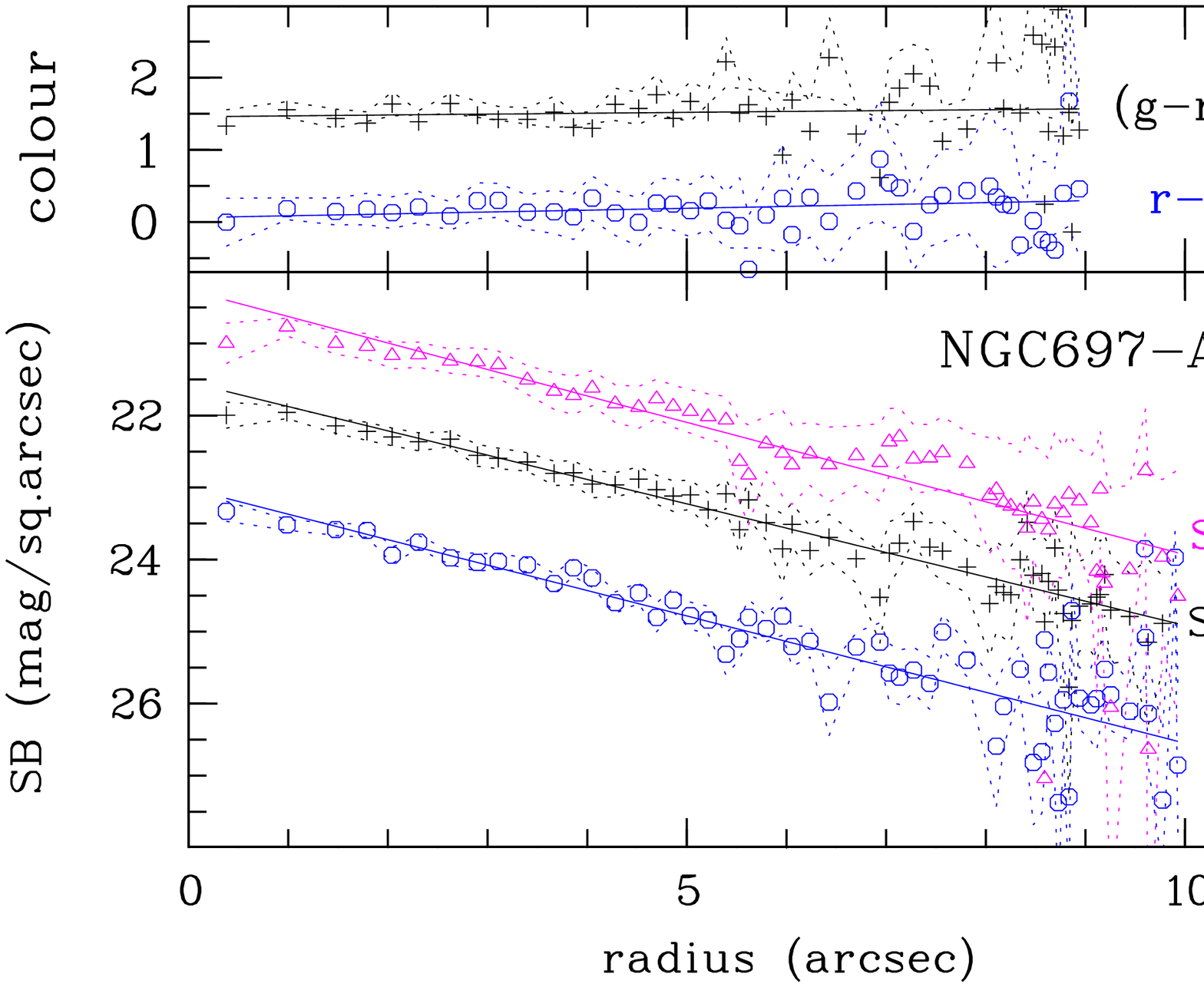}}
\hspace{9mm}
\resizebox{0.27\textwidth}{!}{\includegraphics{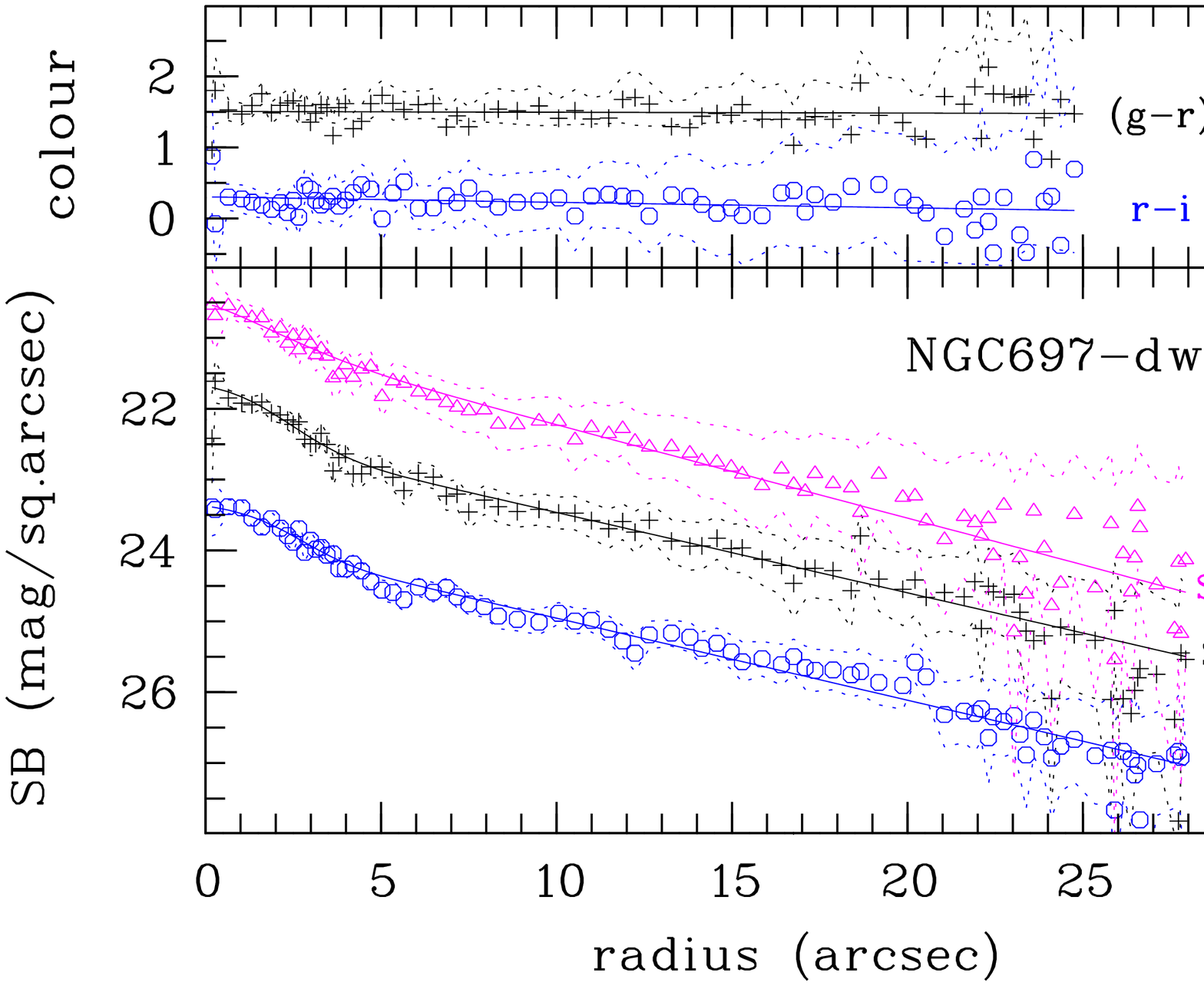}}
\hspace{9mm}
\resizebox{0.27\textwidth}{!}{\includegraphics{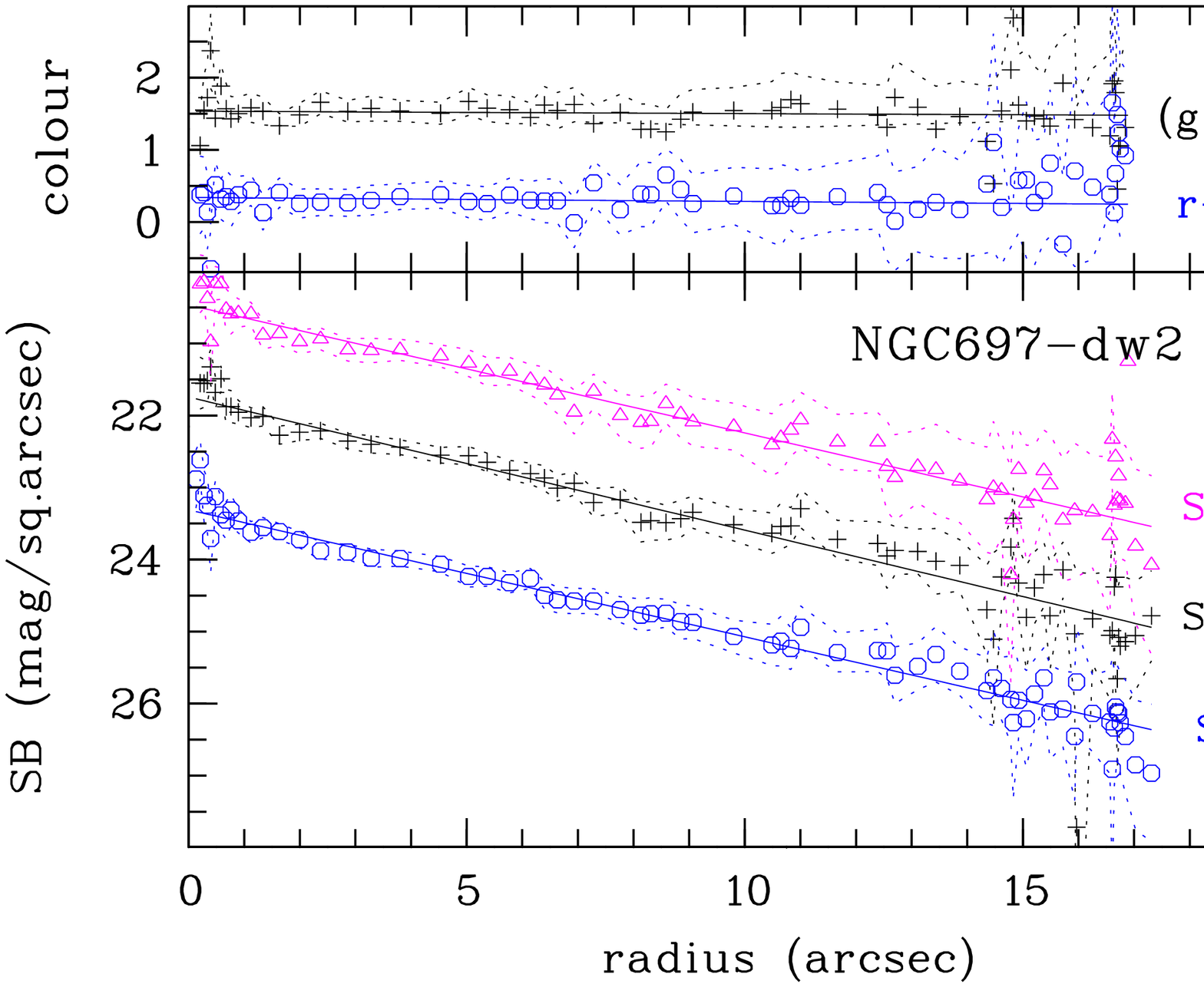}}
\caption{Surface brightness (SB) and colour profiles of new dwarf 
galaxies in studied groups.
{\it Bottom:} the $SB\_r'$-profile is shifted by 1 magnitude (SB\_r+1), 
and the $SB\_i'$-profile is shifted by 2 magnitudes (SB\_i+2);
the Sersic model profiles are indicated by solid lines. 
{\it Top:} the $g'-r'$ colour profile is shifted by 1 magnitude;
the linear regressions are shown with solid lines.
The random errors are delineated with dotted lines. Radius is 
the equivalent radius.}
\label{fig:profiles}
\end{figure*}

In the Figs.~\ref{fig:CI-MR} - \ref{fig:scl-MR} we show the 
histograms of the  effective and exponential disk parameters and 
$(B-R_c)$ colours as well as their scaling relations with the 
absolute magnitude $(M_{Rc})$. In the Table~\ref{tab:HET-LV-WHISP} we 
compare the mean and median values of the SB, colour and linear 
scale characteristics of our HET sample and of both reference samples. 
The photometric characteristics of the HET group DGs (absolute 
magnitudes, colours, linear scale) fit nicely into the range, 
defined by both (local) comparison samples, except the disk central 
SBs of the HET DGs being, as a mean, $\sim$0.5 mag fainter. 
The HET 'field' (bkg) gala\-xies are a random sample of (late-type) 
disk galaxies: their disk central SB $\sim$~22.0 (median value) 
is close to that of the canonical Freeman's disks 
$\sim$~21.75~Bmag arcsec$^{-2}$ (Freeman~\cite{freeman70}); 
their mean colours $(B-R_c) \simeq 0.92$ 
and $(R_c-I_c) \simeq 0.43$ are typical for the Scd - Im galaxies 
(Fukugita et al.~\cite{fukugita96}), and their mean disk scale 
length of $<h>$= 3.7 kpc being very close to the average disk scale 
length of non-interacting disk galaxies in the SDSS 
of 3.79$\pm$2.05 kpc, as given in Fathi et al.~(\cite{fathi10}).

The photometric scaling relations are useful tools, e.g. when 
evaluating the formation scenarios of galaxies and/or addressing 
the question of the environmental influence on the evolution of galaxies. 
Several global and local properties of disk galaxies, e.g. the 
nearly exponential stellar density and light profiles, and their 
correlations are a natural consequence of the inside-out disk 
formation scenario under the assumption of constancy of spin 
parameter in time and angular momentum conservation during the 
gas collapse within hierarchically growing dark matter halos. 
(Avi\-la-Reese \& Firmani.~\cite{AReese00}). Their models predict 
the following relation between the luminosity and the 
Holmberg' radius $L_B \propto R_{Ho}^{2.4}$ (equal to 
$\log R_{Ho} \propto -0.167~M_B$). 

The newly observed galaxies fit nicely the distribution of the 
local LTGs in the $\log h - M_R$ and $\mu_0 - M_R$ planes 
(Fig.~\ref{fig:exp0-MR}, \ref{fig:scl-MR}). In the $\log h - M_R$ 
plane the galaxies of the combined data-set of galaxies of our 
and the reference samples are distributed in a narrow stripe with 
a well defined slope of $-0.156\pm0.006$ ($rms = 0.21$), which 
is similar to the relation predicted in the model of 
Avila-Reese \& Firmani~(\cite{AReese00}). 
Sharina et.al.~(\cite{sharina08}) determined for their LV sample 
a relation of $L \propto h^{2.0}$, which is consistent 
with the central SB constancy of Freeman's disks. 
Actually, the disk central SB slightly increases with increasing 
luminosity in the combined data-set (Fig.~\ref{fig:exp0-MR}), 
as $\mu_0 \propto (0.26\pm.03) M_{Rc}$. This slope is valid for 
the dwarf galaxies in the range $-10 > M_R > -19$, too, however 
with a large scatter. 
The scatter in SBs at a given luminosity certainly reflects contributions
by photometry errors, but also may be influenced by physical mechanisms 
of the disk formation (Sharina et al.~\cite{sharina08}, 
Dalcanton et al.~\cite{dalcanton97}). 
Large scatter renders the $SB - L$ correlation uncertain and brings 
different authors to different conclusions about its reality. 
For example, Sharina~\cite{sharina08} found a (strong) correlation 
of $SB \propto 0.5 M_L$, valid both for the disk central SB ($\mu_0$) 
and for the effective SB ($\mu_{ef}$). This correlation is consistent 
with $L \propto R^4$ relation, valid for the LG DGs 
(Dekel \& Silk~\cite{dekel86}) and corresponds to the gas outflow model 
from the DGs (Larson 1974). On the other hand, 
Carrasco et al.~(\cite{carrasco01}) found no correlation in the 
$\mu_0 - M_V$ plane, and a weak correlation in the $\log h - M_V$ plane. 
They conclude that correlations in these two planes may be produced 
by selection effects.

As evident in Figs~\ref{fig:logRef-MR} and \ref{fig:scl-MR} our combined 
data-set shows a strong correlation between the linear scale 
($\log h,\log R_{\rm{ef}}$) and total luminosity in the form 
$L \sim R^{2.65\pm 0.10}$. A statistically significant correlation 
between  the SB and luminosity could be traced in Figs ~\ref{fig:mSBef-MR} and 
\ref{fig:exp0-MR} too, despite of large scatter, and it results 
in a similar luminosity- radius relation (as $L \sim R^{2.6\pm 0.1}$).
  
\section{Summary and conclusions}

The results of the given study can be summarised as follows: \\
 - we have selected a small sample of five X-ray dim, late-type dominated, 
and reasonably isolated groups of galaxies in and around the Virgo 
super-cluster, at the distances where dwarf galaxies could have been 
distinguished. \\
 - we have selected about 60 dwarf galaxy candidates in these groups 
based on their SB and colour characteristics and on their morphology. 
The follow up spectroscopy of the 55 highest priority candidates 
with the HET LRIS has resulted in 48 new redshift determinations. The 
new data have shown that 17 galaxies are true group members, 2 are 
DGs located in a newly detected foreground group; 28 galaxies 
are late-type galaxies in the background of studied groups. 
We conclude a nearly 55\% success rate in morphological 
classification of dwarf galaxies.\\
 - 19 galaxies (i.e. 50\% of all detections) with sufficiently 
high S/N ratios show emission line ratios typical for star-forming 
HII dwarfs; \\
 - detailed surface photometry of the 38 studied galaxies on the 
SDSS $g', r'$, and $i'$ frames has revealed several cases of 
large differences (up to 3 mag), when comparing newly derived magnitudes 
with those in SDSS photometric catalogue, caused by poor object 
de-blending (detection) and shredding by automatic pipeline reduction of 
large, very LSB galaxies; \\
 - photometric effective parameters ($R_{ef}, \mu_{ef}$) and exponential 
disk model parameters ($h, \mu_0^{exp}$) of the studied galaxies fit 
nicely to the distribution of corresponding characteristics of the 
Local Volume dIrr's and of the late-type galaxies of the WHISP sample, 
defining a narrow/tight scaling relation between the exponential 
scale length (also $R_{ef}$) and luminosity $h \propto L^{0.4\pm 0.03}$ 
(or $L \propto R^{2.6\pm 0.1}$). Our studied galaxies show a shallower 
dependency between the central (and also effective) SB and luminosity 
than galaxies in two comparison samples. This is mainly caused be our 
target selection criteria - selecting preferentially LSB galaxies.\\
- the dwarf members of the groups show positive colour gradients, e.g.
getting redder to their rims. This might be indicative for a outside-in
dying out of the star formation activity contrary to what is discussed
for the star formation history of large disk galaxies.

\section*{Acknowledgements}

We are grateful to the anonymous referee whose comments and suggestions
were most helpful in improving the paper.
It is a pleasure to acknowledge the support by the Resident Astronomers
of the Hobby-Eberly Telescope (HET) which made the observations possible
despite the many hardly visible low surface objects involved in the
program. We especially acknowledge the efficient and very helpful
on-line discussion with the 'RAs' during the execution of our tar\-gets within 
the general queue scheduling of the HET.
The Hobby-Eberly Telescope is a joint project of the 
University of Texas at Austin, the Pennsylvania State University, 
Stanford University, Ludwig-Maximilians-Universit\"at M\"un\-chen, 
and Georg-August-Universit\"at G\"ottingen. The HET is na\-med 
in honor of its principal benefactors, William P. Hobby and 
Robert E. Eberly. The Marcario Low Resolution Spectrograph is 
named for Mike Marcario of High Lonesome Optics who fabricated 
several optics for the instrument but died before its completion. 
The LRS is a joint project of the Hobby-Eberly Telescope partnership 
and the Instituto de Astronom\'a de la Universidad Nacional Autonoma de
Mexico. The research of JV has been supported by the Estonian Science
Foundation grants 6106 and 7765, by the Estonian Ministry for Education and Science 
research project SF0060067s08, and by the European Structural Funds grant for 
the Centre of Excellence "Dark Matter in (Astro)particle Physics and Cosmology" TK120. 
This study has made use of the 
NASA/IPAC Extragalactic Database (NED) which is operated by the 
Jet Propulsion Laboratory, California Institute of Technology, 
under contract with the National Aeronautics and Space Administration, 
the STScI Digitized Sky Survey (DSS), and the Sloan Digital Sky Survey (SDSS).

%=========\\

\begin{table*}[t]
\caption{Log of the spectroscopic observations with the HET LRS.}
\protect\label{tab:speclog}
\begin{center}
\begin{small}
\begin{tabular}{lllclcl}
\hline
Galaxy         &R.A.      &DEC  & set-up  & date      & Exp.-time & remarks\\
                &2000      &2000  &         &           & [sec]     &        \\
\hline
UGC 378        &00:37:59.0&+48:11:53.0  & g1\_2.0 & 2008 08 06& 2400 & \\
NGC 278-kkh4  &01:05:49.9&+45:30:53.6  & g1\_2.0 & 2008 09 30& 2700 & second exposure 2008 10 02\\
NGC 278-kk6   &00:37:28.6&+48:10:25.1  & g1\_2.0 & 2008 09 24& 1200 & only one exposure\\
\hline
IC 65-dw1      &01:00:10.1&+47:34:01  & g2\_1.0 & 2005 11 20& 3000 & \\
IC 65-dw2      &01:00:07.7&+47:56:05  & g2\_1.0 & 2007 11 09& 2400 & \\
 -"-              &&   & g2\_2.0 & 2007 09 14& 1200 & \\
IC 65-dw3     &01:01:16.1&+47:44:33  & g2\_2.0 & 2007 09 13& 1800 & \\
IC 65-dw4     &01:01:44.4&+47:52:06  & g2\_2.0 & 2007 09 08& 1200 & \\
PGC 3574       &00:59:56.8&+47:47:10  & g2\_1.0 & 2005 11 20& 1968 & \\
PGC 3684       &01:01:45.3&+47:54:08  & g2\_1.0 & 2005 01 25& 3000 & \\
\hline
NGC 524SW-16  &01:25:08.1&+09:36:27.0  & g2\_2.0 & 2008 01 31& 1200 & \\
NGC 524SW-18 &01:24:51.7&+09:06:50.0 & g2\_2.0 & 2008 01 14& 1200 & \\
\hline
NGC 691-E      &01:50:51.2&+21:45:54  & g1\_2.0 & 2008 12 31& 2400 & \\
%\hline
NGC 697-dw1    &01:49:52.8&+22:26:35.8  & g1\_2.0 & 2008 10 02& 2400 & \\
NGC 697-dw2    &01:48:50.3&+22:06:50.0  & g1\_2.0 & 2008 10 30& 2400 & \\
NGC 697-dw3    &01:47:48.3&+22:08:31.9  & g1\_2.0 & 2008 10 07& 2400 & \\
NGC 697-A      &01:51:49.8&+22:18:34  & g1\_2.0 & 2009 01 22& 2400 & \\
NGC 697-B      &01:51:43.9&+22:20:00  & g1\_2.0 & 2009 01 24& 2400 & \\
NGC 697-C      &01:52:22.4&+22:10:07  & g1\_2.0 & 2010 01 06& 2011 & only one exposure \\
  -"-             &&   & g1\_2.0 & 2010 01 19& 2400 & \\
NGC 697-D      &01:50:26.4&+21:57:05  & g1\_2.0 & 2009 01 26& 2400 & \\
NGC 697-F      &01:51:31.1&+22:45:40  & g1\_2.0 & 2009 12 11& 2400 & \\
NGC 697-G      &01:50:20.3&+22:25:53  & g1\_2.0 & 2010 02 08& 2149 & \\
NGC 697-J      &01:48:13.3&+22:02:30  & g1\_2.0 & 2010 01 17& 2400 & \\
NGC 697-H      &01:51:56.4&+22:41:42  & g1\_2.0 & 2009 12 14& 2400 & \\
NGC 772-gal1   &01:58:58.5&+18:57:46  & g1\_2.0 & 2009 12 12& 2400 & one slit with NGC 772-gal2\\
NGC 772-gal2   &01:59:07.0&+18:57:40  & g1\_2.0 & 2009 12 12& 2400 & one slit with NGC 772-gal1\\
\hline
NGC 972-kdg17  &02:31:00.4&+27:57:25.7  & g1\_2.0 & 2008 10 06& 2300 & \\
\hline
NGC 5033-15    &13:14:20.7&+36:34:11.0  & g2\_2.0 & 2008 01 14& 2400 & \\
NGC 5033-30N   &13:10:57.5&+36:48:59.0& g2\_2.0 & 2008 02 07& 2400 & \\
NGC 5033-30S   &13:10:57.5&+36:48:59.0& g2\_2.0 & 2008 02 07& 2400 & \\
NGC 5033NE-Adw &13:17:07.6&+37:57:34 & g1\_2.0 & 2008 12 21& 2400 & \\
NGC 5033WW-LSB &13:04:22.5&+36:29:40  & g1\_2.0 & 2009 01 25& 2400 & \\
NGC 5033S-dw1  &13:14:23.2&+35:43:41  & g1\_2.0 & 2009 01 29& 2400 & \\
NGC 5033-23    &13:11:06.5&+37:10:40  & g1\_2.0 & 2009 02 01& 2400 & \\
%\hline
NGC 5112-MAPS  &13:23:05.9&+39:09:25  & g1\_2.0 & 2009 02 02& 2400 & \\
NGC 5112-dw1   &13:22:13.0&+38:44:55  & g1\_2.0 & 2009 02 27& 2400 & \\
NGC 5112-dw2   &13:21:58.4&+38:50:54  & g1\_2.0 & 2009 02 28& 2400 & \\
\hline
NGC 6278-kkr31  &16:58:33.5&+23:12:20  & g1\_2.0 & 2009 02 02& 2200 & \\
NGC 6278-A      &17:01:31.4&+22:44:41  & g1\_2.0 & 2009 03 02& 2400 & \\
NGC 6278-5      &17:00:33.2&+22:51:58  & g1\_2.0 & 2009 03 04& 2450 & \\
NGC 6278-C      &17:00:39.4&+23:03:08  & g1\_2.0 & 2009 03 05& 2400 & \\
\hline
NGC 6962-B10   &20:46:56.8&+00:12:37  & g1\_2.0 & 2010 07 20& 2400 & \\
NGC 6962-Ec1   &20:50:18.4&+00:45:20.0  & g1\_2.0 & 2010 06 06& 2400 & \\
NGC 6962-L02   &20:47:01.3&+00:26:56  & g1\_2.0 & 2010 07 13& 1896 & \\
NGC 6962-L08   &20:48:04.9&+00:35:13.0  & g1\_2.0 & 2010 05 20& 2400 & \\
NGC 6962-L12   &20:47:22.3&+00:21:42.0  & g1\_2.0 & 2008 11 04& 2400 & \\
NGC 6962-N02   &20:48:23.7&+00:23:00.0  & g1\_2.0 & 2008 06 05& 2400 & \\
NGC 6962-N06   &20:48:29.1&+00:05:41  & g1\_2.0 & 2010 06 11& 2400 & \\
NGC 6962-N16   &20:46:32.1&+00:33:18.0  & g1\_2.0 & 2008 06 11& 2400 & \\
NGC 6962-N19   &20:51:53.6&+00:20:13  & g1\_2.0 & 2010 06 16& 2280 & \\
NGC 6962-N20   &20:46:49.9&+00:39:44.0  & g1\_2.0 & 2008 10 27& 1801 & \\
NGC 6962-SE14  &20:50:55.8&-00:11:07.3  & g1\_2.0 & 2010 06 14& 2400 & \\
\end{tabular}
\end{small}
\end{center}
\end{table*}

\begin{table*}[t]
\caption{Spectroscopic results.
Columns contain the following data: (1) galaxy sequence number; 
(2) galaxy name; (3) galaxy identification in the 
SDSS DR9 PhotoObject catalogue (generally the brightest part, when 
shredded objects); (4, 5) the heliocentric velocity and its error, 
based on the mean of individual line measurements; 
(6) type of spectrum: 'HII' is for spectrum dominated by the emission 
lines of HII regions, 'abs' is for absorption lines only, 'HII/abs' - 
when spectrum shows both emission and absorption lines, 'HII/AGN' refers 
to relatively broad emission lines, 'no HII' indicates that no emission 
lines are visible and the faint continuum flux does not allow 
any template fitting; (7) the morphological type as estimated on the 
SDSS frames,using Visual Tools; (8) the group assignment based on 
observed redshift ('bkg' stands for galaxies, located in the
background of the target group); (9) remarks on the data quality, 
other name or redshift of the galaxy, internal (knotty) structure, etc.
}
\protect\label{tab:speresults}
\begin{tabular}{rllrrccll}
\hline
Nr  & Galaxy     & ID-SDSS             & v$_{\odot}$& $\epsilon_V$& spec & morph   & Group&  remarks\\
    &              &            & \multicolumn{2}{c}{[km s$^{-1}$]} &   &     &      &          \\
\hline
(1)  & (2) & (3) & (4) & (5) & (6) & (7) & (8) & (9) \\  %& (10) & (11) & (12) & (13) & (14)  & (15) \\\hline
\hline

1  & UGC 378           &  & 4493 &  22 & HII & Sc  & bkg & Z/Z$_\odot$ $\sim$ 0.3 ?\\ %J003758.99+481152.9
2  & NGC 278-kkh4     & J010549.63+453057.6 & 3339 &  48 & HII & Irr & bkg & very low SB \\
3  & NGC 278-kk6      &   & 4715 &  47 & HII &    & bkg & with UGC~378?\\ % J003728.59+481025
\hline
4  & IC 65-dw1         & &  568 &  11 & HII & dIrr & N278 &A1000+4734$^{1)}$ \\ %J010010.1+473400.9
5  & IC 65-dw2         &     & 2736 &  14 & HII & dIrr & IC65 &A1000+4756$^{1)}$ \\ % J010007.7+475605
%IC 65-dw2        & J010007.7+475605 &01:00:07.7&+47:56:05  &          &        &         &    &                  \\
6  & IC 65-dw3         &  & 2756 &  16 & HII & dIrr & IC65 &A0101+4744$^{1)}$ \\ % J010116.09+474432.9
7  & IC 65-dw4        &  &  565 &  17 & HII & dIrr & N278 &A0101+4752$^{1)}$ \\ % J010144.4+475205.9
8  & PGC 3574          &  &14090 & 100 & abs & S0   & bkg  & \\ % J005956.8+474709.9
9  & PGC 3684          &   & 5718 &  18 & HII/abs & E: & bkg & \\ % J012508.1+093626.9
\hline
10 & NGC 524SW-16   &J012508.08+093628.2  & 4618 & 100 & abs & S... & bkg & poor signal \\
11 & NGC 524SW-18   & J012451.77+090649.3 &   - & 300 & no HII & S...  &    &too faint \\
\hline
12 & NGC 691-E         & J015050.97+214553.2  &20959 &  30 & HII/abs & S0/a & bkg & a) \\
%\hline
13 & NGC 697-dw1       & J014952.94+222634.4 & 9834 &  44 & HII & Scd:  & bkg & b) \\
14 & NGC 697-dw2       & J014850.16+220650.9 & 2997 &  30 & HII & dSph: & N697 & bright knot in SE \\
15 & NGC 697-dw3       & J014748.40+220831.9 & 2940 &  28 & HII & S...  & N697 & \\
16 & NGC 697-A         & J015150.07+221831.3 & 2901 &  50 & HII & Scd  & N697 & bright knot in NW \\
17 & NGC 697-B         & J015144.23+221957.3 & 3314 &  50 & abs & S0-a & N697 & post-STB? \\
18 & NGC 697-C         & J015222.40+221007.0 & 2647 &  45 & HII & Irr  & N697 & 2010 01 06\\
                 & &                     & 2664 &  45 & HII &    & N697 & 2010 01 19\\
19 & NGC 697-C Back    & J015223.01+221010.2 &68598 &  46 & HII & S... & bkg & 2010 01 06\\
                 & &                     &68587 & 166 & HII &   & bkg & 2010 01 19\\
20 & NGC 697-C Back2   &      &252909&     & HII &    & bkg & 2010 01 19\\ % J015224+221021
21 & NGC 697-D         & J015027.14+215702.5 &26945 &  30 & HII & S0-a & bkg & \\
22 & NGC 697-F         & J015131.10+224540.2 &14032 &  56 & HII & S... & bkg & \\
23 & NGC 697-G         & J015020.28+222553.5 & 9621 &  83 & HII & S-Irr & bkg & b), bright eccentric patch \\
24 & NGC 697-J         & J014813.29+220233.6 &37520 & 114 & abs & E-S0 & bkg & \\
25 & NGC 697-H         & J015156.41+224142.1 & 9913 &   4 & HII & S-Irr & bkg & b) \\
%\hline
26 & NGC 772-gal1      & J015858.60+185740.2 &19905 & 230 & abs & E  & bkg & a), LEDA 212884 \\ % T=2.1 \\
27 & NGC 772-gal2      & J015907.21+185736.4 &19204 & 140 & abs & E-S0  & bkg & a) \\
\hline
28 & NGC 972-kdg17     &   & 1464 &  26 & HII &  & N972 & \\ % J023100.4+275725.6
\hline
29 & NGC 5033-15       & J131420.58+363407.0 & 1175 &  20 & HII & dIrr  & N5033 & DR9: cz=1737 \\
30 & NGC 5033-30N      & J131058.75+364943.8 & 1365 & 300 & abs & dIrr & N5033 
& \\
31 & NGC 5033-30S      & J131058.38+364811.3 &81716 & 150 & HII & S-Irr & bkg  & bright knot in SE\\
32 & NGC 5033NE-Adw    & J131704.86+375708.8 &  366 &  24 & HII & S0-a & field?  & PGC 046257\\
33 & NGC 5033WW-LSB    & J130419.75+362925.9 &  -   & 300 & no HII& Scd &     & too faint \\
34 & NGC 5033S-dw1     & J131422.85+354340.7 &  867 &  25 & HII & Irr & N5033 & 2nd bright knot in NE\\
35 & NGC 5033-23       & J131106.51+371041.1 &  -   & 300 & no HII& Irr &      & too faint, 3 more knots \\
%\hline
36 & NGC 5112-MAPS     & J132303.79+390939.3 & 2391 &  25 & HII & Irr  & bkg  & many knots \\
37 & NGC 5112-dw1     & J132212.86+384456.1 &  -   & 300 & no HII& S... &     & too faint\\
38 & NGC 5112-dw2     & J132157.87+385052.9 &  790 &  19 & HII & S-Irr & N5033 & \\
\hline
39 & NGC 6278-kkr31    & J165833.31+231223.0 & - & 300 & no HII& Irr &     & too faint and knotty\\
40 & NGC 6278-A        & J170131.42+224441.4 &10611 &  32 & HII & Sbc & bkg  & \\
41 & NGC 6278-5        & J170033.18+225157.5 &  -   & 300 & no HII& S... &  &  cz=2725(NED) \\ %z=0.00415(DR9); NED/DR4:cz=2725\\
42 & NGC 6278-C        & J170039.43+230308.4 &16983 &  20 & HII/AGN & Sbc & bkg & \\
43 & NGC 6278-C246     &                     &32359 &  19 & HII/AGN & & bkg & \\
\hline
\end{tabular}
  \end{table*}
\newpage
\begin{table*}
\begin{tabular}{lllrrccll}
{\bf Table~4,} continued \\
\hline
(1)  & (2) & (3) & (4) & (5) & (6) & (7) & (8) & (9) \\  
\hline

44 & NGC 6962-B10      & J204656.97+001238.4 & 7958 &  50 & HII & Sb-c & bkg & 2nd bright knot in SW\\
45 & NGC 6962-Ec1      & J205018.39+004519.6 &15526 &  68 & HII & S...  & bkg & \\
46 & NGC 6962-L02      & J204701.32+002656.2 &  -   &  60 & no HII& S0-a &  & too faint\\
47 & NGC 6962-L08      & J204804.85+003513.1 & 4616 &  50 & HII & Scd & N6962 & \\
48 & NGC 6962-L12      & J204722.3+002142    & 4456 & 300 & abs & S... & N6962 & no object in SDSS!\\
49 & NGC 6962-N02      & J204823.71+002300.5 & 3751 &  25 & HII & Sbc: & N6962 & 2nd bright knot in NW\\
50 & NGC 6962-N06      & J204829.07+000540.7 & 6638 &  50 & HII & S-Irr & bkg  & \\
51 & NGC 6962-N16      & J204632.08+003317.6 & 4156 &   6 & HII & S0-a & N6962 &  asym., several knots in NE\\
52 & NGC 6962-N19      & J204535.66+004332.0 &38723 & 200 & HII & S... & bkg & \\
53 & NGC 6962-N20      & J204649.90+003944.4 & 4358 &  24 & HII & S-Im & N6962 & DR9: z=0.01368 (STB)\\
54 & NGC 6962SE-14a    & J205055.83-001107.2 &60047 &  30 & HII/abs & Sab & bkg & \\
55 & NGC 6962Se-14b    & J205056.07-001108.5 &57214 &  61 & HII & S... & bkg & \\
\hline
%objects projected onto the slit
\end{tabular}
\\
$^{1)}$ galaxy name in Vennik \& Hopp~(2008) \\ %\cite{vh08}
a) NGC691-E, NGC772-gal1, and NGC772-gal2 seem to belong to a common background structure\\
b) NGC697-dw1, NGC697-G, and NGC697-H indicate  - together with 3 NED galaxies - a background group at 9794 km/s, see text for details. \\
\end{table*}

%============\\

\begin{table*}[t]
\caption{Photometric data.
Columns contain the following data: 
(1) galaxy seq. number in the Table~\ref{tab:speresults},
(2) luminosity distance computed with the NED cosmology 
calculator (Wright \cite{cosmo-calc}),
(3) the asymptotic $g'$-magnitude, 
(4) the absolute $g'$-magnitude, (5, 6) the $g'-r'$ and $r'-i'$ 
colour indices, 
(7, 8, 9) the central SB and the SB at the half-light radius, 
and the exponential model central SB in the $g'$ filter, respectively;
(10) the half-light radius in arcseconds and in kiloparsecs;
(11) the exponential model scale-length in arcseconds and in kiloparsecs.
All magnitudes were corrected for the Galactic extinction, as obtained 
from the SDSS database, and are K-corrected according to
Chilingarian et al.~(\cite{K-cor}). Surface brightnesses were corrected 
for cosmological SB dimming $(1+z)^4$. The magnitudes were not 
corrected for the inclination effects. The magnitude and colour 
errors correspond to the 1$\sigma$ sky errors.
}
\protect\label{tab:phot-gri-err}
\begin{center}
\begin{tabular}{lrlcllccccc}  
\hline
Nr & $D_{\mathrm {lum}}$ & $g'_{\mathrm T}$ & $M_{g'}$  & $g'-r'$ & $r'-i'$ & $\mu_{\mathrm {0},g'}$ & $\mu_{\mathrm {ef},g'}$ & $\mu_{\mathrm {0},g'}^{\mathrm {exp}}$ & $R_{\mathrm {ef}}$ & $h$ \\
 & [Mpc]  & [mag] & [mag] & [mag] & [mag] & \multicolumn{3}{c}{[mag arcsec$^{-2}$]} & \multicolumn{2}{c}{[arcsec(kpc)]} \\ 
\hline
(1) & (2) & (3) & (4) & (5) & (6) & (7) & (8) & (9) & (10) & (11) \\  
\hline
\\
 2 &  49.9 & 18.48$\pm$.15 & -15.01 & 0.05$\pm$0.23 & 0.15$\pm$.34 & 24.00 & 25.32 & 23.72 & 7.4(1.7) & 4.9(1.2) \\ %  2 N278-kkh4
12 & 312.7 & 17.49$\pm$0.12 & -19.98 & 0.49$\pm$0.18 & 0.26$\pm$0.21 & 21.88 & 23.25 & 21.85 & 5.5(7.3) & 3.6(4.8) \\% 13 N691-E
13 & 143.6 & 17.27$\pm$0.07 & -18.51 & 0.35$\pm$0.10 & 0.16$\pm$0.12 & 22.75 & 23.59 & 21.81 & 6.7(4.3) & 3.5(2.3) \\% 14 N697-dw1
14 &  44.0 & 17.25$\pm$0.09 & -15.97 & 0.44$\pm$0.11 & 0.26$\pm$0.11 & 22.49 & 24.46 & 22.92 & 8.8(1.8) & 6.1(1.3) \\% 15 N697_dw2
15 &  43.2 & 16.72$\pm$0.02 & -16.46 & 0.40$\pm$0.02 & 0.13$\pm$0.02 & 22.97 & 24.81 & 23.38 & 13.8(2.8) & 9.3(1.9) \\% 16 N697_dw3
16 &  42.6 & 18.45$\pm$0.08 & -14.70 & 0.45$\pm$0.09 & 0.14$\pm$0.11 & 22.98 & 24.19 & 22.67 & 4.6(0.9) & 3.1(0.6) \\% 17 N697_A
17 &  48.6 & 17.14$\pm$0.08 & -16.30 & 0.53$\pm$0.09 & 0.29$\pm$0.10 & 21.21 & 23.39 & 22.31 & 5.0(1.1) & 4.0(0.9) \\% 18 N697_B
18 &  39.0 & 18.21$\pm$0.07 & -14.74 & 0.37$\pm$0.10 & 0.18$\pm$0.10 & 22.92 & 23.92 && 4.6(0.8) &  \\% 19 N697-C
21E$^1$ & 407.3 & 18.15$\pm$0.05 & -19.90 & 0.30$\pm$0.07 & 0.36$\pm$0.07 & 21.71 & 22.91 & 21.29 & 3.2(5.3) & 2.2(3.6) \\% 22 N697-DE
21W$^2$ & 407.3 & 17.59$\pm$0.09 & -20.46 & 0.46$\pm$0.13 & 0.24$\pm$0.12 & 22.40 & 23.93 & 22.39 & 6.7(11.2) & 4.7(7.9) \\% 23 N697-DW
22 & 206.4 & 17.30$\pm$0.04 & -19.28 & 0.10$\pm$0.06 & 0.25$\pm$0.11 & 21.07 & 23.42 & 22.16 & 4.9(4.4) & 4.1(3.7) \\% 24 N697-F
22SE$^3$ & (206.4)&    19.33$\pm$0.07 & (-17.2)  & -0.26:  & 0.37:  & 22.39   & 23.72  & 22.08 & 2.2(2.0) & 1.5(1.3) \\% 25 N697-FSE
22SW$^4$ & (206.4)&    20.10$\pm$0.07 & (-16.5)  & 0.56$\pm$0.10 & 0.49$\pm$0.12 & 22.97  & 23.82 & 22.48 & 1.7(1.6) & 1.3(1.2) \\%26 N697-FSW
23 & 140.4 & 18.43$\pm$0.07 & -17.31 & 0.30$\pm$0.09 & 0.20$\pm$0.09 & 22.60 & 23.34 & 21.92 & 3.3(2.1) & 2.2(1.4) \\% 27 N697-G
24 & 580.4 & 17.42$\pm$0.05 & -21.40 & 0.50$\pm$0.07 & 0.28$\pm$0.06 & 20.87 & 23.02 & 21.22 & 4.9(11.0) & 3.0(6.6) \\% 28 N697-J
25 & 144.7 & 18.20$\pm$0.05 & -17.60 & 0.21$\pm$0.06 & 0.16$\pm$0.09 & 22.61 & 23.74 & 22.23 & 4.4(2.9) & 2.9(1.9) \\% 29 N697-H
26 & 296.1 & 15.39$\pm$0.08 & -21.97 & 0.81$\pm$0.11 & 0.44$\pm$0.09 & 18.16 & 22.35 && 5.7(7.2) &           \\% 30 N772-gal1
27 & 285.2 & 16.20$\pm$0.14 & -21.08 & 0.74$\pm$0.18 & 0.37$\pm$0.17 & 18.86 & 23.04 && 5.5(6.7) &          \\%   * 31 N772-gal2
29 &  17.3 & 16.82$\pm$0.01 & -14.37 & 0.29$\pm$0.01 & 0.14$\pm$0.01 & 23.24 & 24.52 & 23.02 & 11.2(0.9) & 7.7(0.6) \\% 33 N5033_15dw1
30 &  20.0 & 17.77$\pm$0.07  & -13.71  & 0.36$\pm$0.10 & 0.32$\pm$0.12 & 23.96  & 25.79 & 24.00 & 11.8(1.1) & 7.2(0.7) \\% 34 N5033-30dw1N
31 & 1379.3& 18.61$\pm$0.16 & -22.09 & 0.41$\pm$0.20 & -0.06$\pm$0.23 & 22.85 & 24.35 & 22.77 & 7.0(29.0) & 4.7(19.4) \\% 35 N5033-30dw1S
32 &   5.9 & 15.17$\pm$0.01 & -13.68 & 0.35$\pm$0.01 & 0.17$\pm$0.01 & 21.45 & 22.95 & 21.87 & 10.3(0.3) & 7.8(0.2) \\% 36 N5033NE-Adw
34 &  12.9 & 18.62$\pm$0.08 & -11.93 & 0.43$\pm$0.10 & 0.02$\pm$0.11 & 22.40 & 24.79 & 23.47 & 5.2(0.3) & 4.4(0.3) \\% 38 N5033Sdw1
36 &  34.7 & 17.17$\pm$0.02 & -15.53 & 0.13$\pm$0.03 & 0.04   & 22.74 & 24.04 & 22.40 & 7.7(1.3) & 4.6(0.8) \\% 40 N5112-MAPS
38 &  12.0 & 18.61$\pm$0.08 & -11.78 & 0.32$\pm$0.13 & 0.03$\pm$0.28 & 23.15 & 24.90 & 23.39 & 6.2(0.4) & 4.0(0.2) \\% 42 N5112dw2
40 & 155.6 & 18.62$\pm$0.10 & -17.34 & 0.35$\pm$0.13 & 0.17$\pm$0.20 & 21.29 & 23.09 & 21.24 & 2.4(1.7) & 1.4(1.0) \\% 44 WLB629-A
42 & 251.8 & 18.36$\pm$0.03 & -18.64 & 0.26$\pm$0.04 & 0.24$\pm$0.03 & 21.04 & 22.10 & 20.25 & 2.0(2.2) & 1.1(1.2) \\% 46 WLB629-C
44 & 116.7 & 19.10$\pm$0.05 & -16.24 & 0.17$\pm$0.07 & -0.03$\pm$0.08 & 22.54 & 23.49 & 20.98 & 2.6(1.4) & 1.2(0.6) \\% 48 B10_str82
45 & 229.8 & 19.40$\pm$0.04 & -17.41 & 0.18$\pm$0.08 & 0.04$\pm$0.20 & 23.09 & 23.94 & 21.72 & 2.9(2.9) & 1.4(1.4) \\% 49 Ec1_str82
47 &  68.0 & 18.45$\pm$0.02 & -15.71 & 0.44$\pm$0.08 & 0.27$\pm$0.16 & 23.15 & 24.45 & 22.98 & 4.8(1.5) & 3.1(1.0) \\% 51 L08_str82
48 &  65.7 & 17.81$\pm$0.07 & -16.28 & 0.44$\pm$0.10 & 0.27$\pm$0.12 & 22.55 & 24.19 & 22.41 & 5.6(1.7) & 3.5(1.1) \\% 52 L12_str82
49 &  55.6 & 15.91$\pm$0.11 & -17.82 & 0.49$\pm$0.15 & 0.26$\pm$0.15 & 21.26 & 22.88 & 21.12 & 7.3(1.9) & 4.5(1.2) \\% 53 N02_tihe
50 &  97.3 & 18.10$\pm$0.11 & -16.84 & 0.48$\pm$0.15 & 0.31$\pm$0.17 & 22.69 & 24.21 & 22.95 & 5.0(2.3) & 3.4(1.5) \\% 55 N06_str82
51 &  61.4 & 15.58$\pm$0.01 & -18.36 & 0.54$\pm$0.01 & 0.29$\pm$0.01 & 20.13 & 21.89 & 21.92 & 5.1(1.5) & 5.1(1.5) \\% 56 N16_str82
52 & 601.4 & 19.04$\pm$0.05 & -19.85 & 0.38$\pm$0.08 & 0.29$\pm$0.10 & 22.81 & 24.21 & 22.22 & 4.5(10.3) & 2.2(5.0) \\% 57 N6962-N19
53 &  64.3 & 18.06$\pm$0.08 & -15.98 & 0.42$\pm$0.13 & 0.16$\pm$0.15 & 22.69 & 23.86 & 21.67 & 4.7(1.4) & 2.3(0.7) \\% 58 N20_str82
54 & 974.2 & 19.07$\pm$0.09 & -20.87 & 0.50$\pm$0.10 & 0.29$\pm$0.09 & 21.59 & 22.87 & 20.95 & 2.6(8.7) & 1.3(4.4) \\% 59 N6962-SE14a
55 & 923.1 & 20.42$\pm$0.03 & -19.40 & 0.18$\pm$0.04 & 0.29$\pm$0.04 & 21.09 & 22.26 & 20.29 & 1.0(3.1) & 0.5(1.7) \\% 60 N6962-SE14b
\hline
\end{tabular}
\end{center}
$^{1}$ Eastern part (knot) of the NGC 697-D, J015027.14+215702.5;~~~\\
$^{2}$ Western part (knot) of the NGC 697-D, J015026.32+215706.0 \\
$^{3}$ LSB object to the SE of NGC 697-F, J015131.80+224535.1 (redshift uncertain);~~~~~  \\
$^{4}$ LSB object to the SW of NGC 697-F, J015130.89+224534.0 (redshift uncertain)\\
\end{table*}

%===========\\
\begin{table*}[h]
\caption{Photometric characteristics of studied galaxies. 
Columns contain the following data: (1) sample name, 
(2) number of galaxies, (3) luminosity distance, 
(4) absolute $g'$-magnitude,
(5) total ($g'-r'$) colour index, (6) total ($r'-i'$) colour index, 
(7) exponential model central SB in $g'$ filter, (8) mean SB within
the effective radius in $g'$ filter, (9) exponential model scale length 
in $g'$ filter, (12) effective radius in $g'$ filter.
The magnitudes are in AB-system and are corrected for Galactic 
extinction and are K-corrected. Mean values are in the first row,
medians are in the second row.
}
\protect\label{tab:HET-gri}
\begin{center}
\begin{small}
\begin{tabular}{lccccccccc}
\hline
sample & $n_\mathrm{gal}$ & $dist$ & $M_{g'}$ & $g'-r'$ & $r'-i'$ & $\mu_{\mathrm{0},g'}$ & $<\mu_{\mathrm{ef},g'}>$ &  $h$ &  $R_\mathrm{ef}$ \\  %& $R_{ef}$ \\
           &  & [Mpc]  & [mag] & [mag]   & [mag]     & \multicolumn{2}{c}{[mag~arcsec$^{-2}$]} & [kpc] & [kpc] \\ %\multicolumn{2}{c}{[kpc]} \\
\hline
(1) & (2) & (3) & (4) & (5) & (6) & (7) & (8) & (9) & (10) \\ %& (11) & (12) \\
\hline
\\
group-DGs & 18 & 38$\pm$21 & -15.3$\pm$1.9 & 0.42$\pm$.07 & 0.19$\pm$.09 & 22.61$\pm$.60 & 23.39$\pm$.98 & 0.79$\pm$.50 & 1.16$\pm$.69 \\
 (medians) &  & 42  & -15.3 & 0.43 & 0.17 & 22.55 & 23.54 & 0.7 & 1.2 \\
field-LTGs & 23 & 498$\pm$964 & -18.9$\pm$2.1 & 0.35$\pm$.20 & 0.21$\pm$.13 & 21.81$\pm$.89 & 23.14$\pm$.92 & 3.7$\pm$4.2 & 5.9$\pm$5.9 \\
 (medians)  &  & 208 & -19.3 & 0.35 & 0.24 & 21.85 & 23.18 & 1.9  & 3.9 \\
%\\
\hline
\end{tabular}
\end{small}
\end{center}
\end{table*}

\begin{table*}[h]
\caption{ Photometric characteristics of studied galaxies (group DGs and 
field LTGs) compared to those of the Local Volume
DGs (LV-DGs) and those of the LTGs of the WHISP sample (WHISP-LTGs). 
Mean values are in the first row, medians are in the
second row.}
\protect\label{tab:HET-LV-WHISP}
\begin{center}
\begin{small}
\begin{tabular}{lcccccccc}
\hline
sample & $n_{gal}$ & $dist$ & $M_B$ & $B-R_c$ &  $\mu_{0,R_c}$ &  $<\mu_{ef,R_c}>$ & $h$ &$R_{ef}$ \\  %& $R_{ef}$ \\
           &  & [Mpc]  & [mag] & [mag] &  \multicolumn{2}{c}{[mag~arcsec$^{-2}$]} & [kpc] & [kpc] \\ %\multicolumn{2}{c}{[kpc]} \\
\hline
\\
group DGs & 18 & 38$\pm$21 & -14.6$\pm$1.7 & 0.94$\pm$.15 & 22.00$\pm$.64 & 22.76$\pm$1.04 & 0.79$\pm$.50 & 1.16$\pm$.69 \\
 (medians) &  & 42  & -14.9 & 0.98 &  22.16 & 23.00 & 0.7 & 1.2 \\
field LTGs & 23 & 498$\pm$964 & -18.7$\pm$2.1 & 0.92$\pm$.32 & 21.14$\pm$.89 & 22.60$\pm$1.03 & 3.7$\pm$4.2 & 5.9$\pm$5.9 \\
 (medians)  &  & 208 & -19.3 & 0.89 &  21.15  & 22.56 & 1.9  & 3.9 \\
LV DGs     & 98 & 5.8$\pm$2.3 & -14.2$\pm$1.6 & 0.91$\pm$.23 & 21.53$\pm$1.10  & 22.68$\pm$1.07 &0.45$\pm$.26 & 0.76$\pm$.43 \\
 (medians) &  & 5.4 & -14.2 & 0.90  & 21.56 &  22.67 & 0.4  & 0.7 \\
WHISP LTGs & 173& 12.1$\pm$9.0 & -16.2$\pm$1.5 & 1.02$\pm$.57 & 21.39$\pm$1.12  & 22.59$\pm$1.15 & 1.65$\pm$1.16& 2.11$\pm$1.53 \\
 (medians) & & 9.0 & -16.4 & 0.98 &       21.49  & 22.79 & 1.4 & 1.7 \\
\hline
\end{tabular}
\end{small}
\end{center}
\end{table*}


\begin{thebibliography}{}

\bibitem[2000]{AReese00}
Avila-Reese, V., \& Firmani, C. 2000, RMxAA, 36, 23
\bibitem[2006]{barazza06}
Barazza, F.D., Shardha, J., Rix, H.-W., Barden, E., Bell, E.F., Caldwell, J.A.R., McIntosh, D.H., Meisenheimer, K., 
 Peng, C.Y., \& Wolf, C. 2006, ApJ, 643, 162 
\bibitem[1992]{bh92}
Barnes, J. E., \& Hernquist, L. 1992, ARA\&A, 30, 705
\bibitem[2007]{belokurov07}
Belokurov, V., Zucker, D.B., Evans, N.W., etal. 2007, ApJ, 654, 897
\bibitem[1996]{Sxtr}
Bertin, E., \& Arnouts, A. 1996, A\&A, 117, 393
\bibitem[1994]{bing94}
Binggeli, B. 1994, in Panchromatic View of Galaxies. Their Evolutionary Puzzle, eds. G. Hensler, C. Theis, \& 
J.S. Gallagher, Gif-sur-Yvette:Edition Frontiers, 173 
\bibitem[2005]{VAGC}
Blanton, M.R., Schlegel, D.J., Strauss, M.A., et al. 2005, AJ, 129, 2562
\bibitem[2007]{BRI-Vega}
Blanton, M.R., \& Roweis, S. 2007, AJ, 133, 734.
\bibitem[1998]{bremnes98}
Bremnes, T., Binggeli, B., \& Prugniel, P. 1998, A\&AS, 129, 313
\bibitem[1999]{bremnes99}
Bremnes, T., Binggeli, B., \& Prugniel, P. 1999, A\&AS, 137, 337
\bibitem[2000]{bremnes00}
Bremnes, T., Binggeli, B., \& Prugniel, P. 2000, A\&AS, 141, 211
\bibitem[2013]{bretherton13}
Bretherton, C.F., Moss, C., \& James, P.A. 2013, A\&A, 553, 67
\bibitem[1985]{bret85}
Bretschinger, E. 1985, ApJS, 58, 39
\bibitem[1997]{carl97}
Carlberg, R.G., Yee, H.K.C., \& Ellington, E. 1997, ApJ, 478, 462 
\bibitem[2001]{carrasco01}
Carrasco, E.R., Mendes de Oliveira, C., Infante, L., \& Bolte, M. 2001, AJ, 121, 148
\bibitem[2012]{chernin12}
Chernin, A.D., Teerikorpi, P., Valtonen, M.J., Dolgachev, V.P., Domozhilova, L.M., \& Byrd, G.G. 2012, A\&A, 539, A4
\bibitem[2009]{chib09}
Chiboucas, K., Karachentsev, I.D., \& Tully, R.B. 2009, AJ, 137, 3009  % New DGs in M81 group
\bibitem[2010]{K-cor}
Chilingarian, I.V., Melchior, A.L., \& Zolotukhin, I.Yu. 2010, MNRAS, 405, 1409
\bibitem[2006]{A1367}
Cortese, L., Gavazzi, G., Boselli, A., Franzetti, P., Kennicutt, R.C., O'Neil, K., \& Sakai, S. 2006, A\&A, 453, 847
\bibitem[2009]{cote09}
Cote, S., Draginda, A., Skillman, E.D., \& Miller, B.W. 2009, AJ, 138, 1037  % SF in Cen A group DGs
\bibitem[2009]{crn09} 
Crnojevic, D., Grebel, E.K., \& Koch, A. 2009, AN, 330, 1001    % DGs in Cen A group
\bibitem[2004]{cross04}
Cross, N.J.G., Driver, S.P., Liske, J. et al. 2004, MNRAS, 349, 576 
\bibitem[1997]{dalcanton97}
Dalcanton, J.J, Spergel, D., \& Summers, F. 1997, ApJ, 482, 659
\bibitem[1986]{dekel86}
Dekel, A., \& Silk, J. 1986, ApJ,303, 39
\bibitem[2010]{dong2008}
D'Onghia, E., \& Lake, G. 2008, ApJ, 686, L1
\bibitem[2009]{ez09}
Eigenthaler, P., \& Zeilinger, W.W. 2010, A\&A, 511, 12E 
\bibitem[1974]{je1974}
Einasto, J., Saar, E., \& Chernin, A.D. 1974, Nature, 252, 111
\bibitem[1975]{je1975}
Einasto, J., Kaasik, A., Kalamees, P., \& Vennik, J. 1975, A\&A, 40, 161
\bibitem[1994]{fair94}
Fairall, A.P., Paverd, W.R., \& Ashley, R.P. 1994, in "Unveiling Large-Scale Structures Behind the Milky Way", 
eds. C. Balkowski \& R.C. Kraan-Korteweg, ASP Conf. Ser. 67, 21
\bibitem[2008]{fardal08}
Fardal, M. A., Babul, A., Guhathakurta, P., Gilbert, K. M., \& Dodge, C. 2008, ApJ, 682, L33
\bibitem[2010]{fathi10}
Fathi, K., Allen, M., Boch, T., Hatziminaoglou, E., \& Peletier, R.F. 2010, MNRAS, 406, 1595
\bibitem[2013]{fossati13}
Fossati, M., Gavazzi, G., Savorgnan, G., Furnagalli,M., Boselli, A., Gutierrez, L., Hernandez Toledo, H., 
Giovanelli, R., \& Haynes, M.P. 2013, A\&A, 553, 91
\bibitem[1970]{freeman70}
Freeman, K. 1970, ApJ, 160, 81
\bibitem[1995]{fukugita95}
Fukugita, M., Shimasaku, K., \& Ichikawa, T. 1995, PASP, 107, 945
\bibitem[1996]{fukugita96}
Fukugita, M., Ichikawa, T., Gunn, J.E., Doi, M., Shimasaku, K., \& Schneider, D.P. 1996, AJ, 111, 1748 
\bibitem[2010]{gal2010}
Gallagher, J.S. 2010, ASPC, 423, 3
\bibitem[1993]{LGG}
Garcia, A.M. 1993, A\&AS, 100, 47
\bibitem[2006]{geha06}
Geha, M., Blanton, M.R., Masjedi, M., \& West, A.A. 2006, ApJ, 653, 240
\bibitem[2003]{grebel03}
Grebel, E.K., Gallagher, III J.S., \& Harbeck, D. 2003, AJ, 125, 1926
\bibitem[1988]{midas1}
Grosbol, P., Banse, K., Guirao, C., Ponz, J.D., \& Warmels, R.H. 1988, ESO Messenger, ESO, Garching, 54 , 59 
\bibitem[2005]{gruetz05} 
Gr\"utzbauch, R., Kelm, B., Focardi, P., Trinchieri, G., Rampazzo, R., \& Zeilinger, W.W. 2005, AJ, 129, 1832 % N4756 group
\bibitem[2011]{hart11}
Hartwick, F.D.A. 2011, AJ, 141, 198
\bibitem[1986]{PPS}
Haynes, M.P., \& Giovanelli, R. 1986, ApJ, 306L, 55
\bibitem[2008]{henriques08}
Henriques, B.M., Bertone, S., \& Thomas, P.A. 2008, MNRAS, 383, 1649
\bibitem[1999]{henriksen99}
Henriksen, M., \& Cousineau, S. 1999, ApJ, 511, 595
\bibitem[1998]{LRS}
Hill, G.J., Nicklas, H.E., MacQueen, P.J., Tejada, C., Cobos, D., Francisco, J., \& Mitsch, W. 1998, SPIE, 3355, 375
\bibitem[1986]{horne1986}
Horne, K. 1986, PASP, 98, 609
\bibitem[2004]{he04}
Hunter, D.A. \& Elmegreen, B.G. 2004, AJ, 128, 2170
\bibitem[2001]{ibata01a}
Ibata, R., Irwin, M., Lewis, G.F., \& Stolte, A. 2001, ApJ, 547, L133
\bibitem[2001]{ibata01b}
Ibata, R., Irwin, M., Lewis, G., Ferguson, A.M.N., \& Tanvir, N. 2001, Nature, 412, 49
\bibitem[2000]{jerjen00}
Jerjen, H., Binggeli, B., \& Freeman, K.C. 2000, AJ, 119, 593
\bibitem[1995]{johnston95}
Johnston, K.V., Spergel, D.N., \& Hernquist, L. 1995, ApJ, 451, 598
\bibitem[1999]{kara99}
Karachentseva, V.E., Karachentsev, I.D., \& Richter, G.M. 1999, A\&AS, 135, 221
\bibitem[1999]{ikar99}
Karachentsev, I.D., Makarov, D.I., \& Huchtmeier, W.K. 1999, A\&AS, 139, 97
\bibitem[2002]{kara02}
Karachentsev, I.D., Sharina, M.E., Makarov, D.I., et al. 2002, A\&A, 389, 812
\bibitem[2002]{kara13}  
Karachentsev, I.D., Nasonova, O.G., \& Courtois, H.M. 2013, MNRAS, 429, 2264
\bibitem[2008]{kew08}
Kewley, L. J., \& Ellison, S. L., 2008, ApJ 681, 1183
\bibitem[1996]{kc96}
Kinney, A.L., Calzetti, D., Bohlin, R.C., McQuade, K., Storchi-Bergmann, T.
 \& Schmitt, H.R., 1996, ApJ, 467, 38
\bibitem[1999]{klypin99}
Klypin, A., Kravtsov, A.V., Valenzuela, O., \& Prada, F. 1999, ApJ, 522, 82
\bibitem[2004]{kniazev04}
Kniazev, A.Y., Grebel, E.K., Pustilnik, S.A., Pramskij, T.F., Kniazeva, T.F., Prada, F., \& Harbeck, D. 2004, AJ, 127, 704 
\bibitem[2009]{galev}
Kotulla, R., Fritze, U., Weilbacher, P. \& Anders, P. 2009, MNRAS, 396, 462
\bibitem[1999]{lake99}
Lake, G., \& Moore, B. 1999, in Proc. of the IAU Symp. 186: Galaxy Interactions at Low and
 High Redshift, eds. J.E. Barnes, \& D.B. Sanders, 393
\bibitem[2012]{lisker12}
Lisker, T. 2012, AN, 333, 405
\bibitem[2013]{lisker13}
Lisker, T., Weinmann, S.M., Janz, J., \& Meyer, H.T. 2013, MNRAS, 432, 1162
\bibitem[1993]{afi}
Lorenz, H., Richter, G.M., Capaccioli, M., \& Longo, G. 1993, A\&A, 277, 321
\bibitem[2005]{mah05}
Mahdavi, A., Trentham, N., \& Tully, R.B. 2005, AJ, 130, 1502  
\bibitem[1993]{mamon93}
Mamon, G. A. 1993, in: Gravitational Dynamics and the N-body Problem, eds. F. Combes \&
 E. Athanassoula, Obs. de Paris, 188 [arXiv:astro-ph/9308032]
\bibitem[2007]{mamon07}
Mamon, G. A. 2007, in Proc. of the ESO workshop: Groups of Galaxies
 in the Nearby Universe, eds. I. Saviane, V. Ivanov, \& J. Borissova, ESO Astrophysics
 Symposia, 203 [arXiv:astro-ph/0607482]
\bibitem[1990]{mass1990}
Massey, P., Gronwall, C., 1990, ApJ, 358, 344
\bibitem[2000]{mayer00}
Mayer, L., Governato, F., \& Colpi, M. 2000, ASP Conf. Ser., 215, 42
\bibitem[1998]{moore98}
Moore, B., Lake, G., \& Katz, N. 1998, ApJ, 495, 139
\bibitem[2003]{MDMB}
Mulchaey, J.S., Davis, D.S., Mushotzky, R.F., \& Burstein, D. 2003, ApJS, 145, 39
\bibitem[1990]{oke1990}
Oke, J.B., 1990, AJ, 99, 1621
\bibitem[2004]{GEMS}
Osmond, J.P.F. \& Ponman, T.J. 2004, MNRAS, 350, 1511
\bibitem[2007]{panter07}
Panter, B., Jimenez, R., Heavens, A.F., \& Charlot, C. 2007, MNRAS, 378, 1550 
\bibitem[2002]{parodi02}
Parodi B.R., Barazza F.D., Binggeli B., 2002, A\&A, 388, 29 
\bibitem[1996]{PatThuan96}
Patterson, R.J., \& Thuan, T.X. 1996, ApJS, 107, 103 
\bibitem[2002]{pope99}
Popescu, C. C., Hopp, U., \& Rosa, M. R., 1999, A\&A, 350, 414
\bibitem[2002]{UZC}
Ramella, M., Geller, M.J., Pisani, A., \& da Costa, L.N. 2002, AJ, 123, 2976
\bibitem[1998]{HET}
Ramsey, L.W., Adams, M.T., Barnes, T.G., Booth, J.A., et al. 1998, SPIE, 3352, 34 
\bibitem[2003]{rines03}
Rines, K., Geller, M.J., Kurtz, M.J., \& Diaferio, A. 2003, AJ, 126, 215
\bibitem[1986]{sandage86}
 Sandage, A. 1986, ApJ, 307, 1
\bibitem[2008]{sharina08}
Sharina M.E., Karachentsev I.D., Dolphin A.E., et al. 2008, MNRAS, 384, 1544
\bibitem[2007]{shetrone07}
Shetrone, M., Cornell, M.E., Fowler, J.R., et al. 2007, PASP, 119, 556
\bibitem[1993]{Skillman1993}
Skillman, E.D., Kennicutt, R.C., 1993 ApJ, 411, 655
\bibitem[2010]{smith2010}
Smith, B., Bastian, N., Higdon, S.J.U., \& Higdon J.H. (eds.):
'Galaxy Wars: Stellar Populations and Star Formation in Interacting
Galaxies', ASP Conference Series Vol. 423, proceedings of a conference
held 19-22 July 2009 at East Tennessee State University, Johnson City,
Tennessee, USA, San Francisco, Astronomical Society of the Pacific, 2010
\bibitem[2002]{smith02}
Smith, J.A., Tucker, D.L., Kent, S., et al. 2002 ApJ, 123, 2121
\bibitem[2002]{WHISP}
Swaters, R.A., \& Balcells, M. 2002, A\&A, 390, 863
\bibitem[2008]{DR5groups}
Tago, E., Einasto, J., Saar, E., Tempel, E., Einasto, M., Vennik, J., \& M\"uller, V. 2008, A\&A, 479, 927
\bibitem[2010]{tanaka10}
Tanaka, M., Chiba, M., Komiyama, Y., Guhathakurta, P., Kalirai, J.S., \& Iye, M. 2010, ApJ, 708, 1168
\bibitem[2008]{teeri08}
Teerikorpi, P., Chernin, A.D., Karachentsev, I.D., \& Valtonen, M.J., 2008, A\&A, 483, 383
\bibitem[2009]{tolstoy}
Tolstoy, E. \& Hill, V. \& Tosi, M. 2009, ARAA, 47, 371
\bibitem[1972]{toomre1972}
Toomre, A., \& Toomre, J. 1972, ApJ, 178, 623
\bibitem[2001]{trent01}
Trentham, N., Tully, R.B., \& Verheijen, M.A.W. 2001, MNRAS, 325,385
\bibitem[2009]{tt09a} 
Trentham, N., Tully, R.B. 2009 MNRAS, 398, 722    
\bibitem[1982]{LScl}
Tully, R.B. 1982, ApJ, 257, 389 
\bibitem[1987]{LVoid}
Tully, R.B., \& Fisher, J.R. 1987, Nearby Galaxies Atlas, (Cambridge Univ. Press 1987)  
\bibitem[1988]{NBG}
Tully, R.B. 1988, Nearby Galaxies Catalogue, (Cambridge University Press 1988)
\bibitem[2009]{tt08}
Tully, R.B., \& Trentham, N. 2008 AJ, 135, 1488    
\bibitem[1994]{vader94}
Vader, J.P., \& Chaboyer, B. 1994, AJ, 108, 1209
\bibitem[2001]{zee01}
van Zee, L. 2001, AJ, 121, 2003
\bibitem[1987]{veilost87}
Veilleux, S., \& Osterbrock, D. E., 1987, ApJS, 63, 295
\bibitem[1984]{jv84}
Vennik, J. 1984, TarTO, 73, 1  
\bibitem[1996]{v96}
Vennik, J., Richter, G.M., Th\"anert, W., \& Biering, C. 1996, AN, 317, 289 
\bibitem[2007]{vt07} 
Vennik, J., \& Tago, E. 2007, in Groups of Galaxies in the Nearby Universe, ed. I. Saviane, V. Ivanov, \& J. Borissova, ESO Astrophysics Symposia, 119
\bibitem[2008]{vh08} 
Vennik, J., \& Hopp, U. 2008, A\&A, 481, 79
\bibitem[1991]{midas2}
Warmels, R.H. 1991, The ESO-MIDAS System, in Astronomical Data Analysis Software and Systems I , PASP Conf. Series, Vol. 25, p. 115. 
\bibitem[2013]{wetzel13}
Wetzel, A.R., Tinker, J.L., Conroy, C., \& Bosch, van den F.C. 2013, MNRAS, 439, 2687
\bibitem[1997]{weinberg97}
Weinberg, M.D. 1997, ApJ, 478, 435
\bibitem[1999]{WBL}
White, R.A., Bliton, M., Bhavsar, S.P., Bornmann, P., Burns, J.O., Ledlow, M.J., \& Loken, C. AJ, 118, 2014
\bibitem[2007]{whiting07}
Whiting, A.B., Hau, G.K.T., Irwin, M., \& Verdugo, M. 2007, AJ, 133, 715
\bibitem[2006]{cosmo-calc}
Wright, E.L. 2006, PASP, 118, 1711
\end{thebibliography}
\end{document}